\documentclass[aps,pre,twocolumn,superscriptaddress,floatfix,10pt]{revtex4}
\usepackage{graphicx}
\usepackage{dcolumn}
\usepackage{bm}
\usepackage{latexsym}
\usepackage{amsmath}
\usepackage{amssymb}
\usepackage{color}
\usepackage{soul}
\usepackage{ulem}
\begin{document}
\title{{\bf Flagellar length control in biflagellate eukaryotes:\\ time-of-flight, shared pool, train traffic and cooperative phenomena}}
\author{Swayamshree Patra} 
\affiliation{Department of Physics, Indian
  Institute of Technology Kanpur, 208016, India} 
\author{Frank J\"ulicher{\footnote{E-mail: julicher@pks.mpg.de}}}
\affiliation{Max-Planck Institute for the Physics of Complex Systems, N\"othnitzer Strasse 38, 01187 Dresden, Germany}
\author{Debashish Chowdhury{\footnote{E-mail: debch@iitk.ac.in}}}
\affiliation{Department of Physics, Indian Institute of Technology
  Kanpur, 208016, India}
\affiliation{Max-Planck Institute for the Physics of Complex Systems, N\"othnitzer Strasse 38, 01187 Dresden, Germany}

\begin{abstract}

Flagella of eukaryotic cells are transient long cylindrical protrusions. The proteins needed to form and maintain flagella are synthesized in the cell body and transported to the distal tips. What `rulers' or `timers' a specific type of cells use to strike a balance between the outward and inward transport of materials so as to maintain a particular length of its flagella in the steady state is one of the open questions in cellular self-organization. Even more curious is how the two flagella of biflagellates, like {\it Chlamydomonas Reinhardtii}, communicate through their base to coordinate their lengths. In this paper we develop a stochastic model for flagellar length control based on a time-of-flight (ToF) mechanism. This ToF mechanism decides whether or not structural proteins are to be loaded onto an intraflagellar transport (IFT) train just before it begins its motorized journey from the base to the tip of the flagellum. Because of the ongoing turnover, the structural proteins released from the flagellar tip are transported back to the cell body also by IFT trains. We represent the traffic of IFT trains as a totally asymmetric simple exclusion process (TASEP). The ToF mechanism for each flagellum, together with the TASEP-based description of the IFT trains, combined with a scenario of sharing of a common pool of flagellar structural proteins in biflagellates, can account for all key features of experimentally known phenomena. These include ciliogenesis, resorption, deflagellation as well as regeneration after selective amputation of one of the two flagella. We also show that the experimental observations of Ishikawa and Marshall are consistent with the ToF mechanism of length control if the effects of the mutual exclusion of the IFT trains captured by the TASEP are taken into account. Moreover, we make new predictions on the flagellar length fluctuations and the role of the common pool. 

\end{abstract}

\maketitle

\section{INTRODUCTION}

In a classic article, titled ``on being the right size'', J.B.S. Haldane \cite{haldane25} first analyzed the physical reasons that explain why ``for every type of animal there is a convenient size''. He focused his analysis on the size of whole organisms. However, the mechanisms that ensure the ``convenient'' size of a cell \cite{marshall16} and sub-cellular structures \cite{marshall02,marshall15,goehring12} have become a very active field of research in recent years. Membrane-bound organelles are prominent among the sub-cellular structures. Flagella of eukaryotic cells (not to be confused with bacterial flagella), which are the organelles of our interest in this paper, appear as long cell protrusions \cite{ginger08} (key features of its structure are summarized in the next section). The short eukaryotic flagella are  often referred to as cilia. In this paper, we'll use the terms ``flagellum'' and ``cilium'' interchangeably. 

From the perspective of organelles size control,  what makes flagella very interesting is not only the one-dimensional nature of the problem but also their highly dynamic lengths. The lengths of flagella change with time in sync with the cell cycle \cite{cavaliersmith74,cross15,quarmby05}.  Even when their growth is complete, flagellar structure remains highly dynamic because each of the flagella continue to incorporate new proteins to make up for the high ongoing turnover,  thereby maintaining a steady balance of the elongation and shortening \cite{marshall01,mirvis18}. So, the first challenging question is how a specific cell maintains this balance at a particular length of a flagellum.  

The number of flagella vary from one species to another. Flagellar length control in biflagellated and multiflagellated cells are more interesting than that in monoflagellates. So, in the context of flagellar length control, the second challenging questions is how  biflagellated and multiflagellated cells coordinate the dynamics of their different flagella. For the sake of simplicity, in this paper, we consider only biflagellated eukaryotes for which the green algae {\it Chlamydomonas reinhardtii} (CR) serves as the most popular model organism \cite{jeanneret16,wemmer07}. 

CR is an interesting organism for studying flagellar length control because CR can loose its flagella in three distinct ways: resorption, deflagellation and selective amputation. Even more interesting is the fact that the CR can successfully regenerate its flagella as well. 
Both the flagella of a CR are gradually retracted into the cell prior to the cell division \cite{bloodgood74}; this phenomenon is usually referred to as ``{\it resorption}''. The flagellar components disassembled during resorption are returned to the cell body \cite{quarmby05}. Flagellar disassembly \cite{liang16} via resorption should be distinguished from ``{\it deflagellation}'' (also known as flagellar excision, flagellar shedding or flagellar autotomy) \cite{quarmby04}.  In the latter process, in response to wide varieties of stimuli, the axoneme is severed resulting in a complete loss of the flagellar components. Deflagellated CR cells can regenerate their flagella when stress causing stimulus disappears \cite{johnson93,engel09}. One of the flagella, or a distal part of it, can be selectively amputated in controlled experiment. All the proteins constituting the severed part of the amputated flagellum are lost by the cell. The regeneration of the amputated flagellum and the concomitant {\it nonmonotonic} variation of the length of its unsevered partner display most vividly the cooperation of the dynamics of the two flagella. The model we develop here describes resorption, deflagellation as well as regeneration of flagella within a single theoretical framework.

Proteins are synthesized in the cell body, and not in the flagella. Therefore, the flagellar structural proteins are transported from the base to the tip of each flagellum by intraflagellar transport (IFT) \cite{kozminski93,kozminski12,rosenbaum02}. Similarly, structural constituents of flagella that turn over are transported back to the cell body. IFT particles, which are multi-protein complexes at the core of the IFT machinery, operate essentially as the ``protein shuttles'' \cite{lechtreck17} (Further details of IFT are given in the next section). 
The directed movement of the IFT particles is powered by molecular motors \cite{howard01,chowdhury13,kolomeisky15}. Note that these motors do not appear explicitly in our model; instead, their role in IFT is captured by assigning the corresponding intrinsic velocities of anterograde (tipward) and retrograde (baseward) movement of each IFT particle in the absence of hindrance.  

An IFT particle may not be able to move with its intrinsic speed in a dense traffic because of steric hindrance caused by other IFT particles in front of it on the same track. Similar traffic-like collective phenomena in many other subcellular processes (see ref.\cite{frey04,lipowsky06,lipowsky10,leduc10,roland15,scnbook} for reviews) have been treated in the past as appropriate variants of the totally asymmetric simple exclusion process (TASEP) \cite{derrida98,schuetz01,mallick15}. In the same spirit, the collective movement of the motor-driven IFT particles is represented in our model as a TASEP.  

The regulation of transport of the structural proteins by IFT can determine the overall dynamics of the length of a flagellum. The length-dependent regulation of IFT requires feedback based on the flagellar length. Even for a single flagellum, it is challenging to understand how the cell `knows' or `senses' the length of its flagellum. Since none of the cells has a `ruler' for direct measurement of flagellar length, indirect mechanisms are believed to be used by a cell for getting a constant feedback about its flagellar length \cite{ludington15}. 
Here we present the theoretical formulation of a generic model based on the ``\textit{time of flight}'' (ToF) mechanism to explore the consequences of such a feedback mechanism on the flagellar length dynamics \cite{ludington15,ishikawa17}. 

The flagellar structural proteins to be transported are loaded as cargoes on the IFT particles; our model explicitly distinguishes between IFT particles and molecular cargoes that the IFT particles transport. Length-sensing by ToF allows a mechanism of ``differential loading'' \cite{wren13,craft15} (see also \cite{wemmer19}) of the flagellar structural proteins on the IFT trains at the flagellar base before they begin their anterograde journey. The longer is the flagellum, the fewer incoming IFT trains are loaded with cargoes and the slower is the rate of growth of the flagellum.

We begin with a model for length control of a single flagellum that incorporates all the following key features: (i) a ToF mechanism for length sensing \cite{ishikawa17}, (ii) a mechanism of differential-loading of flagellar structural proteins as cargo on the IFT trains \cite{wren13,craft15}, (iii)  a TASEP-based representation of the collective traffic-like movement of IFT trains \cite{scnbook}, (iv) a flagellar elongation rate that is proportional to anterograde flux of the flagellar structural proteins at the flagellar tip, (v) a flagellar shortening rate that is independent of the flagellar length, but dependent on the extent of IFT density at the flagellar tip, and (vi) synthesis and degradation of flagellar structural proteins in the cell body. Thus, to our knowledge, this is the most comprehensive model of length control of a single flagellum. By a combination of analytical treatment and computer simulations of both the stochastic and deterministic kinetic equations of this model, we examine the roles of all these  ingredients of the model in controlling the length of a single flagellum. 

The main aim of this paper, however, is to explore the mechanisms of coordination of the dynamics of the lengths of the two flagella in bi-flagellated eukaryotic cells and their consequences. Our stochastic kinetic model, that retains all the six key features of the model listed above for the kinetics of each individual flagellum, postulates coupling of their dynamics through sharing of the common pool of structural proteins at the base of the flagella. The key differences between our theory and another recently published work \cite{fai18} on flagellar length control in biflagellates will be discussed later in this paper. Utilizing some of the known properties of TASEP, we present an alternative interpretation of the experimental observations of Ishikawa and Marshall \cite{ishikawa17}. We show that a ToF mechanism of length regulation is consistent with their experiments. We also predict new results that can, in principle, be tested experimentally.

This paper is organized as follows: in the section \ref{sec-biolinfo} we present brief summary of the structure of flagella and the phenomenon of IFT. The ToF mechanism is explained in section \ref{sec-TOF}. The stochastic model for the length control of a single flagellum is formulated in section \ref{sec-singlemodel} and the corresponding main results are given in section \ref{sec-singleresult}. Experimental supports for the model are claimed in section \ref{sec-expt}. The model and results for biflagellates are presented in sections \ref{sec-doublemodel} and \ref{sec-doubleresult}, respectively. Detailed comparison of our model with those developed earlier is presented in section \ref{sec-modelcomp} thereby highlighting the novel features of our model. Finally, the main conclusions drawn from our model are summarized and suggestions for experimental tests of the new predicted are indicated in section \ref{sec-summary}.


\section{Structure of flagella and intraflagellar transport}
\label{sec-biolinfo}

\subsection{Structures of Flagella}

Eukaryotic flagella are hair like appendages which emerge from the surface of the cell. 
The typical length of fully grown flagella in, for example, unicellular  algae  {\it Chlamydomonas reinhardtii} (CR) is about 12 $\mu$m. However, various experimental methods have been developed to manipulate the flagellar length \cite{lefebvre86} that produce even longer or shorter flagella in the steady state.  

The structure of a flagellum is based on a cytoskeletal arrangement known as axoneme.  It acts both as a scaffold as well as an axle which facilitates beating of the flagellum. The axonemal structure is assembled on a basal body and projects out from the cell surface \cite{fisch11}. The major structural components of all axonemes are microtubule (MT) doublets; each MT being  essentially a tubular stiff filament. Each doublet consists of an A-microtubule (A-MT) and a  B-microtubule (B-MT). Nine doublet MTs, arranged in a cylindrically symmetric fashion form an axoneme; it extends from the base to the tip. Most axonemes have a 9+2 arrangement of MTs, where nine outer doublets surround a coaxial central pair. Some other axonemes lack a central pair and have a 9+0 arrangement of MTs. 
The MT doublets are cross-linked by axonemal dynein motors that drive relative sliding of the MT doublets. This sliding, in turn, causes beating of the flagella of eukaryotic cells \cite{lindemann10,sartori16,geyer16,wan18}.

\subsection{Intraflagellar Transport (IFT)}

In eukaryotic cells a MT serves as a track for two `superfamilies' of cytoskeletal molecular motors, called kinesin and dynein, which move naturally in opposite directions by consuming chemical fuel ATP \cite{howard01,chowdhury13,kolomeisky15}. These motors transport cargo which plays a crucial role not only during growth, but also in the maintenance and shrinkage of flagella \cite{taschner16}. This phenomenon of effective relocation of materials by the active motorized transport machinery is known as intraflagellar transport (IFT) \cite{kozminski93,lechtreck15,prevo17}. The crucial role of IFT in the construction of a growing flagellum was established experimentally by demonstrating the obstruction of flagellar growth upon disruption of IFT \cite{pazour00,rosenbaum02}. Because of their superficial similarities with cargo trains hauled along railway tracks, chain-like assemblies formed by IFT particles are called IFT trains \cite{ishikawa17,buisson13,stepanek16,lechtreck17}. IFT trains consist of two protein complexes (IFT-A and IFT-B) which have multiple protein-protein interaction domains \cite{taschner12, bhogaraju13}. The molecular components of the IFT machinery have also been catalogued in detail \cite{cole09,hao09}. More recently, direct evidence for transport of structural proteins and of signalling proteins as cargo of IFT trains has been reported \cite{wren13,craft15,taschner12}.  The different protein-protein interaction domains in the IFT particles allow different cargos hitchhiking on them.

The IFT trains are pulled by motor proteins walking on the MTs that form the axoneme and cycle between the flagellar tip and base \cite{iomini01,buisson13}. During each leg of their journey the IFT trains remain constrained in the narrow space between the outer surface of the axoneme and the inner surface of the flagellar membrane.  IFT-B and kinesin are associated to anterograde transport and only use B-MT for moving from base to the tip. In contrast, IFT-A and dynein participate in the retrograde transport and use A-MT for moving from tip to the base \cite{stepanek16,huet14}. However, the number of motors per IFT train is not known. Because of the use of the A-MT and B-MT for moving in opposite directions on a MT doublet, anterograde IFT trains do not collide with the retrograde IFT trains.  The IFT particles switch their direction of movement only at the base and the tip of the flagellum. This indicates the plausible existence of a regulatory mechanism for differentially activating and inactivating the appropriate IFT motors at the base and tip to facilitate the directional switching. Recently it has been reported that IFT27 (a component of the IFT train) is responsible for integrating the retrograde machinery (IFT-B complex) into the IFT trains and remodelling of the trains at the tip for the retrograde trip back to the base \cite{huet14}. 

Broadly, three different types of proteins perform distinct functions in IFT. Axonemal proteins (mainly tubulins) and other structural proteins are transported as cargoes within flagella. These cargoes are loaded onto IFT particles \cite{wren13,craft15} which are also proteins. Not all IFT particles are loaded with cargo before they begin their journey. Both the empty and loaded IFT particles are hauled along the narrow space between the axoneme and the flagellar membrane by motor proteins that walk along the MT tracks.Since the number of motors per IFT train is not known, we do not describe the motion of the motors explicitly in the model. Instead, the stochastic movement of the IFT trains along the MT tracks are described in terms of kinetic equations.

Why is IFT required in fully grown flagella? This mystery was unveiled when it was observed that there is an ongoing turnover of axonemal proteins at the tip of a fully grown flagellum. Unless replenished by fresh supply of these proteins in a timely manner the flagellum will keep shortening. Therefore, IFT is necessary even in fully grown flagella to maintain the dynamic balance between the rate of growth and disassembly in order to maintain the length at a stationary value \cite{marshall01}.

\section{Time of flight for measuring length: `ruler' is a `timer'} 
\label{sec-TOF}
In this brief section we introduce the time-of-flight (ToF) mechanism on which our model of flagellar length control is based \cite{lefebvre09,ishikawa17,ludington15}. Let us imagine that either the IFT particle itself, or a timer molecule bound to it, is prepared in a specific `chemical' or `conformational' state $S_{+}$.  The timer enters the flagellum in the state $S_{+}$. However, the state $S_{+}$, being transient, decays spontaneously, and irreversibly, into the state $S_{-}$ at the rate $k$. 
Upon return at the base of the flagellum, the current state of a timer indirectly indicates the length the flagellum because the longer the flagellum, the longer is the duration of its travel and, hence, the higher is the likelihood of change of its state during the travel. Thus, the `ruler' used for measuring the length of the flagellum is actually a `timer'. 

The ToF mechanism is based on the idea that, on returning back to the base, the current state of the timer molecule decides whether flagella building material (tubulin) will be loaded onto the IFT particle  \cite{wren13,craft15,wingfield17} for the next round of journey. If the timer returns in state $S_+$, it indicates smaller flagellum and directs loading of cargo into the IFT particle departing into the flagellum. On the other hand, the timer returning in state $S_-$ conveys that no more precursor is needed at the flagellar tip for further assembly and allows dispatching of empty IFT particles only. So, only those molecules are suitable for the role of timer whose timescale of switching states is comparable to the time taken by IFT trains to commute around the flagellum \cite{ishikawa17,cao09}.

ToF is based on the simple idea that, for a given velocity, the distance travelled is proportional to the time of flight of a particle or a wave. In the context of flagellar length control, a mechanism based on the concept of ToF was formulated first by Marshall and coworkers although the possibility was conjectured by Lefebvre in 2009 \cite{lefebvre09}. Switching of the state of the timer could be a protein modification like, say, phosphorylation \cite{wang14}. There are already other examples in molecular cell biology where Nature uses the trick of converting time into length. To our knowledge, the most celebrated example is that of the segmentation clock that exploits temporal oscillations to create periodic spatial patterns \cite{pourquie03,webb16}. 

Ishikawa and Marshall hypothesised that the timer could be a small GTPase bound to a molecule of GTP as it begins its anterograde journey \cite{ishikawa17}; the rate of GTP hydrolysis by the GTPase would be the rate $k$ of switching of the timer. Two possible candidates for timer are IFT22 and IFT27 which are components of IFT trains \cite{bhogaraju13}. These small Ras-like GTPases \cite{bhogaraju13,bhogaraju11} function as switch molecules which cycle between an active GTP-bound form and an inactive GDP-bound form. Huet et al. \cite{huet14} investigated the role of IFT27 in \textit{Trypanosome} and found that IFT27  enters the flagellum only in GTP-bound state. The cells in which IFT27 is in GTP-locked state, IFT trains enter into the flagellum and build a flagellum of slightly smaller than the normal length flagellum. But, if IFT27 is in GDP-locked state the trains are unable to enter the flagellum thereby preventing its formation. They concluded that the GTP-GDP cycle is essential for maintaining the correct length of the flagellum \cite{huet14}. For {\it Chlamydomonas}, it has been reported that partial knocking down of IFT27 affects the elongation of the flagella and a complete knockdown is lethal for the cell \cite{qin07}. So, these observations are indications that IFT27 is a possible candidate for timer. But more experiments have to be done to clearly establish whether IFT27 is really a timer that can control the loading of precursor proteins into the IFT trains.

\begin{figure*}
\begin{center}
\includegraphics[width=0.8\textwidth]{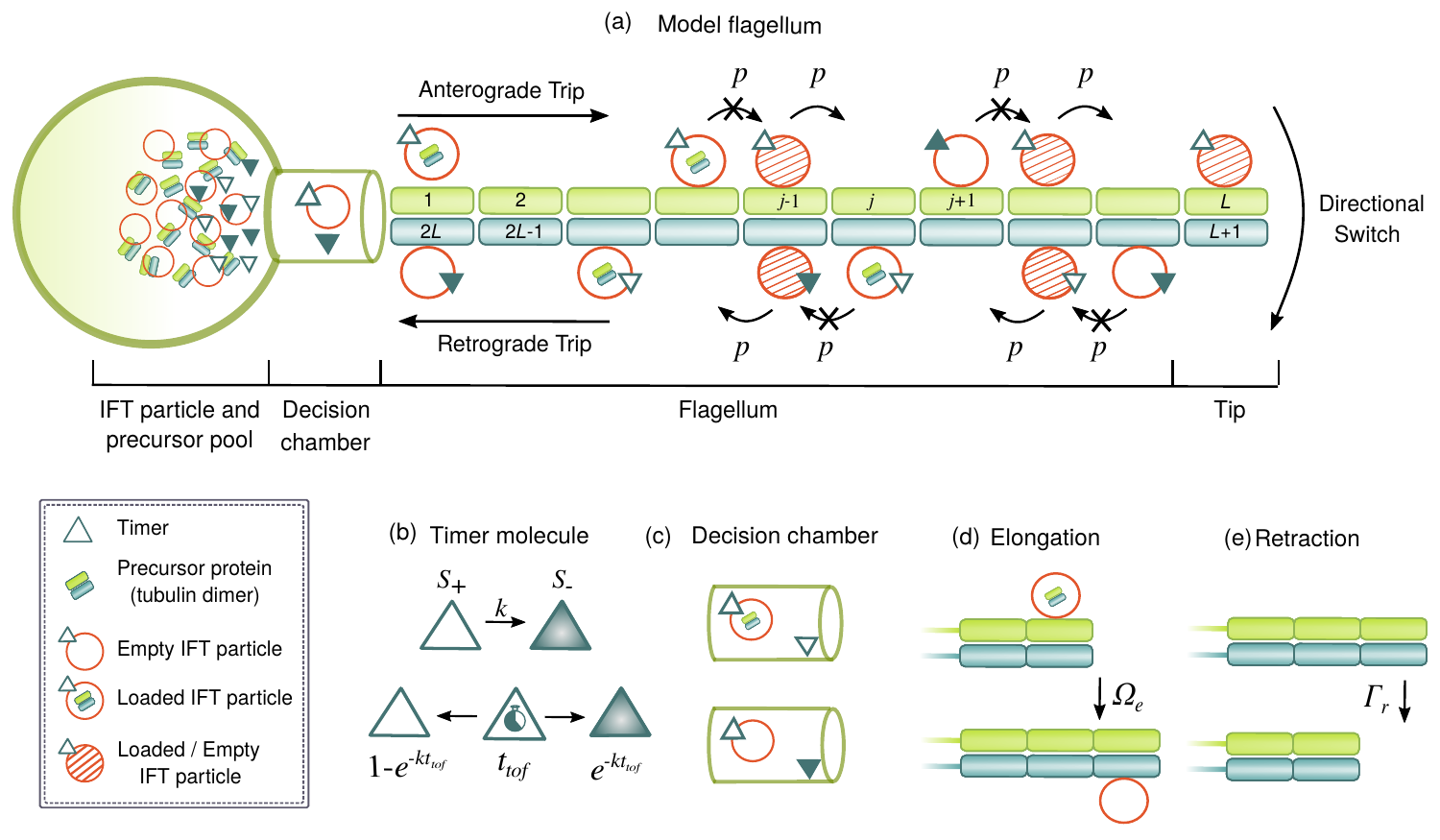}
\end{center}
\caption{ \textbf{ Schematic description of the model for length control of a single flagellum:} (a) Two lattices of equal length, arranged parallely, represent a microtubule doublet in a flagellum extending from the spherical cell body. The cell body has a pool of flagellar precursors (blue-green lattice units),IFT particles (red balls) and timer molecules (triangles). The  IFT particles can either be empty (hollow balls) or loaded (balls with a lattice unit inside them). Whether to dispatch a loaded IFT particle or an empty IFT particle, the decision is taken in the decision chamber which is a small compartment connecting flagellum with the precursor pool in the cell body. Each red ball filled with red lines represents an IFT particle which can be either empty or loaded. On the green lattice, IFT particles move unidirectionally from the cell body towards the tip (anterograde movement) and on the blue lattice, IFT particles move unidirectionally from tip towards the cell body (retrograde movement) with average velocity $v$. The sites on these chains are denoted with integer index $j$ ($j=1, 2, \dots, L$ from the flagellar base to the tip) and $j$ ($j=L+1, L+2, \dots, 2L$ from the tip to the base), respectively. At the tip (i.e, at site $j=L$), an IFT particle on the anterograde (green) lattice simply switches direction by hopping  to the adjacent site on the retrograde (blue) lattice (i.e, into site $j=L+1$,) if the target site is empty. Both the loaded and empty IFT particles obey the exclusion principle, i.e., no site can be occupied by more than one particle simultaneously.  (b) The timer in state $S_+$ switches to state $S_-$ with rate $k$. The timer enters the flagellum in state $S_+$; at the instant of its exit from the flagellum, the probability of finding it in the states $S_+$ and $S_-$ are $e^{-k t_{tof}}$ and $1-e^{-k t_{tof}}$, respectively, where $t_{tof}$ is the total time of flight inside the flagellum. (c) The  timer from the latest IFT particle which comes back from the tip detaches and gets attached to the decision chamber. If the timer is in state $S_+$, loaded IFT particle is dispatched into the flagellum and if the timer attached to the decision chamber is in $S_-$ state, an IFT particle is dispatched into the flagellum. All the trains entering the flagellum carry timer in state $S_+$. (d) Before switching direction at the flagellar tip, a loaded IFT particle can either (i) elongate the flagellum by adding a single lattice site to both the green and blue lattices, with probability $\Omega_e$, and return to the base empty, or (ii) return to the base carrying its undelivered cargo, without elongating the flagellum, with probability $1-\Omega_e$. (e) If there is no IFT particle on the distal tips of both the green or blue lattices, the flagellum can shorten by the chipping of those two sites with the rate $\Gamma_r$.} 
\label{gen-model}
\end{figure*}

                                                                                                                                                                                                                                                                                                                                                                                                                                                                                                                              \section{Stochastic model for length control of a single flagellum}
\label{sec-singlemodel}

First we consider the time-dependence of the length of a single flagellum. 
In this section, we will build the model step by step by clearly justifying all the simplifications. Thereafter, we will formulate the master equations for the qualitative description of our stochastic model and the corresponding Fokker-Planck equation and rate equations.

\subsection{Model}

The entire elongation and resorption dynamics of a flagellum can be effectively captured by a single MT doublet  which, in our model, is represented by two parallel linear chains of equal length $L$ (green and blue lattice chain in Fig. \ref{gen-model}(a). For the convenience of labelling the sites on these two chains with a single integer index $j$, the sites on the green (anterograde) chain  are labelled by $j=1, 2, \dots, L$ from the base to the tip and those on the blue (retrograde) chain are labelled by $j=L+1, L+2, \dots, 2L$ from the tip to the base. Because of this labelling scheme, the sites $j=L$ and $j=L+1$ are adjacent to each other at the tips of the two MT tracks for anterograde and retrograde transport, respectively (See Fig.\ref{gen-model}(a)).  Each lattice site on both the blue and green lattices  represent a tubulin dimer; free dimers in the pool at the base are referred to as precursor proteins. 

The precursor proteins are transported as cargoes on IFT trains. Each IFT train is made up of an array of IFT particles. 
Fusion of IFT particles into IFT trains and fission of IFT trains have been observed experimentally. However, for simplicity, we assume that all the IFT trains consist of a single IFT particle; fusion and fission of the IFT particles do not dominate the phenomena of our interest here. In a flagellum, each IFT train is pulled by several molecular motors. Since the number of motors per IFT train is not known, we do not describe the motion of the motors explicitly in the model. Rather, in our model, each of the self driven hard-core particles (red balls in Fig. \ref{gen-model}(a)) represents a motor-driven single IFT particle; the motors remain implicit. At any given instance, a site of the lattice can be occupied by only one such particle as this mutual exclusion captures the hard-core steric interaction between the IFT particles. An IFT particle at site $j$ moves by hopping to the target site $j+1$ with rate $p$ if and only if the target site is not occupied by any other IFT particle (see Fig.\ref{gen-model}(a)).

Each IFT train may have the capacity to bind (and carry) multiple cargoes at a time \cite{bhogaraju14}. The IFT particles may be loaded up to maximum capacity during the initial stages of flagellar growth whereas their capacity may remain underutilized in full-length flagella in steady-state \cite{bhogaraju14}. However, for simplicity, we assume that an IFT particle in our model can either be empty (empty red balls in Fig.\ref{gen-model}(a)) or carry one unit for flagellar structural building material (red balls each filled with a precursor in Fig.\ref{gen-model}(a)) which is assumed to be a tubulin dimer (Fig.\ref{gen-model}(a)). The red balls filled by red lines  in Fig.\ref{gen-model}(a) denote those IFT particle that can be either empty or loaded with precursor. We have used these to emphasize the `exclusion' principle, i.e., any site occupied by an IFT particle, irrespective of whether or not it is carrying a precursor protein, is not available to the following IFT particle. 
Moreover, every IFT particle switches its direction of movement, from anterograde to retrograde, at the distal tip of the flagellum. 

Whether loaded or empty, every IFT particle that enters the flagellum carries one timer molecule which is in state $S_+$ (see figure \ref{gen-model}(b)). Let the rate at which it can switch stochastically, and irreversibly, to state $S_-$ be $k$. The master equation for the stochastic process 
\begin{equation}
S_{+} \mathop{\longrightarrow}^{k} S_{-} 
\end{equation}
is  given by
\begin{equation}
\frac{dP^+_{tm}(t)}{dt}=-k P^+_{tm}(t)
\label{DC_I}
\end{equation}
where $P^+_{tm}$ is the probability density that the timer remain in the state $S_+$ at time $t$. The time-dependent solution of (\ref{DC_I}), corresponding to the given initial condition $P^+_{tm}(t=0)=1$, is given by 
$P^+_{tm}(t) = e^{-kt}$.
Therefore, if the time spent by the timer inside the flagellum is $t_{tof}$ (see figure \ref{gen-model}(b)), then the probability $P^+_{tm}$ that the timer will remain in state $S_+$ at the moment of its return to the base is given by 
\begin{equation}
P^+_{tm}(t=t_{tof})=e^{-k t_{tof}}
\label{p_s1}
\end{equation} 

In general, the length covered by a IFT particle during its anterograde journey may not be identical to that covered during its retrograde journey along the same flagellum because of the elongation or shortening of the flagellum during that period. The complete journey of an IFT particle in a fully grown flagellum is of the order of ten seconds whereas ciliogenesis requires a time of the order of tens of minutes. Because of this {\it separation of timescales}, the length of the flagellum remains practically unchanged during the time of a single flight of a timer and its ToF is taken simply as   
\begin{equation}
t_{tof}=\frac{2 L(t)}{v}
\end{equation} 
and, hence, from (\ref{p_s1})
\begin{equation}
P^+_{tm}(t=t_{tof})=e^{-2k L(t)/v}
\label{p_s10}
\end{equation} 

Note also that the average velocity $v$ of the timer (which is identical to that of the IFT particles) depends on the number density (i.e., number per site) $\rho$ of IFT particles in the traffic inside the flagellum. Because of the separation of timescales, we assume that the number density $\rho$ and the flux $J$ of the IFT particles in the flagellum always take corresponding values in the steady-state (time-indepedent) of the TASEP that represents their traffic. The $\rho$-dependence of the flux $J$ and mean velocity $v$ in the steady state are given by \cite{derrida98,schuetz01,mallick15}
\begin{equation}
J(\rho)=p \rho(1-\rho)
\label{TASEP-J}
\end{equation} 
and 
\begin{equation}
v(\rho)=p(1-\rho)
\label{TASEP-v}
\end{equation} 
which are fundamental results of TASEP  \cite{derrida98,schuetz01,mallick15}. So far TASEP has been applied to understand both vehicular traffic and molecular motor traffic \cite{roland15,scnbook}.

We assume that whether or not a precursor will be loaded on an IFT particle just before it begins its journey in the flagellum is decided by the state of the timer associated with the latest train to return to the base after shuttling inside the flagellum. In other words, the passengers (precursor proteins) need `ticket' (the state $S_+$ of the timer) to gain access to the train \cite{fort14}. This fact is the foundation of {\it differential loading model} \cite{wren13,craft15}. In this model, the timer dissociates after completing the retrograde trip and then the timer sticks to the decision chamber (see Fig. \ref{gen-model}(a)). The decision chamber is a hypothetical element connecting the precursor pool and the flagellum which we introduce for our convinience. The immediate neighborhood of the basal bodies of {\it Chlamydomonas}, which play several key functions as `flagella organizing centers' \cite{wingfield18}, is a possible candidate for the decision chamber' introduced here. The timer sticks to the decision chamber untill the next timer returns back from the tip with the next train. The decision chamber locks the final state of the timer. The final state of the timer then determines  the decision  of the cell whether to dispatch trains loaded with precursor proteins (if timer is in $S_+$ state) or just empty train (if the timer is in $S_-$ state) into the flagellum (see Fig. \ref{gen-model}(c)). The timer is then  reset  into $S_+$ state and can get utilised by the  trains which are about to enter the flagellum (see Fig. \ref{gen-model}(c)).

 IFT27, the small GTPases component of the IFT train, detaches from the retrograde IFT trains and remains distributed around the flagellar base  \cite{wang09,wang14}. This observation suggests that the flagellar base may be serving as the decision chamber. Moreover, IFT27 only enters the flagellum if in the GTP-bound state. This observation supports our idea that the  timer resets to the $S_+$ state before starting the journey inside the flagellum \cite{huet14}. When IFT27 is in the GDP bound state, it doesn't allow the interaction of many proteins with IFT trains and this supports our assumption that when the timer in $S_-$ state no precursor is able to hitchike the IFT trains which enters the flagellum \cite{huet14}. 

Suppose, at the time of entry of IFT particles into the flagellum, the average number of flagellar precursor proteins at the base is $\langle N(t) \rangle$ (the operational meaning of this averaging will be clarified later in this section).  The probability of the timer to be in $S_+$ (sticked to the decision chamber) is given by equation (\ref{p_s10}). Then, the probability of loading a flagellar precursor onto the IFT particle is 
          \begin{equation}
          \alpha_{tu}=\frac{\langle N(t) \rangle}{N_{max}} P^+_{tm}(t_{tof})=\frac{\langle N(t) \rangle}{N_{max}} e^{-2kL(t)/v}
          \label{alpha_tu}
          \end{equation}
where $N_{max}$ denotes the maximum capacity of the precursor protein pool in terms of the number of precursor proteins. In other words, synthesis and degradation of flagellar precursors happen in such a way that the average of the precursor population of the pool does not exceed $N_{max}$. This can be achieved by choosing synthesis rate as $\omega^+[1-({\langle N(t) \rangle}/N_{max})]$ and degradation rate as $\omega^-\langle N(t) \rangle$. Note that both synthesis and degradation rates depend on the population of precursors in the pool.
 
If the total flux of IFT particles reaching at the tip is $J$, then the flux of loaded trains reaching the tip is $\alpha_{tu} J$. On reaching the tip along the anterograde track, a loaded IFT particle can elongate the tracks by one tubulin unit with probability $\Omega_e$ and this IFT particle (now empty after delivering its cargo) hops to the newly formed site at the tip of the retrograde lane and begins its return journey to the base (see Fig.\ref{gen-model}(d)). Because of the scheme of labelling the sites on two lattices by a single index, as described above, two extra sites are inserted between the two special sites $j=L$ and $j=L+1$ thereby increasing the range of $j$ from $1 \leq j \leq 2L$ to $1 \leq j \leq 2L+2$. Thus, the effective elongation rate of the flagellum is $\alpha_{tu}J \Omega_e$

In addition to polymerization/elongation mediated by loaded anterograde trains, the axoneme can undergo spontaneous shortening, with the rate $\Gamma_{r}$, by the simultaneous removal of both the sites $j=L$ and $j=L+1$ at the tip provided both are empty at that instant of time (see Fig.\ref{gen-model}(e)). As the probability of simultaneously finding both the sites empty is $(1-\rho)^2$ under mean-field approximation, the effective shortening rate is $(1-\rho)^2 \Gamma_r$.

                                                                                                                                                                                                                                                                                                                                                                                                                                                                                                                                                



\subsection{Master equations for a single flagellum}

In this subsection we treat ciliogenesis as a stochastic process where the stochastic kinetics of the length $L(t)$ of the flagellum and that of $N(t)$, the population of the precursors in the common pool, are assumed to be Markovian.
Let $P_L(j,t)$ be the probability that the flagellar length at time $t$ is $L(t)=j$. 
The master equation governing the stochastic kinetics of the flagellar length is given by
\begin{eqnarray}
&&\frac{dP_{L}(j,t)}{dt}   \nonumber \\ 
&=&\underbrace{\lambda^L_{j-1,j}P_{L}(j-1,t)}_{\substack {\text{Gain by elongation} \\ \text{from $L(t)=j-1$ to $L(t)=j$}}}+\underbrace{\mu^L_{j+1,j}P_{L}(j+1,t)}_{\substack {\text {Gain by resorption} \\ \text{from $L(t)=j+1$ to $L(t)=j$}}}\nonumber\\
&-& \underbrace{\lambda^L_{j,j+1}P_{L}(j,t)}_{\substack {\text{Loss by elongation} \\ \text{from $L(t)=j$ to $L(t)=j+1$}}}  -\underbrace{\mu^L_{j,j-1}P_{L}(j,t)}_{\substack {\text{Loss by resorption} \\ \text{from $L(t)=j$ to $L(t)=j-1$}}} \nonumber \\
\label{master_eq_i}
\end{eqnarray}
where $\lambda^L_{j,j+1}$ denotes the rate of elongation of the flagellum from state $j$ to $j+1$ while $\mu^L_{j,j-1}$ denotes that of shortening of flagellar length from state $j$ to $j-1$.

\begin{eqnarray}
\lambda^L_{j,j+1}=\underbrace{\bigg{[}\frac{\langle N(t) \rangle}{N_{max}} e^{-2k j/v}\bigg{]}}_{\alpha_{tu}} J \Omega_e \nonumber\\=\bigg{[}\frac{\sum_{n=0}^{N_{max}} n P_{N}(n,t)}{N_{max}} e^{-2k j/v}\bigg{]} J \Omega_e
\label{lambda}
\end{eqnarray}
where the expression of $\alpha_{tu}$ is taken from Eq.(\ref{alpha_tu}). 
Similarly, the rate $\mu^L_{j,j-1}$ of shortening of flagellar length from state $j$ to $j-1$ is given by 
\begin{equation}
\mu^L_{j,j-1}=(1-\rho)^2 \Gamma_r.
\label{mu}
\end{equation}
Note that, unlike the transition rates $\lambda^L_{j,j+1}$ (equation (\ref{lambda})), $\mu^L_{j,j-1}$ (equation (\ref{mu})) are independent of $j$. The crucial consequence of this difference in the $j$-dependence of $\lambda$ and $\mu$ will be demonstrated by the results that follow from a quantitative analysis.

The precursor synthesis and degradation by the cell, precursor loading onto the IFT particles and addition of precursor chipped from the tip back into pool can change the precursor population from the current state $N(t)=n$ to $n-1$ or $n+1$.  Let $P_N(n,t)$ denote the probability of finding $N(t)=n$ free precursors in the pool at time t. So, the master equation governing the evolution of the pool population $N$ is given by 
\begin{widetext}
\begin{eqnarray}
& \frac{dP_N(n,t)}{dt}  = \nonumber\\ \nonumber\\
& \underbrace{\omega^+ \biggl[1-\frac{(n-1)}{N_{max}}\biggr]P_N(n-1,t)-\omega^+ \biggl[1-\frac{n}{N_{max}}\biggr]P_N(n,t)}_{\text{Population dependent synthesis of flagellar precursor by the cell }}  \nonumber\\ \nonumber\\
& \underbrace{+[\omega^-{(n+1)} P_N(n+1,t)-\omega^- {n} P_N(n,t)]}_{\text{Population dependent degradation of flagellar precursor by the cell }}   \nonumber\\ \nonumber\\
&\underbrace{+\biggl[\sum_{j=0}^{L_{max}}J \Omega_e e^{-2k j/v}P_{L}(j,t)\biggr]\biggl[\frac{(n+1)}{N_{max}} P_N(n+1,t)-\frac{(n)}{N_{max}}P_N(n,t)\biggr]}_{\text{Contribution of pool towards assembly of the flagellum }} \nonumber\\ \nonumber\\
&  \underbrace{+(1-\rho)^2 \Gamma_r[P_N(n-1,t)-P_N(n,t)]}_\text{Precursors returned to the pool by disassembly of the flagellum }
\label{eq-NL}
\end{eqnarray}

\end{widetext} 
The last two terms in (\ref{eq-NL}) have been written under mean-field approximation that ignores correlations between the $L$ and $N$ variables.

As stated earlier, the traffic flow of the IFT particles is represented in our model as a totally asymmetric simple exclusion process (TASEP) \cite{derrida98,schuetz01,mallick15}. Two primary quantities that characterize the steady state of a TASEP are (i) average particle density $\rho$, and (ii) the average particle flux $J(t)$; these are also the only two properties of TASEP that enter directly in our model through the rates $\lambda^L$ and $\mu^L$ (see Eqns.(\ref{lambda}) and (\ref{mu})). In our numerical plots we'll choose values of $\rho$ and $J$ that correspond to one of the three dynamical phases of immediate interest in that analysis (Further details are given in the section \ref{sec-expt}).

To convert the dimensionless length $L(t)$ to actual length (measured in $\mu$m) we multiply $L$ with $\delta L$=
0.008 $\mu$m, where $\delta L$ is the size of a single tubulin dimer. To convert the dimensionless flux $J$ (i.e, number of particles per unit time passing through a particular point), velocity $v$ (i.e, the distance covered by an IFT particle per unit time) and the other dimensionless rate constants $k, \ \Gamma_r, \ \omega^+$ and $\omega^-$ to actual quantities, we divide them with appropriate $\delta t$ whose specific values are mentioned in the caption of each figure. 
The parameter values have been chosen in such a way that the numerical value of the flagellar length in the steady state is about 12 $\mu$m, which is comparable to that of each flagellum of {\it Chlamydomonas reinhardtii}.

\subsection{Fokker-Planck Equation for a single flagellum} 

Next we take the continuum limit in which the length of the flagellum is represented by a continuous variable $x$. In this limit the probability $P_{L}(j,t)$ reduces to $P_{X}(x,t)$ which denotes the probability that flagellar length is $x$ at time $t$. Carrying out the standard Kramers-Moyal expansion of the master equation (\ref{master_eq_i}) governing the length of the flagellum, we obtain the corresponding Fokker-Planck equation 
\begin{eqnarray}
\frac{\partial P_{X}(x,t)}{\partial t}&=-\frac{\partial}{\partial x}[\{ {\lambda}(x)-{\mu}(x) \} P_{X}(x,t)]\nonumber\\
&+ \frac{\Delta L}{2} \cdot  \frac{\partial^2}{\partial x^2}[\{ {\lambda}(x)+{\mu}(x) \} P_{X}(x,t)] 
\label{eq-FP1}
\end{eqnarray}
where 
\begin{eqnarray}
\lambda(x)&=&\frac{\langle N(t) \rangle}{N_{max}}  J \Omega_e exp{(-2kx/v)} \nonumber\\
\mu(x)&=&(1-\rho)^2 \Gamma_r
\end{eqnarray}
and $\Delta L=1$. The Eq.(\ref{eq-FP1}) describes the stochastic kinetics of the length of the flagellum essentially as a combination of $x$-dependent drift and diffusion of the flagellar tip where $\lambda(x)-\mu(x)$ and $\lambda(x)+\mu(x)$ are proportional to the the effective drift velocity and diffusion constant, respectively.

\subsection{Rate Equations for a single flagellum}

From the master equations for the stochastic time evolution of the length of a single flagellum, we derive the corresponding rate equation (see appendix \ref{app-master2rate4single} for the details)
\begin{equation}
\frac{d\langle L(t) \rangle}{dt}=\underbrace{\bigg{[}\frac{\langle N(t) \rangle}{N_{max}}e^{-2k\langle L(t) \rangle/v}\bigg{]}}_{\alpha_{tu}} J \Omega_e -(1-\rho)^2 \Gamma_r 
\label{rate_l}
\end{equation}
that describes the deterministic time evolution of the mean length 
\begin{equation}
\langle L(t) \rangle =  \sum_{j=0}^{\infty} j P_{L}(j,t) 
\end{equation} 
of the flagellum. 
Similarly, from the master equations for population of the precursor pool, we get 
\begin{equation}
\frac{d \langle N(t) \rangle}{dt} =\omega^+\bigg{[}1-\frac{\langle N(t) \rangle}{N_{max}}\bigg{]} -\omega^-\langle N(t) \rangle   - \frac{d\langle L(t) \rangle}{dt}
\label{rate_pool}
\end{equation}
which describes the deterministic temporal evolution of the average population of the precursors 
\begin{equation}
\langle N(t) \rangle = \sum_{n=0}^{\infty} n  P_{N}(t). 
\end{equation}

\section{Results on length control of a single flagellum}
\label{sec-singleresult}


\subsection{Steady-state of a flagellum: a ``balance point''}

The steady state of the system is defined by the condition $d\langle L(t) \rangle/dt = 0 = d\langle N(t) \rangle/dt$;  the corresponding average length of the flagellum and the average population of precursors in the pool are denoted by $\langle L_{ss} \rangle$ and $\langle N_{ss} \rangle$, respectively. From (\ref{rate_l}), in the steady state, we get $\alpha_{tu} J \Omega_e = (1-\rho)^2 \Gamma_r$ and using the expression for $\alpha_{tu}$ that follows from eq. (\ref{alpha_tu}) in the steady-state , we get 
\begin{equation}
\langle L_{ss} \rangle = \frac{v}{2k}\ell og \bigg{[} \frac{J \Omega_e }{(1-\rho)^2 \Gamma_r}  \frac{\langle N_{ss} \rangle}{N_{max}} \bigg{]}
\label{eq-Lss1}
\end{equation}
and
\begin{equation}
\langle N_{ss} \rangle = \frac{\omega^+}{\omega^-+\frac{\omega^+}{N_{max}}} .
\label{eq-Nss}
\end{equation}
For future convenience, we introduce the symbols 
\begin{equation}
A=J \Omega_e, 
\label{eq-A} 
\end{equation} 
\begin{equation}
B=(1-\rho)^2 \Gamma_r 
\label{eq-B} 
\end{equation} 
and
\begin{equation}  
C=\frac{2k}{v}. 
\label{eq-C} 
\end{equation}
In terms of $A, B, C$, the steady state flagellar length is expressed as 
\begin{equation}
\langle L_{ss} \rangle = C^{-1}  \ell og \bigg{[} \frac{A}{B}  \frac{\langle N_{ss} \rangle}{N_{max}} \bigg{]}
\label{eq-Lss2}
\end{equation}

The factor within the square bracket on the right hand side of (\ref{eq-Lss1}) (or, equivalently, (\ref{eq-Lss2})) corresponds to the ratio of the rates of elongation and shortening of the flagellum. However, these rates affect the steady state length of the flagellum only logarithmically. The length of the flagellum is essentially determined by $v/k$ which is a characteristic length set by the ratio of two properties of the timer. Thus, the faster the timer moves (and/or the slower is its conversion to the state $S_{-}$) the longer is the magnitude of $\langle L_{ss} \rangle$. In the steady-state the flagellum neither enriches nor depletes the population of the precursors in the pool. The only variation in the population of precursors in the pool arises from the synthesis and degradation of the precursors. Not surprisingly,  in large $N_{max}$ limit, the steady-state population $\langle N_{ss} \rangle$ is determined by the ratio $\omega^{+}/\omega^{-}$; the larger is the rate of production (and/or the smaller is the rate of degradation) the higher is the population $\langle N_{ss} \rangle$.

As mentioned in section IV.A, in our model, the effective {\it assembly} rate $J \Omega_e exp(-2k \langle L(t) \rangle /v) \langle N(t) \rangle/N_{max}$ is length-dependent whereas the {\it disassembly} rate $(1-\rho)^2 \Gamma_r$ is independent of length. More precisely,  the ToF mechanism leads to monotonic decrease of the assembly rate with increasing length ; the steady-state is a ``{\it balance point}'' where the assembly rate just balances the rate of disassembly (see Fig.\ref{plot_1}(a)). This result is consistent with the concept of ``balance-point'' proposed by Rosenbaum, Marshall and others \cite{marshall01,marshall05,engel09}. 

Because of the intrinsic stochastic nature of the kinetics, as descibed by the full master equations, the flagellar length $L(t)$ keeps fluctuating around the average length $\langle L_{ss} \rangle $ even in its steady-state. Using the master equation as well as the Fokker Planck equation (see appendices  A, B and  \ref{app-SSlengthDistFP} for the detailed derivations) we have calculated the steady state distribution that, as shown in Fig.\ref{plot_1}(b), is peaked at $\langle L_{ss} \rangle$. This distribution of the flagellar length in the steady-state is very similar to the distribution of the steady-state lengths of cytoskeletal filaments obtained earlier by following a master equation approach (see, for example, Fig.5(c) of ref.\cite{mohapatra16} and Fig.2 of \cite{erlenkamper09}).

\subsection{Ciliogenesis: controlled assembly of a single flagellum}

\begin{figure}[h!]
\begin{center}
\includegraphics[width=0.440\textwidth]{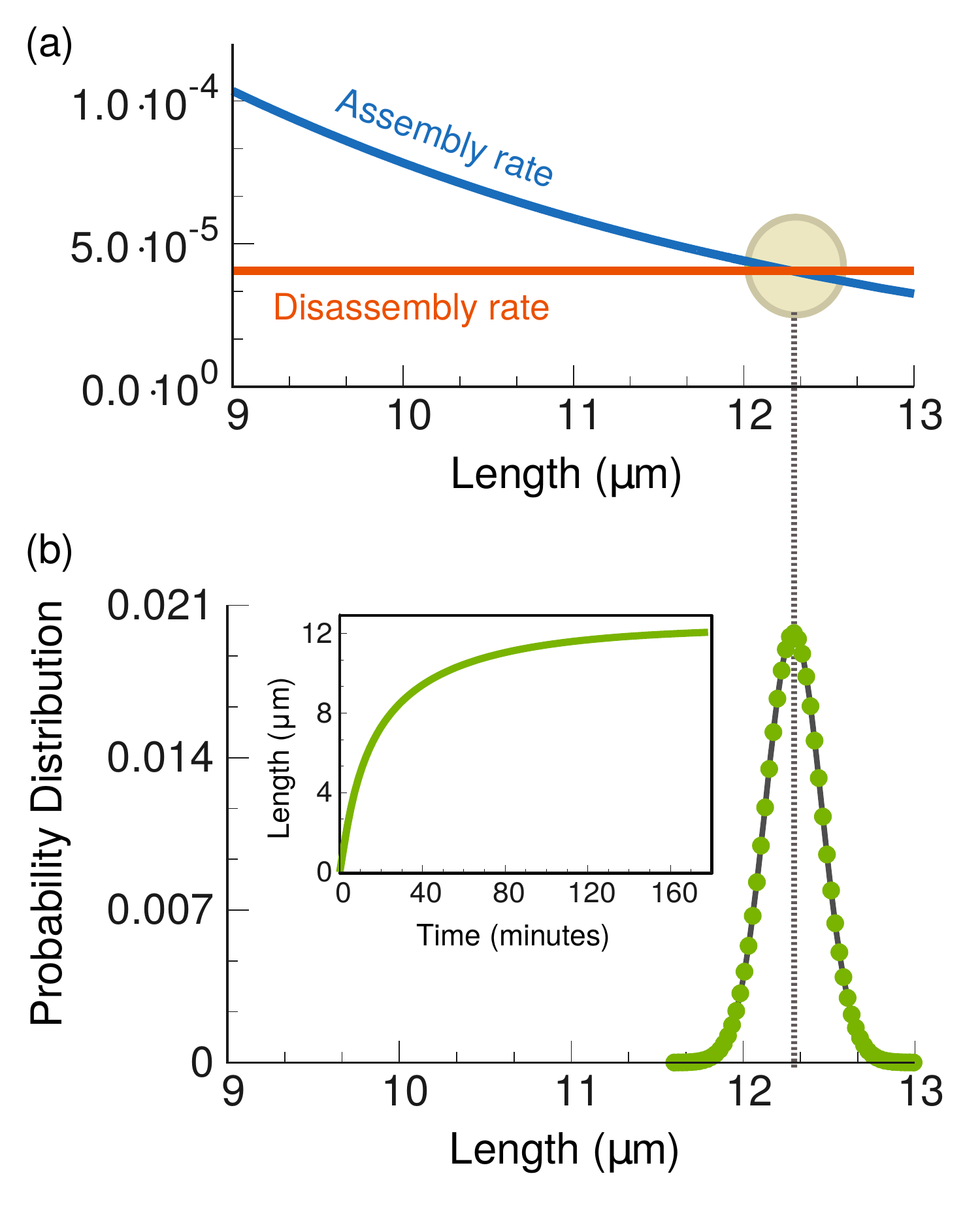}
\end{center}
\caption{ \textbf{ToF mechanism for flagellar length control, ciliogenesis and length fluctuation}:  (a) Assembly rate ($J \Omega_e (\langle N(t) \rangle/N_{max}) \ exp(-2k \langle L(t) \rangle /v)$) and Disassembly rate ($(1-\rho)^2 \Gamma_r$) are plotted as functions of  flagellar length. (b) Distribution of flagellar length in the steady state (green dots denote the predictions from master equation while the black line shows the corresponding predictions of the  Fokker-Planck equation). The plot of $\langle L(t) \rangle$ vs $t$ in the inset depicts how a new flagellum elongates with time eventually attaining its steady-state length; this process is called ciliogenesis.   
 (Parameters: $\rho=0.1$, $J=0.09$, $v=0.9$, $k=1.1 \times 10^{-3} $, $\Omega_e=0.5$, $\Gamma_r=5.0 \times 10^{-5} $, $\omega^+=2.0 \times 10^{-3}  $, $\omega^-=1.0 \times 10^{-5} \ $, $N_{max}=5000$, $\delta t =3.6 \times 10^{-4}  s$ ). }
\label{plot_1}
\end{figure}

The process of assembly and disassembly of flagella is referred to as ciliogenesis \cite{avasthi12}.  
All quantitative studies of ciliogenesis normally begin by probing the time-dependent growth of a flagellum. With the same aim, we solved the coupled rate equations (equation (\ref{rate_l}) and \ref{rate_pool}), subject to the initial conditions $\langle L(t=0) \rangle =L(0)$ and $\langle N(t=0) \rangle =N(0)$, respectively, for a set of values of the model parameters; the results are plotted in Fig.\ref{plot_1}(b). The rate of growth of the mean length  slows down with time as $\langle L(t) \rangle$ approaches its steady-state value $\langle L_{ss} \rangle$ asymptotically as $t \to \infty$. This qualitative trend of variation of $\langle L(t) \rangle$ with $t$ is very similar to those observed earlier in experiments \cite{rosenbaum69}.

Various time scales in the problem have been analyzed in appendix \ref{sec-Timescales}. 
In the special limiting situation where $\langle N(t) \rangle$ attains steady state value $\langle N_{ss} \rangle$ much faster than $\langle L(t) \rangle$ such that the quantity $N(t)/N_{max}$ remains practically constant throughout the evolution of the flagellar length, we can approximate the Eq(\ref{rate_l}) by
\begin{equation}
\frac{d\langle L(t) \rangle}{dt}= \biggl[\frac{\langle N_{ss} \rangle}{N_{max}}e^{-2k \langle L(t) \rangle /v}\biggr] J \Omega_e -  (1-\rho)^2\Gamma_r
\end{equation}
whose solution is given by 
\begin{eqnarray}
\langle L(t) \rangle   = \frac{1}{C} \ {\ell og \bigg{[} {\frac{N_{ss}}{N_{max}} \frac{A}{B}- \left(\frac{N_{ss}}{N_{max}} \frac{A}{B}- e^{{C} {L_0}}\right)e^{-{B} {C} t}}\bigg{]}}\nonumber \\
\label{eq-expL}
\end{eqnarray}
where $L_0$ is the initial length of the flagellum. From this solution, we conclude that, in this special limit, $\langle L(t) \rangle$ relaxes to its steady-state value $\langle L_{ss} \rangle$ exponentially with the corresponding relaxation time $\tau=1/(BC)$. 
In the general case, the correlation between the shapes of the curves $\langle L(t) \rangle$ and $\langle N(t) \rangle$ will be discussed in detail in the next subsection.

\subsection{Effects of precursor pool on length of a flagellum}

Although the initial amount of precursor $N(0)$ doesn't affect $L_{ss}$ and $N_{ss}$, it does affect how steady state is achieved by the flagellar length  $L(t)$. In this subsection we systematically study the effects of the time-dependence of $\langle L(t) \rangle$ on $N(0)$. In other words, we systematically explore the interplay of the population kinetics of the precursors $\langle N(t) \rangle$ and growth of the flagellum $\langle L(t) \rangle$ during ciliogenesis. For this purpose, we vary the numerical value of the parameter $N(0)$ over about three orders of magnitude; the values of other model parameters are such that $\langle N_{ss} \rangle \simeq 1000$. We have chosen the interesting regimes of $\omega^{+} \gg \omega^{-}$ and $J \Omega_{e} \gg \Gamma_{r}$. We present results for three regimes, namely, 
$N(0) \gg \langle N_{ss} \rangle$,  $N(0) \simeq \langle N_{ss} \rangle$, and $N(0) \ll \langle N_{ss} \rangle$.

\begin{figure}
\begin{center}
\includegraphics[width=0.380\textwidth]{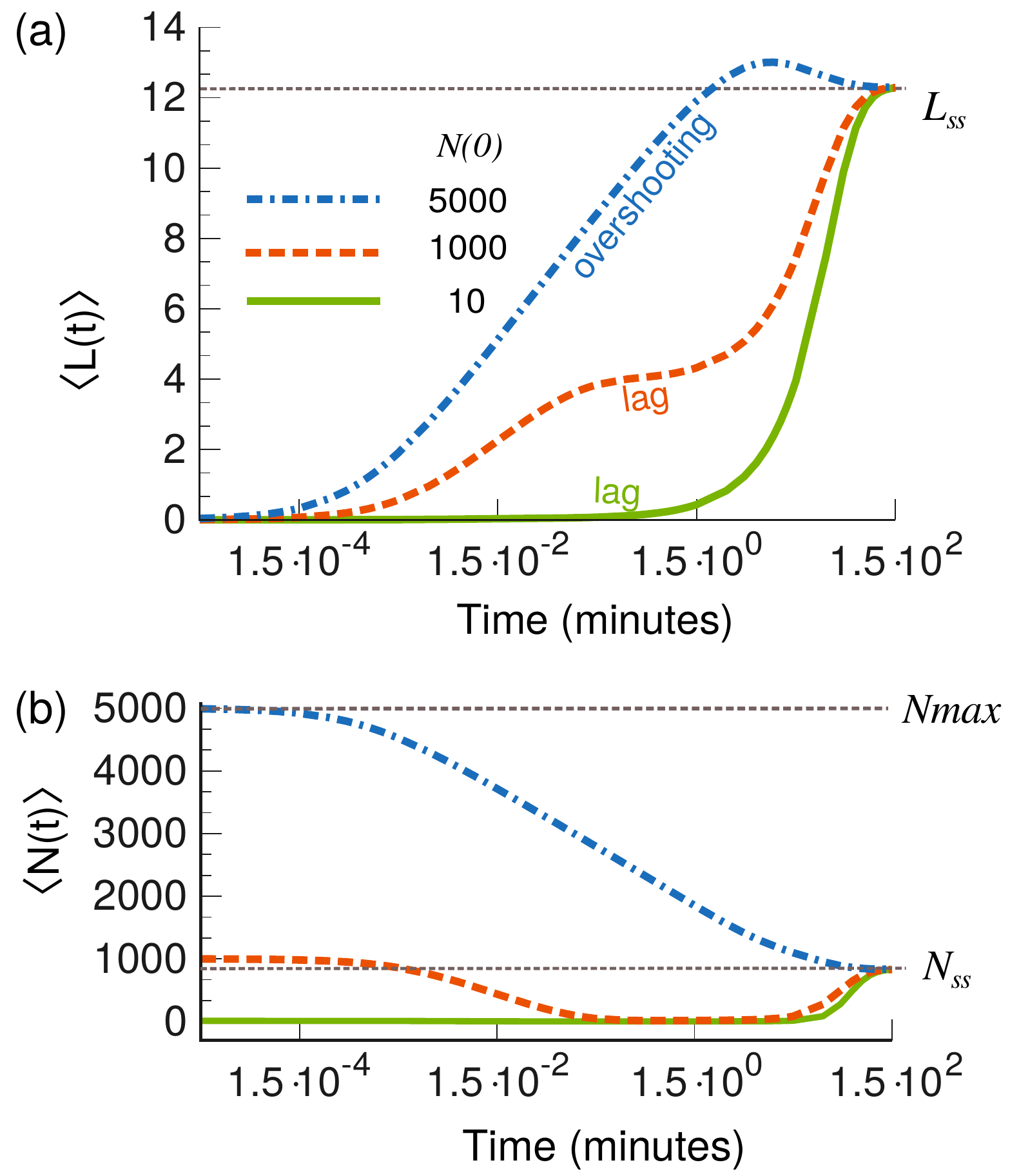}
\end{center}
\caption{ {\bf Dependence of ciliogenesis on the initial precursor population $N(0)$}: (a) Semi-log plot of $\langle L(t) \rangle$ vs $t$, and (b) $\langle N(t) \rangle$ vs $t$, both for three different values of $N(0)$. The length of a growing flagellum can overshoot beyond its steady-state length $L_{ss}$, before relaxing to $L_{ss}$, if $N(0)$ is sufficiently high. (Parameters: $\rho=0.1$, $J=0.09$, $v=0.9$, $k=2.0\times 10^{-3} $, $\Omega_e=0.5$, $\Gamma_r=1.0  \times 10^{-5}$, $\omega^+=1.0  \times 10^{-5}$, $\omega^-=1.0  \times 10^{-8}$, $N_{max}=5000$, $\delta t=9.0 \times 10^{-6} \ s$. Other quantities: $L_{ss}=12.3 \ \mu$m, $N_{ss}=833.3$)}
\label{pool}
\end{figure}

In the $N(0) \gg \langle N_{ss} \rangle$ regime the most remarkable observation is that $\langle L(t) \rangle$ can overshhot beyond $\langle L_{ss} \rangle$, before shortening and eventually relaxing to$\langle L_{ss} \rangle$ (see Fig \ref{pool}(a) uppermost curve). As the flagellum grows, and finally relaxes to its stead-state length $\langle L_{ss} \rangle$, the population of the precursors in the pool also relaxes to the corresponding value $\langle N_{ss} \rangle$   (see Fig \ref{pool}(b) uppermost curve).  To our knowledge, this effect has not been reported so far in the experimental literature, perhaps, because the value(s) of one or more of the parameters or $N(0)$ in the experiments have never been in the range required to observe this phenomenon. 

In the opposite limit $N(0) \ll \langle N_{ss} \rangle$ the most remarkable feature of ciliogenesis is the ``lag period''. Although the population kinetics of the precursors is switched on at $t=0$, the growth of the flagellum becomes significant only after a `lag period' (see the lowermost curve in Fig \ref{pool}(a)). The precursor population can contribute to sustained growth of the flagellum only after the precursor population in the pool itself begins to rise beyond a critical level (see the lowermost curve in Fig \ref{pool}(b)). 

For the intermediate value of $N(0) \simeq \langle N_{ss} \rangle$, initially the flagellar growth exhibits practically no lag period (see the middle curve in Fig \ref{pool}(a)). But, the precursors supplied during this initial growth and those lost by natural decay are not replenished at a comparable rate resulting in a fall in the precursor population (see the middle curve in Fig \ref{pool}(b)). This low population of precursors, in turn, reduces the flagellar growth to almost vanishingly small level (see the middle curve in Fig \ref{pool}(a)). This situation continues, just like the `lag period' discussed before, till fresh synthesis of precursors enlarges the pool population to levels that can resume sustained growth of both the flagellar length as well as its own population, eventually, reaching the steady state ((see the middle curves in Fig \ref{pool}(a) and (b)).

\subsection{Interplay of traffic, timer and polymerization}
\begin{figure}
\begin{center}
\includegraphics[width=0.350\textwidth]{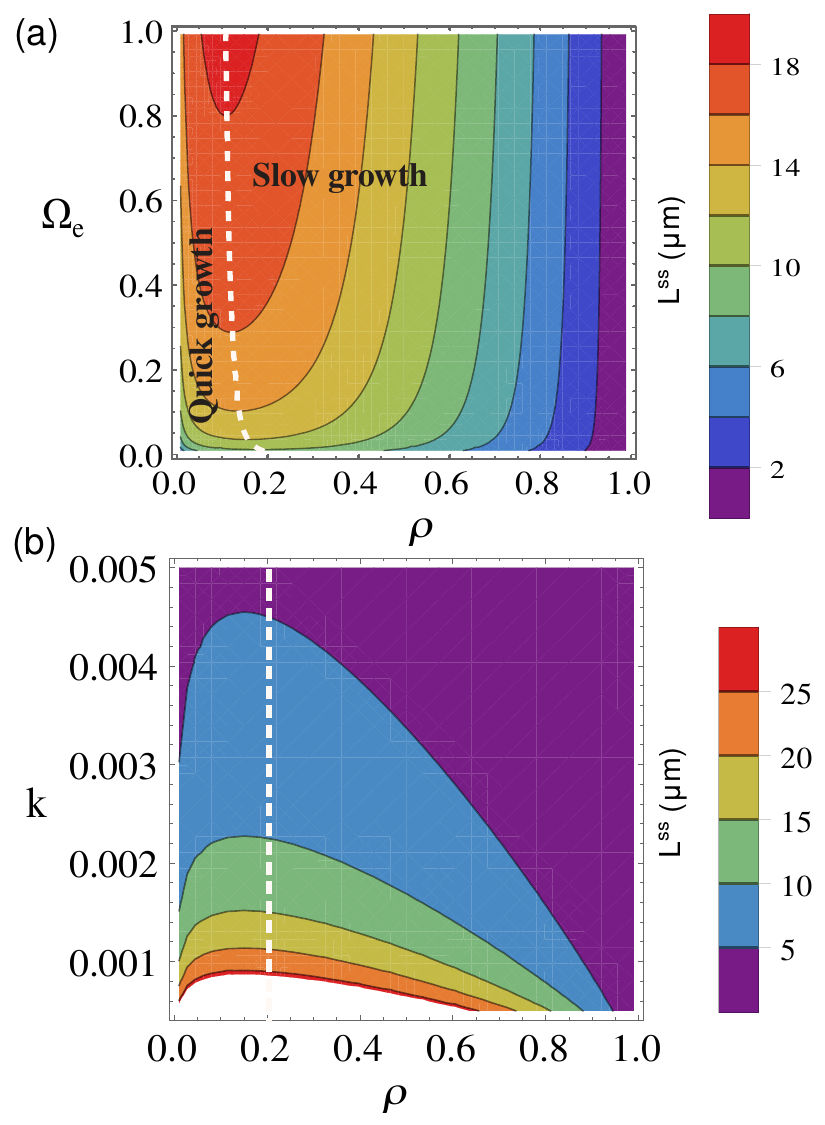}
\end{center}
\caption{ {\bf Contour Plots for $L^{ss}$}: (a) Contours of constant $L^{ss}$ on the $\rho$-$\Omega_e$ plane, keeping $k$ and $\Gamma_r$ constant. (Parameters: $k=0.0018$, $\omega^+=5 \times 10^{-6}$, $\omega^-=5 \times 10^{-8}$ and $\Gamma_r=10^{-6}$) (b) Contours of constant $L^{ss}$ in the $\rho$-$k$ plane, keeping $\Omega_e$ and
$\Gamma_r$ constant. (Parameters:$J=\rho(1-\rho)$, $v=1-\rho$, $k=0.0018$, $\omega^+=5 \times 10^{-6}$, $\omega^-=5 \times 10^{-8}$, $N_{max}=1000$, $\Omega_e=0.5$ and $\Gamma_r=10^{-5}$).}
\label{con_plot}
\end{figure}

As we show in this subsection, the density $\rho$ gives rise to interesting features of the flagellar length dynamics.
We have explored the combined effect of $\rho$, $k$ and $\Omega_e$ on the $L_{ss}$ (by using equation (\ref{eq-Lss1})) through contour plots. From equation (\ref{TASEP-J}), (\ref{TASEP-v}) and (\ref{eq-Lss1}), the dependence of $\langle L_{ss} \rangle$ on $\rho$, $k$ and $\Omega_e$ is given by
\begin{equation}
\langle L_{ss} \rangle = \frac{p(1-\rho)}{2k} \ell og \bigg{[} \frac{p  \rho \ \Omega_e }{(1-\rho) \Gamma_r}  \frac{\langle N_{ss} \rangle}{N_{max}} \bigg{]}
\label{eq-Lss-rho}
\end{equation} 
For a particular density $\rho$, a higher value of $\Omega_e$ results in a longer  $L_{ss}$  (see figure \ref{con_plot} (a)). On the other hand, for a fixed value of $\rho$, $L_{ss}$ decreases with increasing $k$ (see figure \ref{con_plot} (b)).

$L_{ss}$ exhibits non-monotonic variation with $\rho$. For a fixed $\Omega_e$, as we increase $\rho$, the steady state flagellum length $L_{ss}$ increases  with $\rho$. But for the values of $\rho$, which lie on the right side of the white-dotted line in the contour plot in figure \ref{con_plot} (a), the flagellum length $L_{ss}$ keep on decreasing with increasing $\rho$. Similar trend is seen in the second contour plot as well (figure \ref{con_plot} (b)). When $\rho$ is on the left of the dotted white line (figure \ref{con_plot} (b)), with increasing $\rho$, flux increases which, in turn, increases the supply of precursor at the tip and thus results in longer flagellum with time. On the other hand, when $\rho$ further increases, both the velocity $v$ and the flux $J$ decrease due to congestion of the IFT particles (which can be verified from equation (\ref{TASEP-J}) and (\ref{TASEP-v})). Therefore, the timer has to spend more time in the slow-moving congested traffic thereby increasing the probability that it is in the state $S_{-}$ when it returns to the base. In such situations the timer conveys the wrong message that the flagellum is long enough and prevents additional loading of precursor onto the IFT particles.

The properties of the TASEP, that represents the traffic of IFT particles, also provides a means of testing the ToF hypothesis. The ToF mechanism works satisfactory provided the average velocity of the IFT trains remain practically constant. However, if for any reason the rate of entry of the IFT particles onto the anterograde track exceeds a limit imposed by TASEP, the IFT particles may find themselves in the high-density phase in which the IFT particles would take a very long ToF and would erroneously signal against loading of the IFT particles with tubulins. Consequently, the HD phase of the TASEP would result in a shorter than usual $L{ss}$.

\section{Experimental supports for the model}
\label{sec-expt}

The adoption of the TASEP for modeling the intraflagellar traffic of IFT particles is a key new ingredient of our model. 
We summarize here the key features of TASEP, particularly in the context of IFT, before detailed discussion on the interpretation of experimental results from the perspective of TASEP. Irrespective of its load status, an anterograde IFT particle hops to the next site in the forward direction with the rate $p$ only if the target site is empty. Similarly, a retrograde IFT particle hops with the rate $p$ to the target site only if the latter is empty. Thus, the traffic flow of the IFT particles is modelled as a totally asymmetric simple exclusion process (TASEP) [47-49]. This process is  completely characterised by three parameters (see Fig.\ref{fig-TASEP4IFT}): $\alpha$ (rate with which a particle hops into the lattice at one end), $\beta$ (rate with which a particle hops out of the lattice at the other end) and $p$ (rate with which a particle hops into its nearest neighbor lattice site if the target site is empty). The three primary quantities that completely characterize the steady state of such processes are (i) average particle density $\rho$ (or, more precisely, the density profile), (ii) the average particle flux $J$ and (iii) the mean particle velocity $v$ which is defined as the average total number of sites hopped per unit time. The three different (non-equilibrium) phases can be realized on the track in the steady state of the system: (i) sparsely crowded low density (LD) phase, (ii) highly crowded high density (HD) phase and (iii) a phase with the optimal flow known as maximal current (MC) phase. The primary quantities as a function of $\alpha$, $\beta$ and $p$ in three different phases are summarised in table \ref{tab-1}.

The density $\rho$ of the IFT particles depends on the dynamical phase, i.e., whether the traffic of the IFT particles is in the LD, HD or the MC phase. We consider the IFT particle traffic to be always in the LD phase. In the LD phase, if we have the number density $\rho(\alpha,\beta,p)=\rho$, the corresponding flux $J$ and mean velocity $v$ are  unique and can be expressed as function of $\rho$ only. The flux and the mean velocity are given by 
(\ref{TASEP-J}) and (\ref{TASEP-v}), respectively.  

\begin{widetext}

\begin{figure}[h!]
\includegraphics[scale=1.0]{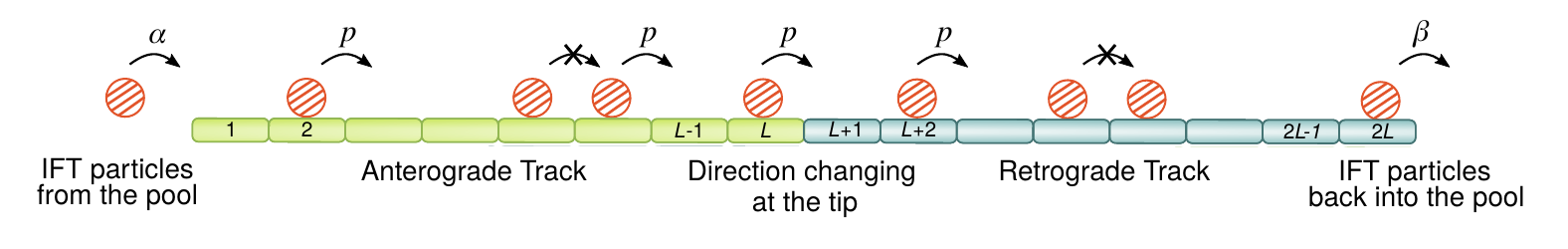}
\caption{{\bf Traffic of IFT particles is modelled as TASEP}: The pair of antiparallel anterograde (green lattice) and retrograde track (blue lattice) can be viewed as a single track connected at the tip. IFT particles enter the anterograde track from the precursor pool with rate $\alpha$  if the first site on the anterograde track is empty. And they move out of the retrograde track from the last site into the pool with rate $\beta$. In the bulk (in both anterograde and retrograde track) they hop with rate $p$ to the next neighboring site on their right if the target site is empty.}
\label{fig-TASEP4IFT}
\end{figure}
\begin{table}[h!]
\begin{tabular}{ |p{4.0cm}|p{3.0cm}|p{3.0cm}| p{3.0cm}| }
\hline
Phase  & Particle density $\rho$ & Particle flux $J$ & Particle velocity $v$ \\
\hline
LD \ ($\alpha < \beta$, $\alpha<\frac{p}{2}$) &  $\frac{\alpha}{p}$ & $\alpha(1-\frac{\alpha}{p})$ & $p(1-\frac{\alpha}{p})$\\
HD \ ($\beta < \alpha$, $\beta<\frac{p}{2}$) &   $(1-\frac{\beta}{p})$  & $\beta(1-\frac{\beta}{p})$ & $\beta$ \\
MC \ ($\alpha>\frac{p}{2}$, $\beta>\frac{p}{2}$) & $\frac{p}{2}$ & $\frac{p}{4}$ & $\frac{1}{2}$ \\
\hline 
\end{tabular}
\caption{ {\bf Quantities in different phases in TASEP}: LD (Low Density), HD (High Density) and MC (Maximal Current).}
\label{tab-1}
\end{table} 

\end{widetext}

\subsection{Experimental test for the validities of ToF mechanism and TASEP for IFT} 

A set of experiments was carried out by Ishikawa and Marshall \cite{ishikawa17} to test the validity of the ToF mechanism. In {\it Chlamydomonas reinhardtii} cells with mutant dyneins the retrograde transport was slower than that in wild type cells, as expected. But, contrary to their expectation, Ishikawa and Marshall \cite{ishikawa17} observed that the slowing down of the retrograde IFT lead to an increase in the flux of the anterograde IFT, instead of  a decrease. Based on their interpretation of the data, they believed that their observations ``rule out the time-of-flight mechanism as a means of controlling injection as a function of length''. But, 
what was missing in their analysis for the interpretation of the data is the role of the principles of TASEP. By re-interpreting their data in this subsection, in the light of the properties of TASEP, we argue that their observation is, contrary to their conclusion, fully consistent  with our ToF-based model developed in this paper.

Like shuttle trains, IFT trains cycle between the base and the tip of a  flagellum. In our model so far the times for anterograde and retrograde travel were considered to be equal although, in reality, the velocities of the IFT particles in the two directions are almost certainly different.  Moreover, after reaching the tip of the flagellum, kinesin-driven IFT particles do not immediately begin their dynein-driven retrograde journey.  Instead, upon arrival at the flagellar tip, a train detaches from the anterograde track (B-microtubule), spends some time $\tau$ at the tip in an unattached state during which it gets `remodelled' \cite{chien17}, then attaches to the retrograde track (A-microtubule) after which it starts moving towards the base from the tip \cite{buisson13}.  During remodeling a loaded IFT particle may unload the cargo (tubulin precursors), an empty IFT particle may get loaded with turned over structural protein, unbind (or deactivate) kinesins and activate dyenins (which are carried as cargo by the anterograde IFT trains). 

In order to explain their key experimental observations, the generalized expression for the time of flight ($t_{tof}$) inside the flagellum considered by Ishikawa and Marshall \cite{ishikawa17}  was  
\begin{widetext}
\begin{equation}
t_{tof}=\underbrace{(L/v_a)}_{\text{time of travel from base to tip}}+\underbrace{(L/v_r)}_{\substack{\text{time of travel from tip to base}}}+\underbrace{\tau}_{\text{time spent at the tip for remodelling}}
\end{equation}
\end{widetext}
Accordingly, the Eq.(\ref{alpha_tu}) would get generalized to  
\begin{equation}
\alpha_{tu} = \frac{\langle N(t) \rangle}{N_{max}} e^{-k\{(L/v_a)+(L/v_r)+\tau\}]}
\label{alpha_tu2}
\end{equation}
where $v_a$ and $v_r$ are the average velocities of IFT particles in the anterograde and retrograde directions, respectively. The Eq.(\ref{alpha_tu2}) implies that any decrease in the retrograde velocity $v_r$ of the IFT particles would cause decrease of $\alpha_{tu}$, i.e., probability of loading of the tubulin into the IFT particles. This observation is consistent with Ishikawa and Marshall's comment that their experimental observations on the increase of the flux of anterograde particles ``do not rule out the possibility that a time-of-flight scheme might regulate cargo loading'' \cite{ishikawa17}.

\begin{figure}
\includegraphics[width=0.95\linewidth]{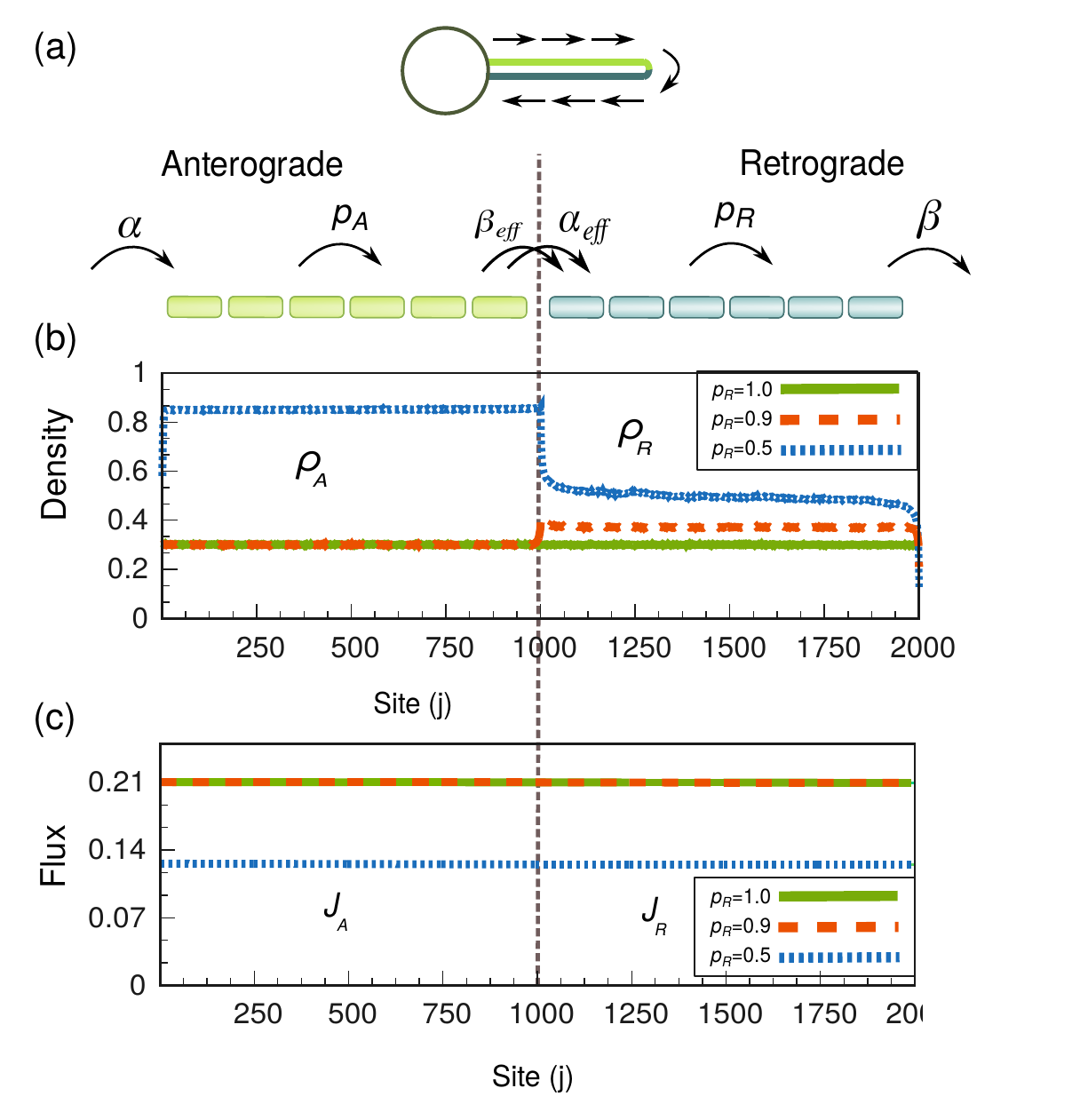}
\caption{{\bf Number density profile and flux in the steady states of the composite TASEP model of IFT}:(a) As explained earlier, the sites on the anterograde lattice are labelled by $j=1, 2, \dots, L$ from the base to the tip and those on the retrograde lattice are labelled by $j=L+1, L+2, \dots, 2L$ from the tip to the base. If the average velocity of the IFT particles during retrograde transport is lower than that during anterograde transport, the IFT becomes composite of two TASEPs, with the respective hopping rates $p_A$ and $p_R$ ($p_R < p_A$) in the anterograde and retrograde segments, respectively. The effective rate $\beta_{eff}$ of exit from the anterograde segment and the effective rate $\alpha_{eff}$ of entry into the retrograde segment must satisfy the condition that the same flux passes through both the segments in the steady state. (b-c) Keeping $\alpha=0.3$, $p_A=1.0$ and $\beta=1.0$ fixed, we plot the density and flux for three different $p_R$.  }
\label{fig-tof_validity}
\end{figure}

Following Ishikawa and Marshall \cite{ishikawa17}, the concept of remodeling time $\tau$ has been introduced in Eq.(\ref{alpha_tu2}) only for the sake of completeness of our discussion. But, in our actual calculation we have used $\tau=0$ and incorporated its effect indirectly through effective rates $\beta_{eff}$ and $\alpha_{eff}$ (see Fig.\ref{fig-tof_validity}) which we have obtained self-consistently by imposing steady-state condition on the flux. The assumption of steady-state condition, in turn, is justified by the fact that neither accumulation nor depletion of IFT particles with passage of time have been observed so far in any experiment. 

Next, we assign different hopping rates to the anterograde and retrograde IFT particles, thereby mimicking different average velocities of the IFT particles in the anterograde and retrograde directions. In such  situations where the rates of hopping of the IFT particles in the anterograde and retrograde transport are unequal, the TASEP effectively becomes a composite of two TASEPs in the two distinct segments that are coupled at the tip of the flagellum. As shown in Fig.\ref{fig-tof_validity}(a), $\beta_{eff}$ is the effective rate of exit of the IFT particles from the first segment (anterograde transport) while $\alpha_{eff}$ is the effective rate of entry of the IFT particles into the second segment (retrograde transport).  

As stated in section \ref{sec-singlemodel}, in the steady-state each TASEP can exist in one of the three possible dynamical phases, namely, LD, HD and MC. Thus, for a composite TASEP, as in Fig.\ref{fig-tof_validity}(a), the phase of the system in the steady state can be denoted by the symbol ${\cal P}_{A}|{\cal P}_{R}$ where ${\cal P}_{A}$ and ${\cal P}_{R}$ refer to the phases of the anterograde and retrograde segments, respectively. Naively, it may appear apriori that the system can exist in nine distinct composite phases ${\cal P}_{A}|{\cal P}_{R}$ where each of ${\cal P}_{\mu}$ ($\mu=A$ or $R$) can be in LD, or HD or MC phase.  Since the same steady state flux has to be sustained in both the segments, not all of the nine phases are physically realizable. Only those composite phases are stable which can maintain a single steady flux through the entire composite system. The physical implications of this principle will be established in this section.

Let us begin our discussion here with the simplest situation $p_A=1.0=p_R=p$. Moreover, we select $\alpha=0.1$ and $\beta=1.0$ so that $\alpha$ is rate limiting. Under this condition, the TASEPs in both the segments (i.e., on anterograde and retrograde direction) are in LD phase, i.e., the composite phase is $LD|LD$. The resulting average density of the IFT particles in both the anterograde and retrograde segments is 
$\rho_{A} = \alpha/p = \rho_{R}$ and the corresponding flux is $J_{A} = \alpha(1-(\alpha/p)) = J_{R}$ (see the curves corresponding to $p_{R}=1.0$ in Figs.\ref{fig-tof_validity}(b) and (c)).

\begin{figure}
\includegraphics[width=0.95\linewidth]{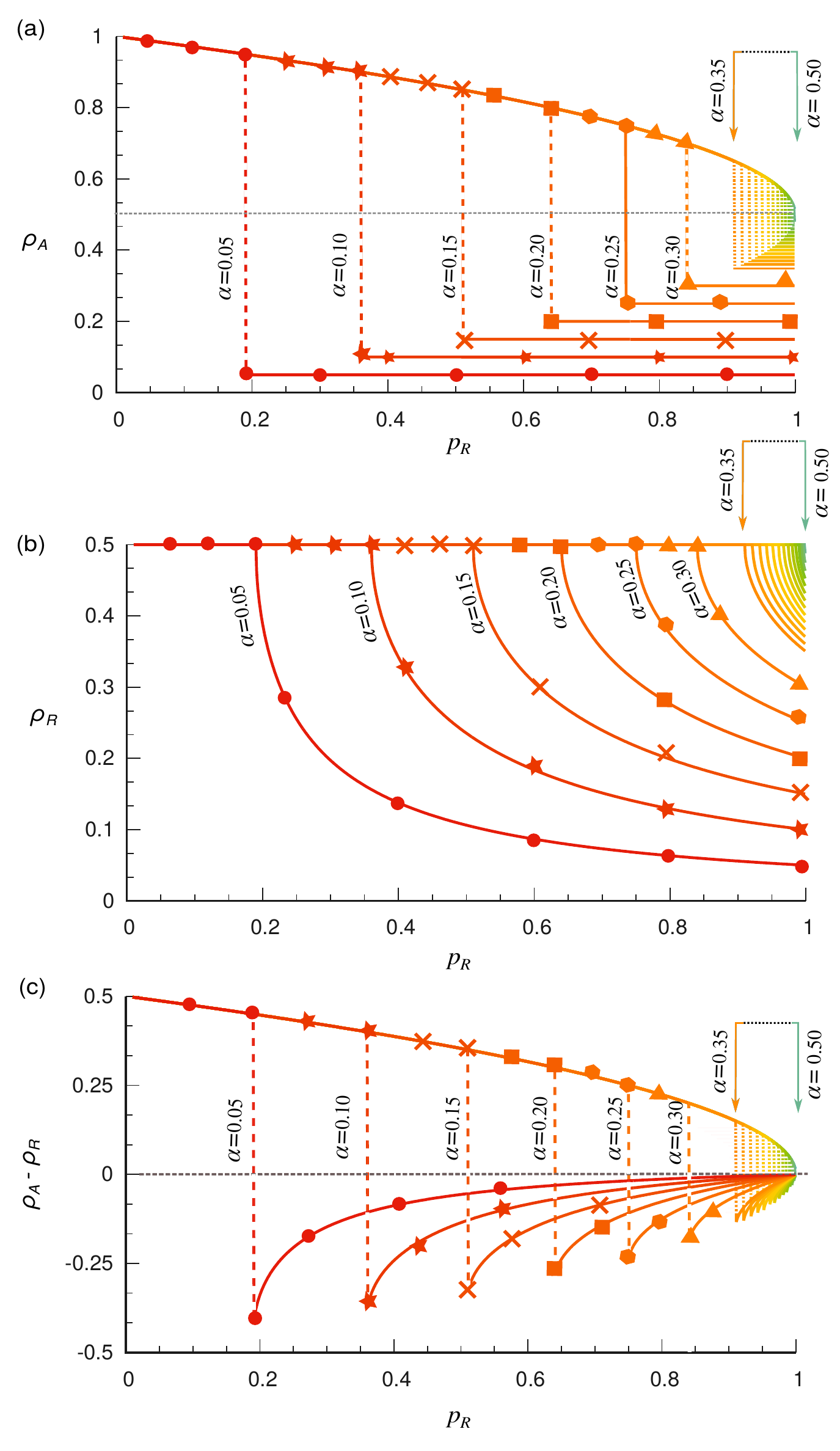}
\caption{{\bf Number densities in the steady states of the composite TASEP model of IFT}: For several different values of the $\alpha$ ($\alpha=0.05$ to $\alpha=0.5$), we plot (a) $\rho_A$ (b) 
$\rho_R$ and (c) $\rho_A-\rho_R$ as functions of $p_R$, keeping $p_{A}=1.0=\beta$ fixed. The system exhibits a transition from the composite phase $LD|LD$ to $HD|MC$ at $p_R* = \frac{(4 \alpha p_A-4 \alpha^2)}{p_A}$. At $p_R=p_R*$, $\rho_{R}$ hits its maximum value (and remains constant with further decrease of $p_{R}$) and $\rho_{A}$ increases by a discontinuous jump resulting in a discontinuous jump also in $\rho_{A}-\rho_{R}$. The magnitudes of these  discontinuous jumps, shown by the dotted vertical lines, in (a) and (c) decrease with increasing $\alpha$ and vanish as $\alpha \to 0.5$.}
\label{fig-Densities}
\end{figure}

As $p_{R}$ decreases, without change in the value of $p_A$, the densities $\rho_{A}$ and $\rho_{R}$ in the two segments change in such a way that the condition $J_{A}=J_{R}$ continue to be satisfied by the two steady-state fluxes $J_{A}$ and $J_{R}$ in the anterograde and retrograde directions. Expressing $J_{A}$ and $J_{R}$ in terms of $p_{A}$, $p_{R}$ and the unknown $\alpha_{eff}$, we get the equation
\begin{eqnarray}
\alpha(1-(\alpha/p_A))=\alpha_{eff}(1-(\alpha_{eff}/p_R)) \nonumber \\
\end{eqnarray} 
whose solution yields 
\begin{eqnarray}
\alpha_{eff}=\frac{p_A p_R-\sqrt{p_A p_R}\sqrt{4 \alpha^2 -4 \alpha p_A+p_A p_R }}{2p_A}
\label{eq-alphaeff}
\end{eqnarray} 
From (\ref{eq-alphaeff}), we find that when $p_{R}$ is decreased, $\alpha_{eff}$ remains real as long as 
$p_{R} > p_{R}^{*}$, with 
\begin{equation}
p_R^{*}  = \frac{(4 \alpha p_A-4 \alpha^2)}{p_A}
\end{equation}
is  satisfied. In such situations, both the segments are in their respective LD phases and the composite system is still in the $LD|LD$ phase although $\rho_{A} \neq \rho_{R}$ because $\alpha \neq \alpha_{eff}$ (see the curves corresponding to $p_{R}=0.9$ in Figs.\ref{fig-tof_validity}(b) and (c)).  

However, if 
\begin{equation}
p_{R} < p_{R}^{*}
\label{p_less}
\end{equation}
the retrograde segment cannot sustain the anterograde flow. The flux in both the segments is controlled by $p_R$ which is now rate limiting. If the condition (\ref{p_less}) is satisfied, the retrograde segment is in the MC phase while 
the anterograde segment is in the HD phase so that the composite system exhibits the $HD|MC$ phase (see the curves corresponding to $p_{R}=0.5$ in Figs.\ref{fig-tof_validity}(b) and (c)). So, now the steady-state condition in terms of the equality of the fluxes $J_{A}$ and $J_{R}$ gives
\begin{eqnarray}
\beta_{eff}(1-(\beta_{eff}/p_A))=p_R/4 \nonumber \\
\end{eqnarray} 
whose solution gives the expression
\begin{eqnarray}
\beta_{eff}=\frac{1}{2}(p_A-\sqrt{p_A^2-p_Ap_R})
\end{eqnarray} 
for $\beta_{eff}$. The average IFT particle density in anterograde segment is now given by 
\begin{equation}
\rho_{A}=1-\beta_{eff}
\end{equation}
while the corresponding flux is $J_{A} = p_{R}/4$. The condition, in terms of $p_{R}$, and the corresponding anterograde density and common flux are summarized in table II. 

\begin{table}
\begin{tabular}{ |c| c| c| }
\hline
 &  &  \\
Condition & $J_A=J_{R}=J$ & $\rho_{A}$ \\
 &  &  \\
\hline
 &  &  \\
 $p_R>\frac{(4 \alpha p_A-4 \alpha^2)}{p_A}$ & $\alpha(1-(\alpha/p_A))$  & $\alpha/p_A$ \\
  &  &  \\
\hline
 &  &  \\
$p_R<\frac{(4 \alpha p_A-4 \alpha^2)}{p_A}$ & $p_R/4$  &  $1-\beta_{eff}$ \\ 
 &  &  \\
\hline   
\end{tabular}
\label{tab-condition_summary}
\caption{Steady state properties of composite two-TASEP model.}
\end{table}

The results plotted in Figs.\ref{fig-tof_validity}(b) and (c) correspond to a specific value of $\alpha$. In order to illuminate the role of $\alpha$, we plot $\rho_{A}$, $\rho_{R}$ and $\rho_{A}-\rho_{R}$ in Fig.\ref{fig-Densities} as functions of $p_{R}$, keeping $p_{A}=1.0=\beta$ fixed. For every given value of $\alpha$ the system exhibits the composite phase $LD|LD$ for all $p_R> p_{R*}$; although $\rho_{A}$ remains unaffected, $\rho_{R}$ continues to increase with the decrease of $p_{R}$ because of the corresponding change of $\alpha_{eff}$. Exactly at $p_R=p_{R*}$ the system makes a transition to the composite phase $HD|MC$ where $\rho_{A}$ increases by a discontinuous jump and $\rho_{R}$ attains its maximum value. With further decrease of $p_{R}$, $\rho_{A}$ continues to increase while $\rho_{R}$ now remains unaffected. 

The most interesting point here is that, for a given $\alpha$, $\rho_{A}-\rho_{R}$ changes sign at $p_R=p_{R*}$ so that for $p_R<p_{R*}$, $\rho_{A} > \rho_{R}$. The higher values of $\rho_{A}$ for $p_R<p_{R*}$ than the value for $p_{R} > p_{R*}$ is consistent with the higher intensity observed by Ishikawa and Marshall \cite{ishikawa17} in the case of IFT with mutant dyneins. We believe that the ``injection intensity'' that Ishikawa and Marshall \cite{ishikawa17} claimed to have measured in their experiment is actually proportional to the average density, rather than flux, of the IFT particles in the anterograde segment. 

With the above interpretation of the experimental observations and comparison with our theoretical predictions, we establish that both the (i) time-of-flight mechanism for length control, and (ii) description of the traffic of IFT particles in terms of TASEP are consistent with experimental observations \cite{ishikawa17}.

\subsection{Role of depolymerases in the `balance-point' scenario} 

\begin{figure}
\begin{center}
\includegraphics[width=0.85\columnwidth]{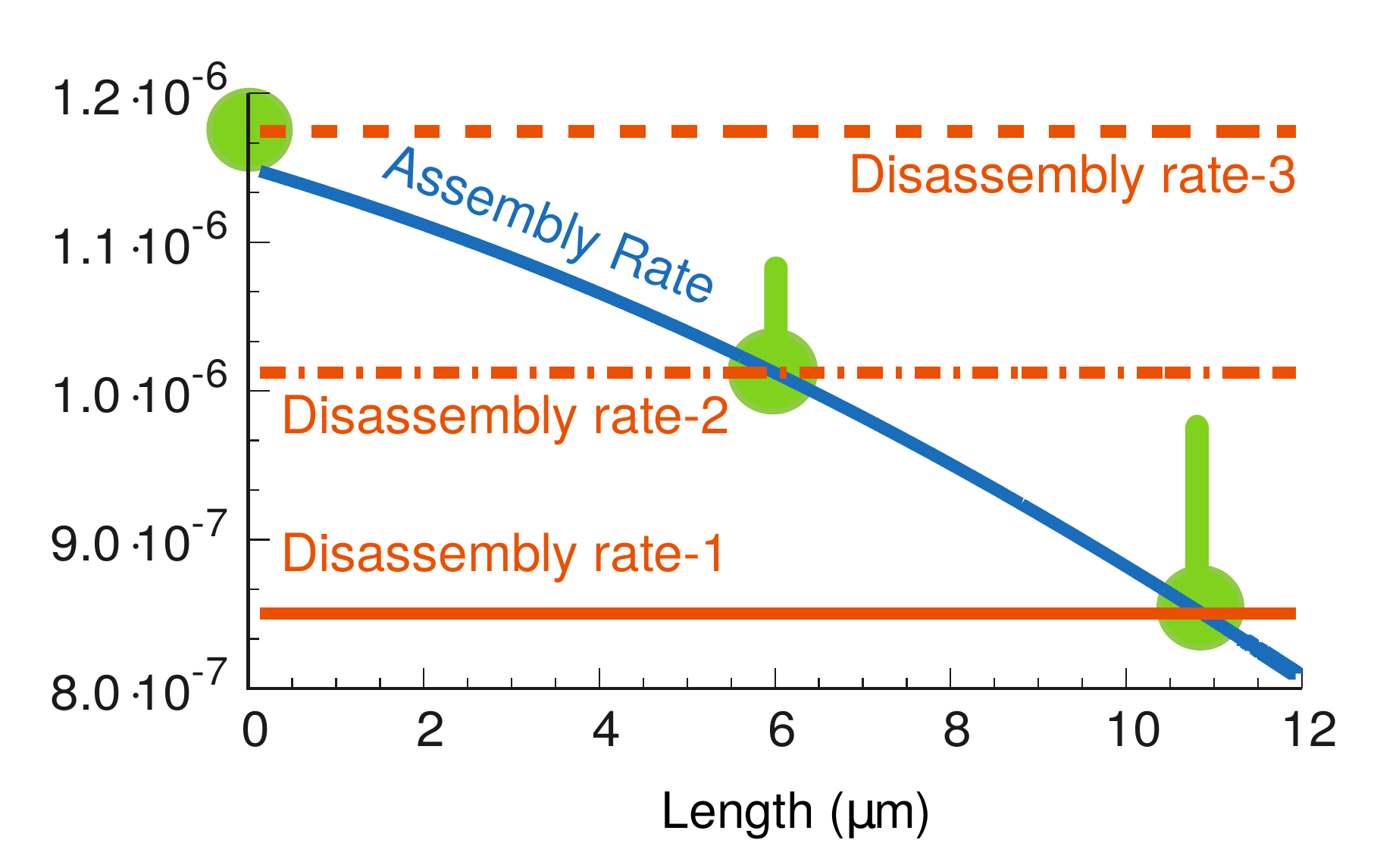}
\end{center}
\caption{{\bf Shift of the balance point with change of the disassembly rate}:.
Common parameter for assembly and disassembly curve: $\rho = 0.1$. For assembly curve: $J = 0.09$; $v =0.9 $;
$ k = 0.00045$; $\Omega_e = 0.10$; $\omega^+= 10^{-6}$; $\omega^-= 5 \times 10^{-7}$; $N_{max} = 3000$; For disassembly curve-1: $ \Gamma_r=1 \times 10^{-6}$, For disassembly curve-2: $\Gamma_r= 1.25 \times 10^{-6}$, For disassembly curve-3: $\Gamma_r= 1.45 \times 10^{-6}$.}
\label{fig-depoly_balance}
\end{figure}

By a series of experiments, Pan and coworkers \cite{piao09,wang13,snell04} established the following facts: \\
(i) Flagellar shortening requires the depolymerases to the extent that the shortening is inhibited in depolymerase-depleted cells. \\
(ii) In the steady state, the depolymerases are almost exclusively located in the cell body and very little traces of it are found in the flagella. However, when flagellar shortening is triggered by internal cues or external signals, the depolymerases are rapidly transported to the flagellar tip where these begin depolymerization of the axonal MTs. \\
(iii) Since the depolymerases in CR do not posses the domains required for active motor-like walk towards the plus-end of the MTs, the only plausible mode of their rapid transport to the flagellar tip is as cargo on anterograde IFT that are driven by other families of processive kinesin motors. 

In our model, continuation of turnover of the tubulins in the steady-state requires depolymerization rate to be non-zero (as for the Disassembly rate-1 in Fig.\ref{fig-depoly_balance}). However, shortening of the flagella during resorption can occur in two different ways. In the first, the polymerization probability $\Omega_{e}$ can be switched off, without altering the depolymerization rate $\Gamma_{r}$, thereby triggering resorption (see Fig.\ref{fig-ciliogen_resorp_regen}). In the second, the depolymerization rate $\Gamma_{r}$ increases abruptly, without any change in the polymerization probability $\Omega_{e}$ (see Fig.\ref{fig-deflag_depolymerase}) thereby shifting the balance points to a shorter length \cite{marshall02} as shown in Fig.\ref{fig-depoly_balance}. In the latter case if the shifted balance point still correspond to a non-zero length, the flagella shorten, but resorption is only partial (as for the Disassemmbly rate-2 in Fig.\ref{fig-depoly_balance}). But, if the increase of $\Gamma_{r}$ is sufficiently large, the resulting shift of the balance points can be so large that the steady-state corresponds to vanishing length of the flagella indicating complete resorption (as for the disassembly rate-3 in Fig.\ref{fig-depoly_balance}). This scenario of depolymerase-induced resorption is consistent with the experimental observations of Pan and co-workers \cite{piao09,wang13}, but quite different from the length-dependent depolymerization proposed recently in ref.\cite{fai18}. The mechanisms of flagellar length control that we have postulated in this paper are also different from that,  proposed for control of length of microtubules, based on a length-dependent feedback on polymerization by kinesin Kip2  \cite{hibbel15}.

\section{Stochastic model for length control in biflagellates}
\label{sec-doublemodel}

In the preceding section we have developed a model for length control of a single flagellum. Analyzing that model and comparing its predictions with known empirical facts, we have established the validity of the hypotheses on which the model is based. In this section we couple two such model flagella to develop a theoretical model for flagellar length control in biflagellates. The emphasis of this section is in the study of cooperative effects of the coupling.

In addition to all the simplifications listed above for the dynamics of a single flagellum, we make one more simplification regarding the coupling of the dynamics of the two flagella in a biflagellate. The dynamics of the two flagella are coupled via the common pool of flagellar protein precursors at the base; we consider explicitly only the tubulins, the building blocks of axonemal MTs, in this pool because those are the most dominant component in it.  The flagella are also assumed to share a common pool of IFT particles. That is why the same flux $J$ of IFT trains appear in the master equations of the two flagella. A timer molecule returning to the base upon completion of a round trip in a flagellum dwells in the decision chamber providing the feedback required for the differential loading of an IFT train that is poised to begin its next journey. Thereafter the timer goes back to the pool at the base, gets re-charged and waits for next hitch-hiking on another IFT train.

\subsection{Master equations for a biflagellate}

Let $P_{L_1}(j,t)$ ($P_{L_2}(j,t)$) be the probability that the length of flagellum $f_1$ ($f_2$) at time $t$ is $L_1(t)=j$ ($L_2(t)=j$). The master equation governing the stochastic kinetics of the length of flagellum $f_1$ and $f_2$,  
given in appendix \ref{appendix-masterBIFLAG}, are appropriate generalizations of the master equations for a single flagellum.

\subsection{Rate Equations for length control in biflagellates}
The equations governing the evolution of average length $\langle L_1(t) \rangle$ and $\langle L_2(t) \rangle$ of flagellum $f_1$ and $f_2$ are
\begin{eqnarray}
\frac{d\langle L_1(t) \rangle}{dt}={\bigg{[}\frac{\langle N(t) \rangle}{N_{max}}e^{-2k\langle L_1(t) \rangle/v}\bigg{]}} J \Omega_e -(1-\rho)^2 \Gamma_r  \nonumber\\
\frac{d\langle L_2(t) \rangle}{dt}={\bigg{[}\frac{\langle N(t) \rangle}{N_{max}}e^{-2k\langle L_2(t) \rangle/v}\bigg{]}} J \Omega_e -(1-\rho)^2 \Gamma_r  \nonumber\\
\label{rate_l1_l2}
\end{eqnarray}
and the equations governing the evolution of the average precursor  population $\langle N(t) \rangle$ in the pool is 
\begin{eqnarray}
\frac{d \langle N(t) \rangle}{dt} & =&\omega^+\bigg{[}1-\frac{\langle N(t) \rangle}{N_{max}}\bigg{]} -\omega^-\langle N(t) \rangle  \nonumber\\  \nonumber\\ 
&-& \frac{d\langle L_1(t) \rangle}{dt}-\frac{d\langle L_2(t) \rangle}{dt} 
\label{rate_pool2}
\end{eqnarray}

\section{Results on length control of a biflagellate}
\label{sec-doubleresult}

\subsection{Ciliogenesis, resorption and subsequent regeneration}

CR cells lose their flagella by one of the two well known mechanisms called (i) {\it resorption}, and (ii) deflagellation  \cite{quarmby05}. During the process of {\it resorption} a flagellum is gradually retracted into the cell. In contrast, {\it deflagellation} refers to the process of shedding of the flagella that involves severing of the entire flagellum from its base \cite{quarmby04}. In this subsection we present results  obtained from our model for the process of resorption, and subsequent regeneration of the flagella. Our results for the process of deflagellation will be presented in the next subsection.

\begin{figure}
\begin{center}
\includegraphics[width=0.420\textwidth]{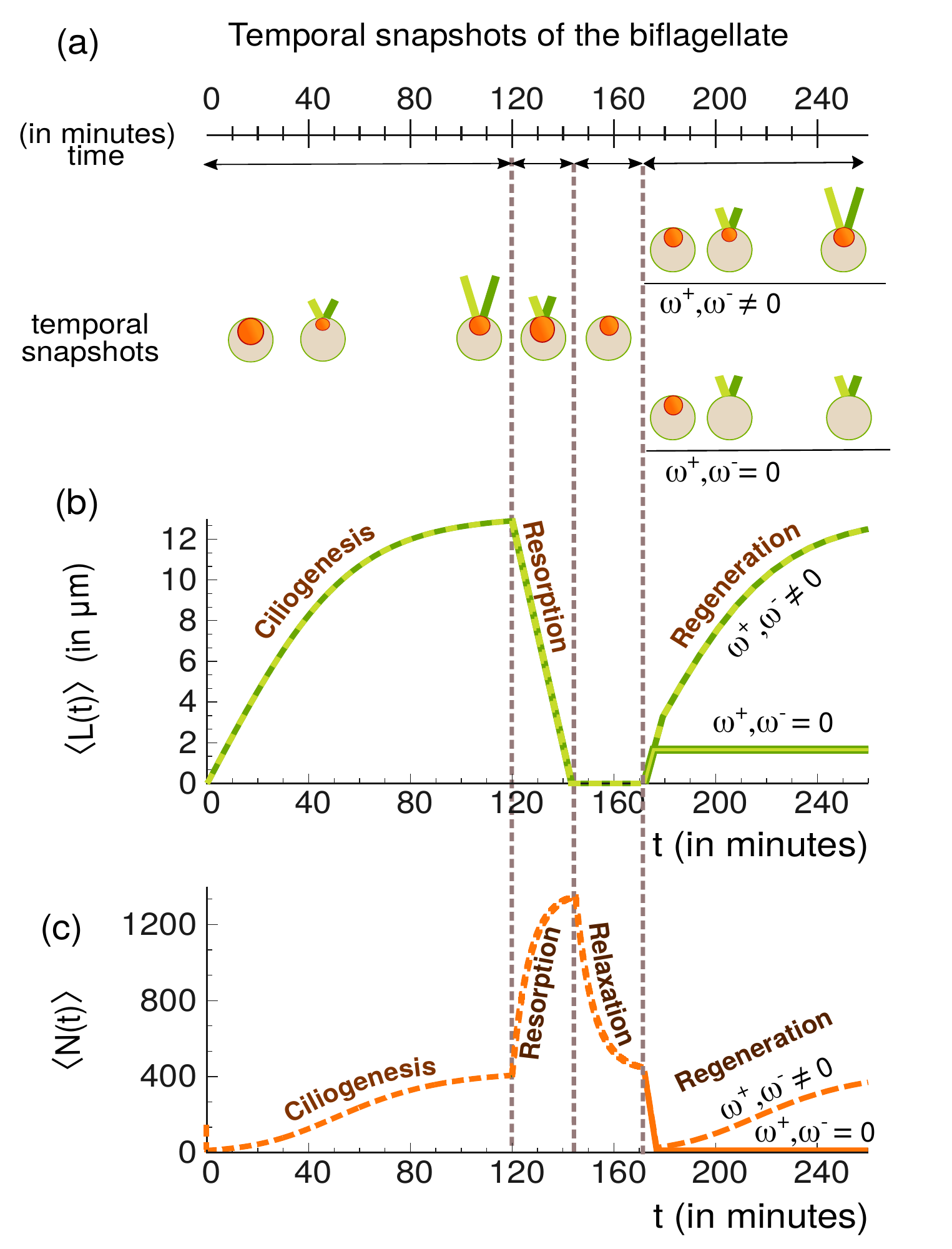}
\end{center}
\caption{ {\bf Ciliogenesis, polymerization-blocked resorption, followed by regeneration, of the flagella}: After completion of ciliogenesis, the rate constant $\Omega_e$ is set to zero to mimic blocking of polymerization of the axonemal MTs, resulting in resorption of both the flagella. After allowing sufficiently long time for relaxation of the precursor population, the rate constant $\Omega_{e}$ is restored to its pre-resorption non-zero value which triggers regeneration of the flagella that eventually attain their steady-state lengths. However, the steady-state lengths  achieved during this regeneration phase depend on the rates of synthesis and degradation of the precursor proteins in the common pool. 
Common parameters used: $\rho=0.09$, $J=0.0819$, $v=0.91$,  $k=0.0011$, $\Gamma_r=4.0 \times 10^{-4}$,  $\omega^+=3.0 \times 10^{-4}$, $\omega^-=5.0 \times 10^{-7}$, $N_{max}=1500$, $\delta t=2.88 \times 10^{-4}$ s. For ciliogenesis and regeneration phase: $\Omega_e=0.75$ and for resorption phase $\Omega_e=0$.}
\label{fig-ciliogen_resorp_regen}
\end{figure}

The plot of lengths $\langle L_{1}(t) \rangle$ and $\langle L_{2}(t) \rangle$ of the two flagella is the simplest, and most direct, way of presenting the empirical data on ciliogenesis. The slope of each of the curves at a given time $t$ indicates the rate $V_{1}(t)$ and $V_{2}(t)$, respectively, of elongation of the corresponding flagella at that instant of time. In most of the systems the rates $V_{1}(t)$ and $V_{2}(t)$ decrease with increasing $t$ and eventually, after a time interval $T$, vanish as the flagella attain their steady-state lengths 
$\langle L_{1}^{ss} \rangle$ and $\langle L_{2}^{ss} \rangle$. These qualitative features of the experimental data are captured very well by the numerical results obtained by solving the rate equations (\ref{rate_l1_l2}), together with the Eq.(\ref{rate_pool2}) for the given initial conditions $\langle L_{1}(0) \rangle =0= \langle L_{2}(0) \rangle $, 
$\langle N(0) \rangle = N_{0}$ (see Fig. \ref{fig-ciliogen_resorp_regen}). The rate constants have been tuned so as to obtain $\langle L_{1}^{ss} \rangle  = 12 \mu m = \langle L_{2}^{ss} \rangle$, which is the typical length of the flagella of wild type CR in the steady-state \cite{ishikawa17}. 


\begin{figure}
\begin{center}
\includegraphics[width=0.420\textwidth]{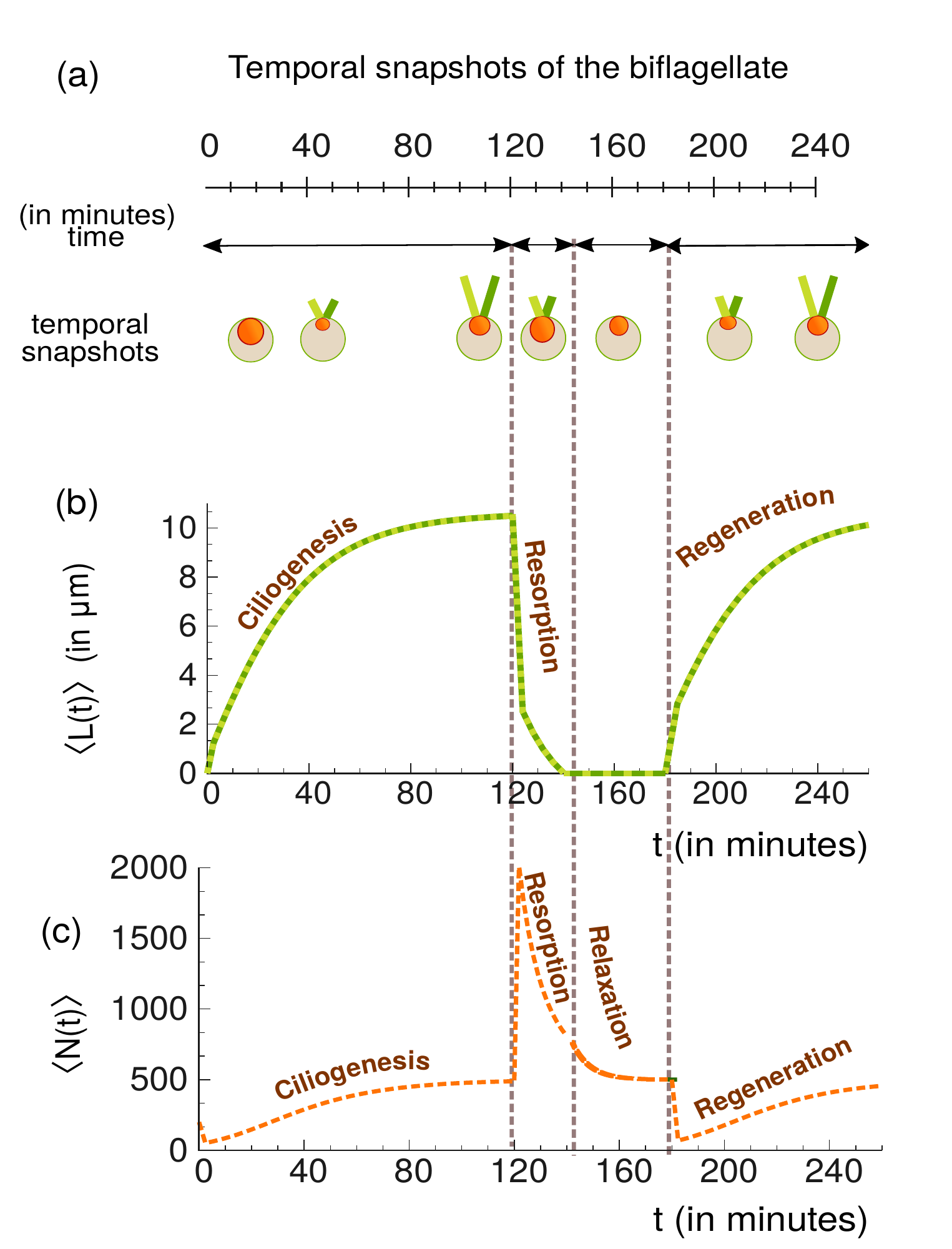}
\end{center}
\caption{{\bf Ciliogenesis, depolymerase-induced resorption, followed by regeneration, of the flagella}: After completion of ciliogenesis, the rate constant $\Gamma_r$ is increased ten-fold to mimic depolymerization of the axonemal MTs by depolymerase motor proteins. This depolymerization results in resorption of both the flagella. Allowing sufficiently long time for relaxation of the precursor population, the rate constant $\Gamma_{r}$ is restored to its pre-resorption value triggering regeneration of the flagella that eventually attain their pre-resorption steady-state lengths. Parameters used for the plots are $\rho=0.09$, $J=0.0819$, $v=0.91$,  $k=0.0011$,$\Omega_e=0.3$, $\omega^+=3.0 \times 10^{-4}$, $\omega^-=5.0 \times 10^{-7}$, $N_{max}=3000$, $\delta t=2.4 \times 10^{-4}$ s.  For ciliogenesis and regeneration phase:  $\Gamma_r=2.0 \times 10^{-4}$, and for resorption phase  $\Gamma_r=8.0 \times 10^{-3}$ (20 times stronger).}
\label{fig-deflag_depolymerase}
\end{figure}

\subsubsection{Resorption in our model: a plausible scenario}

In our numerical studies of the model we mimicked the resorption process by setting the term 
$[\frac{\langle N(t) \rangle}{N_{max}}e^{-2k\langle L_1(t) \rangle/v}] J \Omega_e$ to zero which implies either the rate $\Omega_e=0$ (vanishing of elongation rate), or $J=0$ (vanishing rate of flux of the IFT trains), or $e^{-2k\langle L_1(t) \rangle/v}=0$ (vanishing  rate of tubulin loading). In that situation, because of the nonvanishing $\Gamma_{r}$, the lengths of both the flagella keep decreasing till both $\langle L_{1}(t) \rangle = \langle L_{2}(t) \rangle \to 0$, manifesting as the phenomenon of {\it resorption} as shown in  Fig.\ref{fig-ciliogen_resorp_regen}(a-b). 

Note that  during resorption none of the structural proteins constituting the flagella are lost by the cell; instead, those are actually returned to the basal pool \cite{quarmby04,quarmby05}. Suppose resorption begins when the system is in the steady state. If the synthesis and degradation of the structural proteins were blocked as the resorption begins, then at the end of resorption the net population of structural proteins in the pool would have been 
$\langle N^{ss} \rangle + \langle L_{1}^{ss} \rangle + \langle L_{2}^{ss} \rangle$.  However, if the synthesis and/or degradation of the structural proteins are not blocked and the resorption is not sufficiently rapid, then 
$N_{df} \neq \langle N^{ss} \rangle + \langle L_{1}^{ss} \rangle + \langle L_{2}^{ss} \rangle$ where $N_{df}$ is the population of structural proteins in the pool at the moment of completion of resorption. 

If $[\frac{\langle N(t) \rangle}{N_{max}}e^{-2k\langle L_1(t) \rangle/v}] J \Omega_e$ remains zero for sufficiently long time even after disappearance of the two flagella, the population of the precursors in the common pool relaxes to the new steady-state corresponding to $\Omega_{e}=0$. This {\it relaxation} of the precursor pool population is also shown in Fig.\ref{fig-ciliogen_resorp_regen}(c). 

Resorption does not remove the basal bodies \cite{rosenbaum67}. Therefore, the same basal bodies remain available for regeneration of the flagella. If the elongation rate $\Omega_{e}$ is again switched on at this stage, the {\it regeneration} of the two flagella proceed in a manner qualitativelty similar to that during ciliogenesis (see Fig.\ref{fig-ciliogen_resorp_regen}) and both flagella eventually regain the respective original steady-state lengths 
$\langle L_{1}^{ss} \rangle  = 12 \mu m = \langle L_{2}^{ss} \rangle$.

Rosenbaum et al. \cite{rosenbaum69} found that if CR were deflagellated in cycloheximide, a known inhibitor of protein synthesis, then upon regeneration the flagella can attain only a length of about  6$\mu$m whereas the normal full length of flagella in CR is about 12 $\mu$m. This result established that the CR cells maintain a pool of the essential structural proteins that can be exploited for regeneration of flagella. But, in the absence of fresh synthesis of these proteins, the existing pool is not adequate for regeneration upto the full length of 12 $\mu$m. This feature is also reproduced by our model, as depicted in Fig.\ref{fig-ciliogen_resorp_regen}(a-b). \\

\subsubsection{An alternative scenario of resorption in our model} 

Based on a series of experiments, Pan and collaborators \cite{piao09,wang13,liang16} have suggested that shortening of the  flagella, which requires depolymerization of the axonemal MTs, is dominantly driven by MT depolymerases which belong to distinct families of kinesin motors \cite{howard07,varga09,walczak13}. Those experiments also indicated that under normal conditions the population of the depolymerases in the flagella is negligibly small. However, upon receiving a specific signal, depolymerases rush into the shaft of a flagellum and quickly reach the distal tips of the MTs where they begin MT depolymerazation at a high rate. 

In order to establish that our model is capable of capturing the experimentally indicated role of depolymerases in resorption, we abruptly impleted a ten-fold increase of the rate $\Gamma_{r}$ well after the flagella attained their steady-state values $L^{ss}$, without altering the numerical value of the growth term  
$[\frac{\langle N(t) \rangle}{N_{max}}e^{-2k\langle L_1(t) \rangle/v}] J \Omega_e$. The abrupt increase of 
$\Gamma_{r}$ causes resorption. However, allowing sufficient time for relaxation to the new steady-state, if the numerical value of the parameter $\Gamma_{r}$ is restored to its pre-resorption value, the two flagella again regain their pre-resorption lengths $L^{ss}$ through regeneration process. For this case, the evolution of flagellar length and pool population are shown in Fig.\ref{fig-deflag_depolymerase}(a-c).

\subsection{Flagellar dynamics after deflagellation and subsequent regeneration} 

When subjected to environmental stress like, for example, extreme temperatures or pH or presence of detergents or alcohols in the medium \cite{johnson93,engel09,lefebvre95}, a CR cell itself severs its flagella so that each flagellum abruptly shortens to a length $f \langle L^{ss} \rangle$ where $f=0$ corresponds to shedding of the entire flagellum. That is why deflagellation is also referred to as flagellar excision, flagellar shedding or flagellar autotomy \cite{quarmby04}. 
The dynamics of regeneration of flagella after deflagellation need not be identical to those after resorption because the structural proteins that constitute the severed part of a flagella are lost by the cell during deflagellation whereas the structural proteins are gradually retracted into the common pool during resorption.

\begin{figure}
\begin{center}
\includegraphics[width=0.420\textwidth]{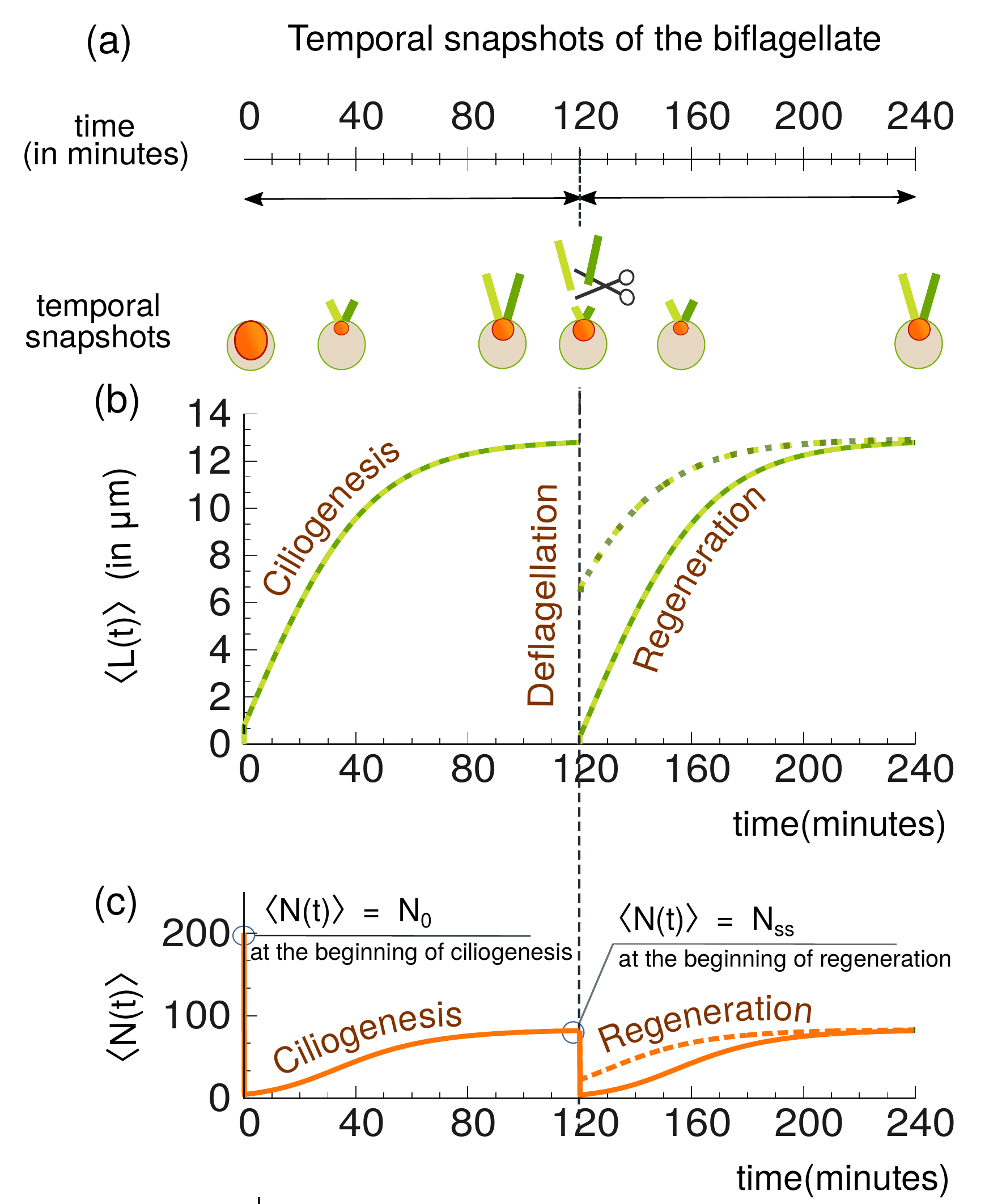}
\end{center}
\caption{ {\bf Ciliogenesis and deflagellation followed by regeneration of both flagella}: After ciliogenesis, the lengths of each flagellum is simultaneously, and abruptly, reduced to a shorter value $f \langle L^{ss} \rangle$ where $ 0 \leq f < 1$; this process mimics {\it deflagellation}. Since numerical values of all the model parameters were kept unchanged during this process, {\it regeneration} of both the flagella begin immediately and finally both attain their original steady-state lengths. The lower and upper curves in the `regeneration' part correspond to $f=0$ (severing of the entire length of each flagellum, and $f=0.5$ (severing of the distal half of each flagellum), respectively. Parameters used for this plot are $\rho=0.09$, $J=0.0819$, $v=0.91$,  $k=0.00105$,$\Omega_e=0.75$,  $\Gamma_r=3.0 \times 10^{-4}$,  $\omega^+=4.5 \times 10^{-4}$, $\omega^-=4.5 \times 10^{-6}$, $N_{max}=500$,  $N_0=200$,  $N_{ss} \approx 81$,  $\delta t=3.6 \times 10^{-4}$ s.}
\label{fig-ciliogen_deflag}
\end{figure}

In the {\it in-silico} experiments with our model, we mimicked deflagellation by abruptly, and instantaneously, reducing the lengths of each of the two flagella to a shorter value $f \langle L^{ss} \rangle$ where $ 0 \leq f < 1$ without altering the numerical value of any of the model parameters. The data for $f=0$ and $f=1/2$ are plotted in Fig.\ref{fig-ciliogen_deflag}(a-c). Immediately after the deflagellation, the existing pool has to provide the much needed initial resources for the regeneration of the flagella. Consequently, in the immediate aftermath of severing of the flagella, the population $\langle N(t) \rangle$ of the precursors in the pool decreases (see Fig.\ref{fig-ciliogen_deflag}). However, in the mean time,  enhanced synthesis of the flagellar components begins; these freshly synthesized proteins not only replenish the depleted pool but also become available for the continued growth of the flagella. IFT particles moving inside a flagellum at the instant of amputation also get lost. But this loss of IFT particles has negligible effect because the pool of IFT particle is generally quite large and only a small fraction of IFT particle participate in shuttling inside the flagellum (roughly 20$\%$) \cite{cole09,silva12}. Both the flagella and the population of the precursors in the common pool  eventually attain their respective original steady-state values irrespective of the value of $f$.

\subsection{Flagellar dynamics after selective amputation and subsequent regeneration} 

In the context of deflagellation, discussed above, both the flagella were severed to equally shorter lengths. In this subsection we consider the more general case where the two flagella are severed unequally. We refer to this process as selective amputation in order to distinguish it from the process of deflagellation. For simplicity, we consider the scenario where one of the two flagella is selectively severed to a length  $f \langle L^{ss} \rangle$ 
($0 \leq f < 1$) while the other flagellum remains intact, the special case of this situation corresponding to $f=0$ is usually referred to as ``long-zero case''. 

The curiosity-driven exploration of the consequences of amputation of flagella of unicellular eukaryotes began almost seventy years ago when regeneration of severed flagella was first observed \cite{chen50}. The first quantitative study of the kinetics of regrowth of the shortened flagella was reported soon thereafter \cite{lewin53}.
Since then the mechanisms of flagellar length regulation under wide varieties of conditions and chemo-physical perturbation have been investigated with many species of flagellated eukaryotes using several different experimental techniques with increasing sophistication  \cite{lefebvre86}. 
In their pioneering works Rosenbaum and coworkers \cite{rosenbaum67,rosenbaum69,coyne70,dentler77} used either paralyzed strains or applied compression through a coverslip to hold the cells under study for direct viewing (see, for example, \cite{rosenbaum69}). Both types of perturbations are likely to affect the objects and processes of interest in this context. In recent times, ingenious experimental methods have been developed that avoid possible adverse effects on the normal physiology of the flagellated cells under investigation \cite{ludington12}.
All those experiments helped in collecting wealth of information not only on the regeneration of the severed flagellum  but also on the effects of this selective amputation and regeneration on the length of the unsevered flagellum. 

In the``long-zero case'', the unsevered flagellum is found to resorb rapidly while the severed one begins to elongate. When the resorbing unsevered flagellum and the regenerating amputated flagellum attain the same length, both elongate at the same rate till regaining their original (equal) steady-state lengths. In principle, a cell could sense the damage/amputation of a flagellum by the loss of a function that crucially depends on the undamaged full-length normal flagellum. However, a paralyzed flagellum, which is disabled to perform its function of driving fluid flow, can still regenerate upon amputation \cite{ludington12}. This experimental evidence indicates that the ``loss-of-function'' is neither a mode of sensing damage/amputation of a flagellum nor the stimulus for fresh synthesis of the flagellar proteins by the cell. Therefore, how the unsevered flagellum senses the amputation of its partner and how it responds to this perturbation by initiating own resorption remains one of the challenging open questions on this phenomenon.

\begin{figure}
\begin{center}
\includegraphics[width=0.40\textwidth]{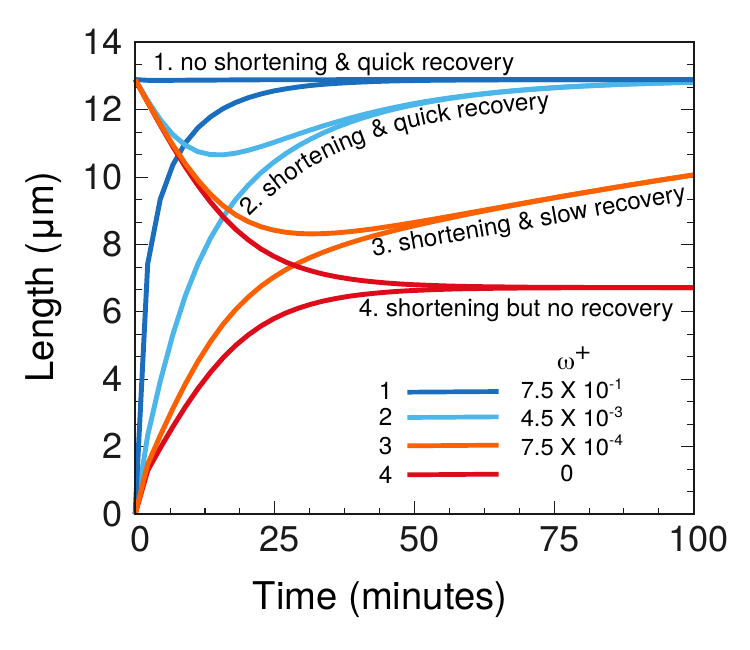}
\end{center}
\caption{ {\bf Regeneration of an amputated flagellum in the `long-zero' case}: The lengths of both the flagella are plotted against time starting from the instant when one of the flagella is amputated from its base leaving the other intact; this type of amputation of flagella of biflagellates is referred to as the `long-zero' case. The four different sets of curves correspond to four different values of $\omega^{+}$. Each pair curves plotted with the same color  correspond to the lengths of the two flagella for same $\omega^{+}$ where the monotonically increasing curve denotes the growing length of the regenerating amputated flagellum. The parameters used for this plot are $\rho=0.09$, $J=0.0819$, $v=0.91$, $\Omega_e=0.75$, $\Gamma_r=3.0\times 10^{-4}$, $k=1.05\times10^{-3}$,$N_{max}=500$, $\delta t=3.6\times 10^{-4}$ s.}
\label{fig-amput_regen}
\end{figure}

We mimicked the long-zero amputation and subsequent regeneration in our model by chosing the initial conditions 
$\langle L_{1}(0) \rangle = \langle L_{1}^{ss} \rangle$, $\langle L_{2}(0) \rangle = 0$, $\langle N(0) \rangle = \langle N^{ss} \rangle$.  The flux $J$ of IFT particles in the two flagella is same because both the flagella share the same pool of IFT particles and amputation of one flagellum doesn't affect the overall population of IFT particles in the pool \cite{cole09,silva12}. The data for four different values of $\omega^{+}$ are plotted in Fig.\ref{fig-amput_regen}. The qualitative trend of variation of $\langle L_{1}(t) \rangle$ and $\langle L_{2}(t) \rangle$ for the two intermediate values of $\omega^{+}$ are consistent with the empirically observed facts; for both the unsevered flagellum shortens initially till equalization of its length with the elongating severed partner and then the two flagella grow together to full recovery. The flagellar proteins released by the shortening flagellum is utilized by the elongating flagellum during the early stages of the latter's regeneration \cite{coyne70}. Subsequently, unless suppressed by inhibitors, fresh synthesis of flagellar proteins provides the material needed for full growth of the two flagella to their pre-amputation original lengths.

\begin{figure}
\begin{center}
\includegraphics[width=1.0\columnwidth]{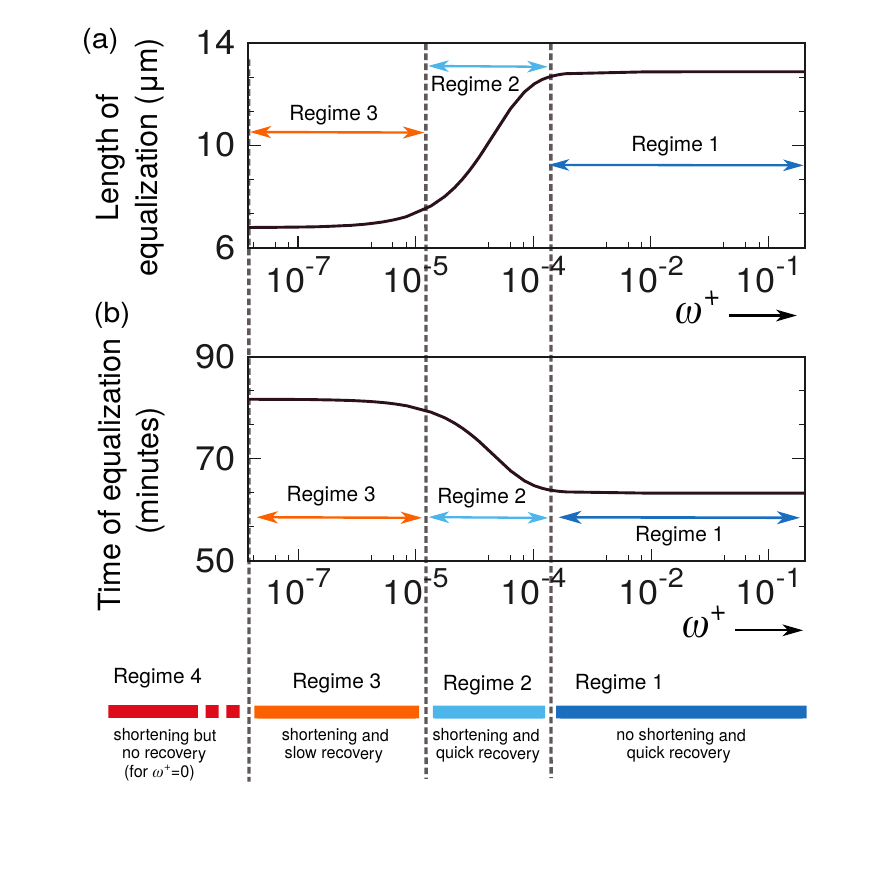}
\end{center}
\caption{{\bf Length and time of equalization as a function of $\omega^+$}: (a) Length at which equalization of the two flagellar lengths happens during regeneration, after amputation in the `long-zero' case. (b) The corresponding time of equalization of the lengths of the two flagella in the `long-zero' case; here time is measured from the instant of amputation. The color code is identical to that in Fig.\ref{fig-amput_regen}. The parameters used for this plot are $\rho=0.09$, $J=0.0819$, $v=0.91$, $\Omega_e=0.75$, $\Gamma_r=3.0\times 10^{-4}$, $k=1.05\times10^{-3}$, $N_{max}=500$, $\delta t=3.6\times 10^{-4}$ s. }
\label{fig-Equalization}
\end{figure}

Moreover,  the larger is the numerical value of $\omega^{+}$ the quicker is the recovery. In fact, in the case of the highest value of $\omega^{+}$ used in Fig.\ref{fig-amput_regen}, the recovery of the amputated flagellum is so quick that practically no shortening of the unsevered flagellum is observed. On the other hand, in the opposite extreme case where $\omega^{+}=0$, only the unsevered flagellum supplied the structural proteins required for the growth of the amputated flagellum; consequently, both the flagella can attain a steady-state length of only $L^{ss}/2 \simeq 6 \mu$m, as observed earlier experimentally. Thus, the nature of the kinetics of regeneration of the amputated flagellum depends on the kinetics of synthesis of the precursor proteins in the common pool (see Fig.\ref{fig-Equalization}).

\begin{figure}
\begin{center}
\includegraphics[width=0.43\textwidth]{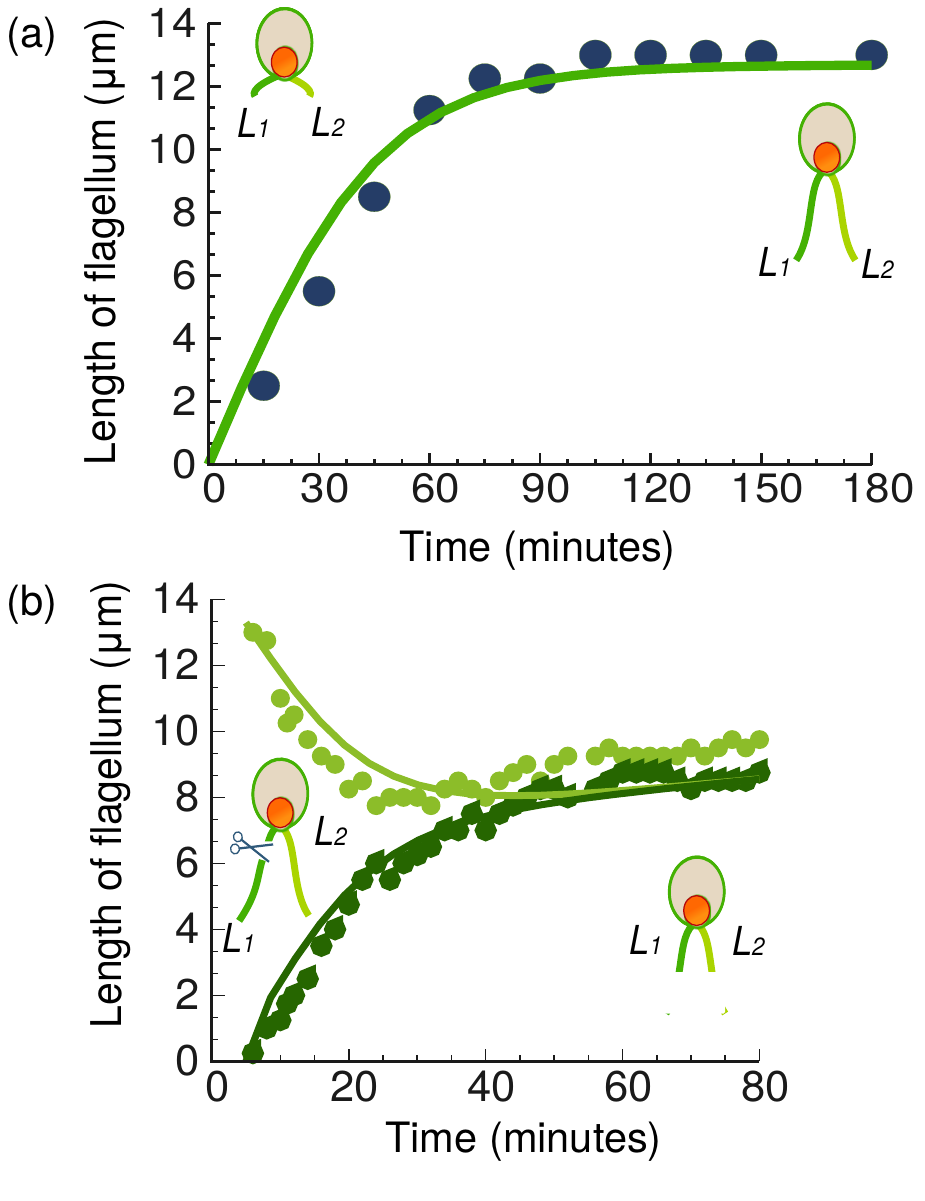}
\end{center}
\caption{ \textbf{Comparison of theory with experimental data on ciliogenesis and flagellar regeneration after selective amputation:} (a) Simultaneous growth of the two flagella during ciliogenesis. The dots denote the experimental data taken from Ishikawa and Marshall \cite{ishikawa17} while the continuous curves have been obtained solving the coupled equations (\ref{rate_l1_l2}-\ref{rate_pool2}). The parameters used for this plot are $\rho=0.08$, $J=0.0736$, $v=0.92$, $\Omega_e=0.65$, $\Gamma_r=3.0\times 10^{-4}$, $k=1.0\times10^{-3}$, $\omega^+=4.5 \times 10^{-4}$, $\omega^-=4.5 \times 10^{-6}$, $N_{max}=500$, $\delta t=3.6\times 10^{-4}$ s. (b) Regeneration after selective amputation of a single flagelum. The dots denote the experimental data taken from Luddington et al. \cite{ludington12} while the continuous curves have been obtained by solving the coupled equations (\ref{rate_l1_l2}-\ref{rate_pool2}). For this plot, $\omega^+=4.5 \times 10^{-5}$, $\omega^-=4.5 \times 10^{-7}$, $\delta t=3.6\times 10^{-4}s$.    }
\label{fig1}
\end{figure}

We tested whether the model explains the experimental observation i.e, the elongation/resorption pattern of the amputated/unamputated flagellum of the CR as reported by Ludington et al.\cite{ludington12} and Ishikawa et al. \cite{ishikawa17}. For this purpose, we selected numerical values of all the parameters to get the best fit between the experimental data \cite{ishikawa17} and our theoretical prediction of the time-dependence of flagellar lengths during ciliogenesis (see Fig.\ref{fig1}(a)). Then, using the same numerical values of all the parameters, except ten times smaller values of $\omega^+$ and $\omega^-$, we could get excellent fit to our theory and the experimental data on the time-dependence of the flagellar lengths following amputation in the long-zero case (see Fig.\ref{fig1}(b)).


\subsection{Beyond mean: fluctuations and correlation}

\begin{figure*}
\begin{center}
\includegraphics[width=0.8\textwidth]{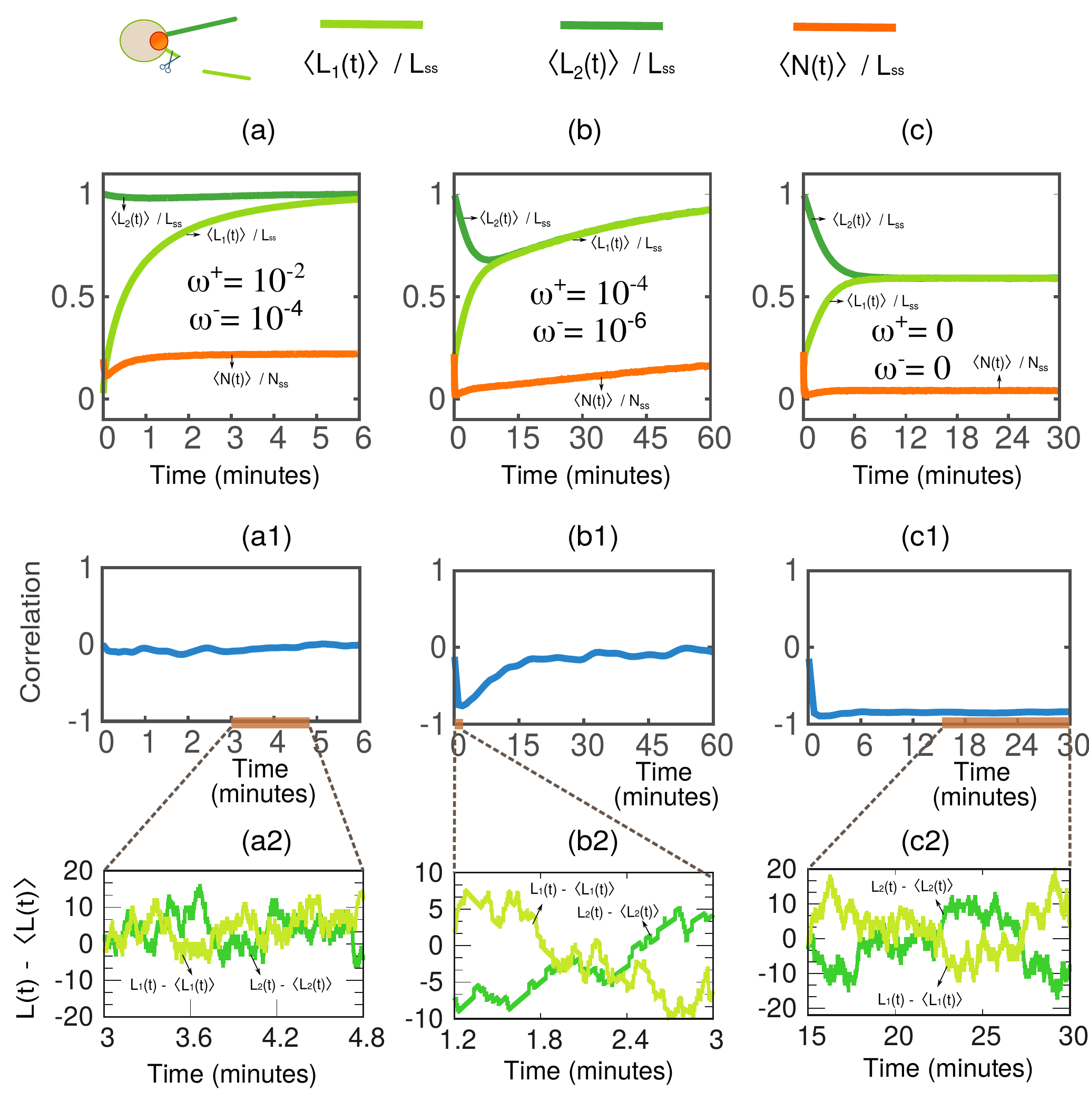}
\end{center}
\caption{{\bf Length fluctuation and Correlation}: The correlation of the fluctuations of the lengths of the two flagella during regeneration in the `long-zero' case are plotted in (a1), (b1) and (c1) under three different conditions shown in (a), (b) and (c), respectively.  The fluctuations in the lengths of the two flagella during short intervals in the plots (a1), (b1) and (c1) are shown explicitly in (a2), (b2) and (c2), respectively. The common numerical values of the parameters used for this figure are $\rho=0.1$, $J=0.09$, $v=0.9$,  $k=0.008$, $\Omega_e=0.5$,  $\Gamma_r=5.0 \times 10^{-4}$,  $N_{max}=100$, $N_0=50$  $\delta t=3.6 \times 10^{-4}$ s.}
\label{fig3}
\end{figure*}


For the numerical computation of the correlations, we begin with the following definitions: 
suppose, the total number of realizations generated is $n$. Let $L_1^i(t)$ and $L_2^i(t)$ denote the length of flagellum-1 and 2 at time $t$ in $i^{th}$ realization.
The instantaneous {\it mean} lengths of the two flagella are defined by 
\begin{equation}
\langle L_1(t) \rangle=\frac{\sum_{i=1}^n {L_1}^i(t)}{n}, ~{\rm and}~ \langle L_2(t) \rangle=\frac{\sum_{i=1}^n {L_2}^i(t)}{n},
\end{equation}
while the corresponding {\it variances} are given by
\begin{eqnarray}
Var(L_1)= \bigg{[}\frac{1}{n-1}{\sum_{i=1}^n(\langle L_1(t) \rangle - {L_1}^i(t))^2}\bigg{]}^{1/2} \nonumber \\
Var(L_2)= \bigg{[}\frac{1}{n-1}{\sum_{i=1}^n(\langle L_2(t) \rangle - {L_2}^i(t))^2}\bigg{]}^{1/2}.  \nonumber \\
\end{eqnarray}
and the \textit{covariance} $Cov(L_1L_2)$ is given by
\begin{eqnarray}
\nonumber\\
\frac{1}{n-1}\bigg{[}{\sum_{i=1}^n(\langle L_1(t) \rangle - {L_1}^i(t))(\langle L_2(t) \rangle - {L_2}^i(t))}\bigg{]}^{1/2}
\end{eqnarray}
In terms of these {\it variances} and the {\it covariance}, 
the \textit{correlation} between the flagellar lengths is defined as 
\begin{eqnarray}
Corr(L_1L_2)=\frac{Cov(L_1L_2)}{Var(L_1)Var(L_2)};
\end{eqnarray}
and it gives a quantitative measure of the correlation of fluctuations in the lengths of the two flagella. 
Then set of $n$  realizations of stochastic trajectories are generated by simulating the model using Monte-Carlo methods as described in Appendix \ref{appendix-MCsteps}.

We studied the $Corr(L_1L_2)$ for three different cases: (i) negligibly small resorption of the unsevered flagellum before equalization of its length with that of regenerating flagellum (Fig.\ref{fig3} (a)), (ii) significant shortening of the unsevered flagellum till equalization of the lengths of the two, followed by recovery of pre-amputation steady-state lengths of both (Fig.\ref{fig3} (b)) and (iii) significant shrinkage of the unsevered flagellum till both the flagella attain a steady-state length of  $\approx L^{ss}/2$ and stop growing further (Fig.\ref{fig3} (c)). 

In the case (i), the correlation remained zero throughout the regeneration process (see figure \ref{fig3} (a1)). Since the proteins required for the regeneration of the amputated flagellum are supplied exclusively by the precursor pool, leaving the unamputated flagellum practically unaffected, there is no correlation between the length fluctuations of the two flagella (see Fig.\ref{fig3} (b1)). 

In case (ii), the correlation exhibited a nonmonotonic behavior;  starting from the initial value zero, it became negative and its absolute value increased with the passage of time till it attained its minimum beyond which it increased gradually to its final value zero (see figure \ref{fig3} (a2)). The correlation was found to be negative during the initial period when the shortening unamputated flagellum made significant contribution to the supply of proteins that sustained the regeneration of the amputated flagellum. This fact is demonstrated clearly by the plots in Fig.\ref{fig3} (b2)). 

In case (iii), the correlation became negative as soon as the amputated flagellum started growing at the expense of the unamputated flagellum and remained negative throughout, even after both the flagella attained their new steady lengths (see figure \ref{fig3} (c1)). Since in this case synthesis and degradation of the precursors were blocked, any increase of the length of one of the two flagella had to be compensated by the corresponding decrease in the length of the other, i.e., the fluctuations in the lengths of the two flagella were always anti-correlated. This is clearly visible in Fig.\ref{fig3} (c2)).

From all these three cases it could be concluded that the correlation between the fluctuation of lengths of the flagellum is strongly related to the precursor population in the pool. Actually, through this precursor pool both the flagellum interact. When sufficient precursor is present in the pool to support the regeneration of the amputated flagellum, the correlation is vanishingly small (Figs.\ref{fig3} (a), (a1), (a2)). But, during those time intervals  when the precursors get depleted, then one flagellum grows at the cost of other, leading to negative correlation (Fig.\ref{fig3} (b), (b1), (b2)). And in those extreme situations where one flagellum can grow only at the expense of the other, correlation remains negative for the whole time (Fig. \ref{fig3} (c), (c1), (c2)).

\section{Comparison with other models}
\label{sec-modelcomp}

A summary of all the known theoretical models of flagellar length control and critical analysis of their implications was reported few years ago by Luddington et al. \cite{ludington15}. Several of those models could be ruled out through their systematic analysis. But, some others, which could not be discarded by the experimental evidence, still remain as plausible, although alternative, scenarios for flagellar length control. One of these is based on a `time of flight' (ToF) mechanism which was considered subsequently by Ishikawa and Marshall \cite{ishikawa17} while analyzing their experimental data. Based on their physical interpretation of the experimental data, Ishikawa and Marshall concluded that their data do not support the ToF mechanism. In contrast, invoking some subtle features of TASEP, which represents IFT in our model, we argue that the experimental observations of Ishikawa and Marshall are consistent with the ToF mechanism. 

Note that eq.(\ref{rate_l}) can be expressed as 
\begin{eqnarray}
\frac{d \langle L(t) \rangle}{dt} = k_{1}~  C_{p}~ T(\langle L \rangle) - k_{2} 
\label{rate_l2}
\end{eqnarray}
with $k_{1}=J \Omega_{e}$, $C_{p} = \langle N(t) \rangle/N_{max}$, $T(\langle L \rangle) = e^{-2k \langle L(t) \rangle /v}$ and $k_{2} = (1-\rho)^2 \Gamma_r$. The form (\ref{rate_l2}) looks exactly like the eq.(1) in the supplementary information of ref.\cite{ludington12}. However, the crucial difference between (\ref{rate_l2}) and eq.(1) in the supplementary information of ref.\cite{ludington12} is that $T(\langle L \rangle)$ in (\ref{rate_l2}) is given by a mathematical expression that follows naturally from the ToF mechanism whereas it was treated as a phenomenological parameter in ref.\cite{ludington12}.

The length-dependent growth and length-independent decay of flagella is at the foundation of Marshall and Rosenbaum's  `balance-point' model \cite{marshall01}. It has been used also in a stochastic model of flagellar length control developed by Bressloff \cite{bressloff06}. In the original version of the balance-point model \cite{marshall01} it was implicitly assumed that each IFT particle carries flagellar structural proteins as cargo. One of the key explicit assumptions of  that version of the balance-point model was that the number of IFT particles and their average speed remain constant in time. Therefore, in that case, the decrease of the flagellar assembly rate with its increasing length could arise only if the rate of the arrival of the IFT particles decreased with the increase of flagellar length. But, this scenario was in direct contradiction with the subsequent experimental observation of Dentler et al. \cite{dentler05}. In the revised balance-point model \cite{engel09} an attempt was made to reconcile the balance-point concept with the experimental observation of Dentler et al. \cite{dentler05} in terms of the variation in the sizes of the IFT trains. 
(see \cite{rathinam19} for an extension of the Bressloff's work to a stochastic version of the revised balance-point model.

The length-dependent effective {\it assembly} rate and a length-independent {\it disassembly} rate of each individual  flagellum in our model (see Fig.\ref{plot_1}(a)) is consistent with the general concept of ``balance-point''  \cite{marshall01,marshall05,engel09}. However, the length-dependence of the effective assembly rate arises in our model from the {\it differential loading} of the IFT particles with flagellar structural proteins. The concept of differential loading was proposed earlier qualitatively \cite{wren13,craft15}; it is  now incorporated quantitatively in our theoretical model.

Marshall and coworkers \cite{hendel18} developed an alternative model where kinesin motor proteins, that diffuse on their way back to the base from tip, serve as `rulers'. In this model the steady state length of the flagellum is given by 
\begin{equation}
L_{ss} = \biggl(\frac{2N~D~\delta L}{d}\biggr)^{1/2} 
\label{eq-LssHendel}
\end{equation}
where $N$ is the number of diffusing motors, $D$ is their diffusion constant, $\delta L$ is the increment of flagellar length when a motor reaches its tip, and $d$ is the rate of shortening (decay) of flagellar length. In spite of the differences in the underlying length control mechanisms the expressions (\ref{eq-Lss1}) and (\ref{eq-LssHendel}) for $L_{ss}$, shares one common feature. The steady-state length of the flagellum is determined by the balance of the competing length-dependent growth rate and length-independent decay rate. 

The more recent model developed by Fai et al.\cite{fai18} is based on Hendel et al.'s postulate \cite{hendel18} that the diffusing kinesins act as rulers for lenth control. 
Fai et al. \cite{fai18} use IFT particles and motors interchangeably throughout the paper without explicitly stating that they do not distinguish between the two. In contrast, the model developed by Hendel et al. \cite{hendel18} does not specifically represent the IFT particles. In fact, Hendel et al.\cite{hendel18} assumed that ``each motor is associated with an IFT particle carrying a fixed quantity of material''. 
The slight difference in the expressions for $L_{ss}$ derived by Hendel et al.\cite{hendel18} and that of Fai et al.\cite{fai18} arises from difference in the scenarios considered by the two. The two assumptions made by Hendel et al. \cite{hendel18} are: (i) ``a constant source of free motor protein at the tip'' and, (ii) ``motors that have reached the base immediately transport back to the tip''. Under these special conditions (i.e., ``no tubulin depletion'' and ``instantaneous ballistic motion'' \cite{fai18}), as Fai et al. point out \cite{fai18}, the more general form of the expression $L_{ss}$ reduces to that of Hendel et al. \cite{hendel18} In this sense Fai et al.'s result is slightly more improved compared to that of Hendel et al. \cite{hendel18}.  
One key feature of Fai et al.'s flagellar length control model is that the rate of shortening of a flagellum is also length dependent. This is in sharp contrast to all the other models of balance-point scenario where shortening rates are independent of the flagellar length. 

As we have discussed above, a balance point in the context of flagellar length control can arise from length-dependent rates of growth or/and shrinkage of axonemal MTs. In other words, at least one of the two competing rates (assembly and disassembly rates) should be length dependent \cite{mohapatra16}. In this way a balance emerges between the assembly and disassembly and gives rise to a time-independent average length of the filament in the steady-state. Specifically, in our model the balance point results from a length-dependent growth and length-independent shrinkage of the axonemal MTs. However, the existence of a balance point is not a unique feature of MTs. This phenomenon occurs also in actin filaments where the rates of attachment and detachment of subunits at the barbed and pointed ends can exactly balance each other provided at least one of them is length-dependent \cite{mohapatra16,erlenkamper09}.

\section{Summary and conclusions}
\label{sec-summary}

In this paper we have developed a rather general theoretical model for eukaryotic flagellar length control. 
This model successfully integrates the following ingredients within a single theoretical framework:
(i) a ToF mechanism for length sensing, (ii) a length-dependent differential loading of the IFT particles  \cite{wren13,craft15}, and (iii) representation of IFT as a totally asymmetric simple exclusion process (TASEP).

We have analyzed the model at two different levels. The intrinsic fluctuations in the quantities of interest are obtained analytically from the master equations and the Fokker-Planck equations, and numerically from MC simulations. Most of these results are new predictions that, in principle, can be tested experimentally. 
The deterministic rate equations derived from the master equations account for the well known time-dependent, as well as the steady-state, properties of the system. 

Next we list the main results of our analysis.
(a) Quantification of the length-dependent growth rate, in terms of the length-dependent differential loading of the precursor proteins, and length-independent shrinkage rate gives rise a mechanism of attaining the steady-state length $L_{ss}$; this scenario is consistent with the concept of balance-point introduced, and elaborated, earlier in the literature \cite{marshall01,marshall05,engel09}.

(b) Our results highlight the important role of the population kinetics of the structural precursor proteins in the common shared pool at the base of the flagella. In some physiologically relevant range of parameters, we demonstrate that during ciliogenesis the elongating flagellum can overshoot beyond its steady-state length $L_{ss}$ before relaxing back to $L_{ss}$. Such overshooting, although not reported so far, is expected to be observed in the parameter range that we propose. 

(c) In the context of the length coordination between the two flagella of biflagellates, it has been known for decades that, during regeneration of the amputated flagellum the unamputated flagellum exhibits a non-motononic variation of its length: initially it  shortens till its length becomes just equal to that of the regenerating flagellum and beyond this point both the flagella grow together maintaining approximately equal length till attaining their pre-amputation steady-state lengths. Our model not only reproduces this non-trivial collective dynamics of the two flagella over a wide range of parameter values, but also reveals new qualitatively different behaviors in parameter regimes that, to our knowledge, have not been explored in laboratory experiments. 

(d) We have carried out a numerical analysis of our model mimicking the conditions under which the {\it in-vivo} experiments were carried out by Ishikawa and Marshall \cite{ishikawa17} to test the validity of the ToF mechanism. We have argued that the experimental observations are not only consistent with the ToF mechanism, but also provide experimental support for the TASEP-based description of the traffic of IFT particles. Moreover, a different numerical study of our model demonstrates that it can capture the experimentally observed \cite{piao09,wang13,liang16} role of depolymerase cytoskeletal motors in the resorption of flagella. 

(e) The stochastic version of our model has made new predictions on the nature of correlations between fluctuations of the lengths of the two flagella in steady-state as well as in states far from the steady-state.  

Thus, in spite of the simplifying assumptions, as listed in section \ref{sec-singlemodel}, the model is remarkably successful in accounting for all the known phenomena in the context of flagellar length control in biflagellated eukaryotes. Moreover, it also makes new predictions on the nature of length fluctuations and on the role of the pool of flagellar structural proteins that, in principle, can be tested experimentally. Furthermore, the stochastic formulations of the model have laid down the foundation on which more detailed structures of the theories can be constructed in future for quadriflagellate and octoflagellate eukaryotes.


{\bf Acknowledgements}: We thank Gaia Pigino for useful discussions. S.P. and D.C. thank the MPI-PKS for hospitality in Dresden where a major part of this work was done. D.C. also acknowledges support by SERB (India) through a J.C. Bose National Fellowship. 

\begin{widetext}
\appendix

\section{Derivation of rate equations for a single flagellum from master equations} 
\label{app-master2rate4single}

Now we will write the master equations governing the evolution of length and pool population in terms of $A$, $B$ and $C$- defined in equation (\ref{eq-A}-\ref{eq-C}). The complete set of master equations governing the length of the flagellum :

 For $j=0$ 
 \begin{eqnarray}
 \frac{dP_L(j,t)}{dt} =  - \biggl[e^{-Cj} \sum_{n=0}^{N_{max}}\frac{n}{N_{max}}P_N(n,t) \biggr]A \  P_L(j,t) 
+ B P_L(j+1,t)
  \end{eqnarray}

 For $j=1$ to $j=L_{max}-1$
 \begin{eqnarray}
 \frac{dP_L(j,t)}{dt} &=& \biggl[e^{-C(j-1)} \sum_{n=0}^{N_{max}}\frac{n}{N_{max}}P_N(n,t) \biggr]A \  P_L(j-1,t)   
- \biggl[e^{-Cj} \sum_{n=0}^{N_{max}}\frac{n}{N_{max}}P_N(n,t) \biggr]A \  P_L(j,t) \nonumber\\ \nonumber\\
 &+& B P_L(j+1,t)-B P_L(j,t) 
\label{eq-masterL}
  \end{eqnarray}
 
  For $j=L_{max}$
 \begin{eqnarray}
 \frac{dP_L(j,t)}{dt} = \biggl[e^{-C(j-1)} \sum_{n=0}^{N_{max}}\frac{n}{N_{max}}P_N(n,t) \biggr]A \  P_L(j-1,t) -   B P_L(j,t)
  \end{eqnarray}

The complete set of master equations governing the precursor population : \\
For $n=0$ 
\begin{eqnarray}
& \frac{dP_N(n,t)}{dt}  =-\omega^+(1-\frac{n}{N_{max}})P_N(n,t)   +\omega^-{(n+1)})P_N(n+1,t)  \nonumber\\ \nonumber\\
&+[\sum_{j=0}^{L_{max}}A e^{-Cj}P_L(j,t)][\frac{(n+1)}{N_{max}}P_N(n+1,t)]  - B P_N(n,t)
\end{eqnarray}
\\

For $n=1$ to $n=N_{max}-1$

\begin{eqnarray}
\frac{dP_N(n,t)}{dt} & =\omega^+(1-\frac{(n-1)}{N_{max}})P_N(n-1,t)-\omega^+(1-\frac{n}{N_{max}})P_N(n,t)  \nonumber\\ \nonumber\\
&+ \omega^-{(n+1)})P_N(n+1,t)-\omega^-({n})P_N(n,t)  \nonumber\\ \nonumber\\
&+[\sum_{j=0}^{L_{max}}A e^{-Cj}P_L(j,t)][\frac{(n+1)}{N_{max}} P_N(n+1,t)-\frac{(n)}{N_{max}}P_N(n,t)] \nonumber\\ \nonumber\\
&+ B[P_N(n-1,t)-P_N(n,t)]
\label{eq-masterN}
\end{eqnarray}

For $n=N_{max}$

\begin{eqnarray}
\frac{dP_N(n,t)}{dt} & =\omega^+(1-\frac{(n-1)}{N_{max}})P_N(n-1,t) -\omega^-({n})P_N(n,t)  \nonumber\\ \nonumber\\
&-[\sum_{j=0}^{L_{max}}A e^{-Cj}P_L(j,t)][\frac{n}{N_{max}}P_N(n,t)]  + B P_N(n-1,t) \nonumber\\
\end{eqnarray}

Some well known probability relations are the following
\begin{eqnarray}
&\Sigma_{j=0}^{\infty} \{ P(j,t) \}  =1  \nonumber\\
&\Sigma_{j=0}^{\infty} \{ j P(j,t) \}  =\langle j(t) \rangle \nonumber\\
&\Sigma_{j=0}^{\infty} \{ j^2 P(j,t) \}  =\langle j^2(t) \rangle \nonumber\\
\end{eqnarray}

Multiplying both the sides of the master equation (\ref{eq-masterL}) with $j$ and summing it over, we get 
\begin{eqnarray}
\sum_{j=0}^{L_{max}} j\frac{d}{dt}({P_{L}(j,t)}) &= \sum_{j=0}^{L_{max}} j \bigg{[} \underbrace{ [ \{ \sum_{n=0}^{N_{max}} \frac{n}{N_{max}} P_N(n,t) \} e^{-C(j-1)} A ] P_{L}(j-1,t)}_{\text{Term-1}} \nonumber\\&-\underbrace{[ \{ \sum_{n=0}^{N_{max}}\frac{n}{N_{max}}P_N(n,t) \} e^{-Cj} A ] P_{L}(j,t)}_{\text{Term-2}} + \underbrace{\{ B \} P_{L_1}(j+1,t) - \{ (1-\rho)^2 \Omega_r \} P_{L1}(j,t)}_{\text{Term-3}} \bigg{]}  \nonumber\\
\end{eqnarray}
where $L_{max}$ is a positive integer and $L_{max}>>L_{ss}$.\\
On simplifying Term -1  we will get
\begin{eqnarray}
\text{Term-1 :} & \sum_{j=0}^{L_{max}} j [  \{ \sum_{n=0}^{N_{max}}\frac{n}{N_{max}}P_N(n,t) \} e^{-C(j-1)} A  P_{L}(j-1,t)]  \nonumber \\  
&=[ \{ \sum_{n=0}^{N_{max}}\frac{n}{N_{max}}P_N(n,t) \}  A] \  [ \sum_{j=0}^{L_{max}} \{  j \  e^{-C(j-1)}  P_{L}(j-1,t) \} ] \nonumber \\  
&=[ \{ \sum_{n=0}^{N_{max}}\frac{n}{N_{max}}P_N(n,t) \}  A] \  [   \sum_{j=0}^{L_{max}} \{ (j-1+1) \  e^{-C(j-1)}  P_{L}(j-1,t) \} ]  \nonumber \\  
&=[ \{ \sum_{n=0}^{N_{max}}\frac{n}{N_{max}}P_N(n,t) \}  A] \  [   \sum_{j=0}^{L_{max}} \{ (j-1+1) \ \underbrace{ e^{-C(j-1)}}_{\text{Expand it}}  P_{L}(j-1,t) \} ] \nonumber \\  
&=[ \{ \sum_{n=0}^{N_{max}}\frac{n}{N_{max}}P_N(n,t) \}  A] \  [   \sum_{j=0}^{L_{max}} \{ (j-1+1) \ (1-C(j-1))  P_{L}(j-1,t) \} ] \nonumber \\  
&=[ \{ \sum_{n=0}^{N_{max}}\frac{n}{N_{max}}P_N(n,t) \}  A] \  [   \sum_{j=0}^{L_{max}} \{ (j-1) \ (1-C(j-1))  P_{L}(j-1,t) \} +\sum_{j=0}^{L_{max}} \{ \ (1-C(j-1))  P_{L}(j-1,t) \} ] \nonumber \\  
&=[ \{ \sum_{n=0}^{N_{max}}\frac{n}{N_{max}}P_N(n,t) \}  A] \  [    \{  \ (\langle L(t) \rangle -{C\langle {L(t)}^2 \rangle})   \} + \{ \ (1-{C\langle L(t) \rangle})  \} ] \nonumber \\  
\end{eqnarray}

Similarly, on simplifying Term -2  we will get

\begin{eqnarray}
\text{Term-2:}  \sum_{j=0}^{L_{max}} j [  \{ \sum_{n=0}^{N_{max}}\frac{n}{N_{max}}P_N(n,t) \} e^{-Cj} A  P_{L}(j,t)] 
=[ \{ \sum_{n=0}^{N_{max}}\frac{n}{N_{max}}P_N(n,t) \}  A] \  [    \{  \ (\langle L(t) \rangle -{C\langle {L(t)}^2 \rangle})   \}  ] \nonumber \\  
\end{eqnarray}

On simplifying Term -3  we will get
\begin{eqnarray}
\text{Term-3} : \sum_{j=0}^{L_{max}} j [\{ B \} P_{L}(j+1,t) - \{ B \} P_{L}(j,t)] \nonumber \\
=\sum_{j=0}^{L_{max}} [\{ B \} (j+1-1)P_{L}(j+1,t) - \{ B \}j P_{L}(j,t)] \nonumber \\
= [\{ B \} (\langle L \rangle -1)- \{ B \}(\langle L \rangle )] =-B
\end{eqnarray}

\begin{eqnarray}
\frac{d\langle L_1(t)\rangle}{dt}=\text{Term 1 -  Term 2 + Term 3  } \nonumber \\
=\underbrace{[ \{ \sum_{n=0}^{N_{max}}nP_N(n,t) \}  A] } \ \  \underbrace{ \ (1-{C\langle L \rangle})   } -B \nonumber \\  
=   \frac{ \langle N(t) \rangle }{N_{max}}  A     e^{-{C\langle L(t) \rangle }} -B   \nonumber \\  
\end{eqnarray}

Now let us consider the master equation for the precursor population at the pool given by equation (\ref{eq-masterN}). Multiplying both the sides with $n$ and summing it over, we get 
\begin{eqnarray}
\sum_{n=0}^{N_{max}}n \frac{dP_{N}(n,t)}{dt}&=\sum_{n=0}^{N_{max}} n \bigg{[} \underbrace{\omega^+(1-\frac{(n-1)}{N_{max}})P_N(n-1,t)-\omega^+(1-\frac{n}{N_{max}})P_N(n,t) }_{\text{Term-1}}    \nonumber\\ \nonumber\\ & +\underbrace{(n+1)P_{N}(n+1,t) \omega^- - nP_{N}(n,t) \omega^- }_{\text{Term-2}}\\ \nonumber 
&+\underbrace{P_N(n+1,t)[\sum_{j=0}^{L_{max}} \{ e^{-Cj} A  P_{L}(m,t) \}]-P_N(n,t)[\sum_{j=0}^{L_{max}} \{ e^{-Cj} A  P_{L}(m,t) \}]}_{\text{Term-3}}  \\ \nonumber 
&+\underbrace{ \{ B \} P_{N}(n-1,t) - \{B \} P_{N}(j,t) }_{\text{Term-4}} \bigg{]} \\ \nonumber 
\end{eqnarray}

On simplifying Term-1:

\begin{eqnarray}
&\sum_{n=0}^{N_{max}} n [ \omega^+ (1-\frac{(n-1)}{N_{max}})    P_{N}(n-1,t) -  \omega^+ (1-\frac{(n)}{N_{max}}) P_{N}(n,t) ] \nonumber \\
&=\sum_{n=0}^{N_{max}}  [(n-1+1) \omega^+ (1-\frac{(n-1)}{N_{max}})    P_{N}(n-1,t) -  \omega^+n (1-\frac{(n)}{N_{max}}) P_{N}(n,t) ] \nonumber \\
&=\sum_{n=0}^{N_{max}}  [\{ ( \langle N(t)\rangle + 1)-\frac{(\langle N^2(t)\rangle + \langle N(t)\rangle)}{N_{max}} \} \omega^+  - \{ \langle N(t)\rangle - \frac{\langle N^2(t)\rangle}{N_{max}} \} \omega^+ ] \nonumber \\
=[1-\frac{\langle N(t)\rangle}{N_{max}}]\omega^+
\end{eqnarray}

On simplifying Term-2:

\begin{eqnarray}
&\sum_{n=0}^{N_{max}} n [\omega^- (n+1)P_{N}(n+1,t)  - \omega^- nP_{N}(n,t) ] \nonumber \\
&=\sum_{n=0}^{N_{max}}  [\omega^-(n+1-1)(n+1)P_{N}(n+1,t)  -  \omega^- n^2P_{N}(n,t) ] \nonumber \\
&= \omega^- \langle N^2 (t)\rangle   - \omega^-\langle N (t) \rangle    -\omega^- \langle N^2 (t)\rangle  \nonumber \\
&=-  \omega^-\langle N(t) \rangle
\end{eqnarray}

On Simplifying Term-3: 
\begin{eqnarray}
\sum_{n=0}^{N_{max}} n ( \frac{(n+1)}{N_{max}} P_N(n+1,t)[\sum_{j=0}^{L_{max}} \{ e^{-Cj} A  P_{L}(j,t) \}]-\frac{(n)}{N_{max}}P_N(n,t)[\sum_{j=0}^{L_{max}} \{ e^{-Cj} A  P_{L}(j,t) \}]  )\\ \nonumber 
=\frac{1}{N_{max}}[\sum_{j=0}^{L_{max}} \{ e^{-Cj} A  P_{L}(j,t) \}] \  [\sum_{n=0}^{N_{max}} n \{  (n+1)P_N(n+1,t) -n P_N(n,t) \} ] \\ \nonumber 
=\frac{1}{N_{max}}[\sum_{j=0}^{L_{max}} \{ e^{-Cj} A  P_{L}(j,t) \}] \  [\sum_{n=0}^{N_{max}}  \{( n+1-1)  (n+1)P_N(n+1,t) -n^2 P_N(n,t) \} ] \\ \nonumber 
=\frac{1}{N_{max}}[\sum_{j=0}^{L_{max}} \{ e^{-Cj} A  P_{L}(j,t) \}] \  [\langle N^2 (t) \rangle - \langle N(t) \rangle -\langle N^2(t) \rangle  ] \\ \nonumber 
=-\frac{\langle  N(t)  \rangle}{N_{max}} \underbrace{  [\sum_{j=0}^{L_{max}} \{ e^{-Cj} A  P_{L}(j,t) \}]}
=-\frac{\langle  N(t)  \rangle}{N_{max}} \ e^{-{C \langle L(t) \rangle }} A  
\end{eqnarray}

On simplifying Term-4:
\begin{eqnarray}
\sum_{n=0}^{N_{max}} n [\{ B \} P_{N}(n-1,t) - \{ B \} P_{N}(n,t) ] \nonumber \\
= \{ B \} \sum_{n=0}^{N_{max}}[ (n-1+1)P_{N}(n-1,t) - n P_{N}(n,t) ]\nonumber \\
= \{ B \} [\langle N(t) \rangle + 1 -  \langle N(t)  \rangle ]
= B 
\end{eqnarray}

On collecting all the terms:
\begin{eqnarray}
\frac{d\langle N(t) \rangle}{dt}=[1-\frac{\langle N(t)\rangle}{N_{max}}]\omega^+-\langle N(t) \rangle  \omega^- -\frac{\langle  N(t) \rangle }{N_{max}}\ e^{-{C \langle L (t)\rangle }} A  +  B \nonumber\\=[1-\frac{\langle N(t)\rangle}{N_{max}}]\omega^+-\langle N(t) \rangle  \omega^- - \frac{d\langle L(t) \rangle}{dt}
\end{eqnarray}


\section{Steady state length distribution from master equation}
\label{app-SSlengthDistMaster}

In steady state, the probabilities become time dependent. So setting $\frac{dP_L(j,t)}{dt}=0$ for the master equations given in equation (\ref{master_eq_i}), we get system of $L_0$ linear equations 
\begin{subequations}
\begin{equation}
\mu^L_{1,0}P_L(1)-\lambda^L_{0,1}P_L(0)=0
\end{equation}
\begin{equation}
\lambda^L_{j-1,j}P_L(j-1)+\mu^L_{j+1,j}P_L(j+1)-(\lambda^L_{j,j+1}+\mu^L_{j,j-1})P_L(j) =0\
 \ \ \ \text{for $j$=1 to $L_{max}-1$} 
\end{equation}
\begin{equation}
\lambda^L_{L_{max}-1,L_{max}}P_L(L_{max}-1)-\mu^L_{L_{max},L_{max}-1}P_L(L_{max}) =0 .
\end{equation}
\label{master-eq-bd-ss}
\end{subequations}

Solving these coupled equations given by equation (\ref{master-eq-bd-ss}) recursively, we obtain all the probabilities $P(j)$ for $j>0$ in terms of P(0). For example, P(1) and P(2) in terms of P(0) are expressed as

\begin{subequations}
\begin{equation}
P_L(1)=\frac{\lambda^L_{0,1}}{\mu^L_{1,0}}P_L(0)
\end{equation}
\begin{equation}
P_L(2)=\frac{(\lambda^L_{1,2}+\mu^L_{1,0})}{\mu^L_{2,1}}P_L(1)-\frac{\lambda^L_{0,1}}{\mu^L_{2,1}}P_L(0)=\frac{(\lambda^L_{1,2}+\mu^L_{1,0})}{\mu^L_{2,1}}\frac{\lambda^L_{0,1}}{\mu^L_{1,0}}P_L(0)-\frac{\lambda^L_{0,1}}{\mu^L_{2,1}}P_L(0)=\frac{\lambda^L_{1,2}\lambda^L_{0,1}}{\mu^L_{2,1} \mu^L_{1,0}}P_L(0)
\end{equation}
\end{subequations}

It can be shown by method of induction that
\begin{equation}
P_L(j)=P_L(0)\prod_{j'=0}^j\frac{\lambda^L_{j-1,j}}{\mu^L_{j,j-1}}
\label{Pj}
\end{equation}

Substituting the expressions of $P(j)$, which are written in terms of $P(0)$, into to the following normalization condition 
\begin{equation}
\sum_{j=0}^{L_{max}}P_L(j)=1
\end{equation}
we solve for $P_L(0)$ which turns out to be a function of the intensities $\lambda^L_{j,j+1}$ and $\mu^L_{j,j+1}$ ($j=0,1,...,L_{max}$)
and is given by
\begin{equation}
P_L(0)=\bigg{[}(1+\sum_{j=0}^{L_{max}} \prod_{i=1}^j \frac{\lambda^L_{i-1,i}}{\mu^L_{i,i-1}})\bigg{]}^{-1}.
\label{P1}
\end{equation}

Hence, substituting the expression of for $P_L(0)$ obtained in equation (\ref{P1}) in the formula for $P_L(j)$ given by equation (\ref{Pj}), all the $P_L{(j)}$ can be  expressed in  terms of transition rates $\lambda^L_s$ and $\mu^L_s$.

\section{Steady state length distribution from Fokker-Planck equation} 
\label{app-SSlengthDistFP}

Carrying out a standard Kramer-Moyal expansion of the master equation, we get the Fokker-Planck equation
\begin{eqnarray}
\frac{\partial P(x,t)}{\partial t}=-\frac{\partial}{\partial x}[ f(x)  P(x,t)]+ \frac{\Delta L}{2} \cdot  \frac{\partial^2}{\partial x^2}[ g(x)  P(x,t)]
\end{eqnarray}
with 
\begin{equation}
f(x)= \lambda(x) - \mu(x)
\end{equation}
and
\begin{equation}
g(x)=\lambda(x) + \mu(x) 
\end{equation}
where 
\begin{eqnarray} 
\lambda(x)&=&\frac{\langle N(t) \rangle}{N_{max}}e^{-2kx/v}J \Omega_e \nonumber \\
\mu(x)&=&(1-\rho)^2 \Gamma_r.
\end{eqnarray}

In terms of $A=J \Omega_e$ $\langle N(t) \rangle/{N_{max}}$,  $B=(1-\rho)^2\Omega_r$, and $C =\frac{2k}{v}$  $f(x)$ and $g(x)$ can be re-written as:
\begin{eqnarray}
f(x)=Ae^{-Cx}-B \nonumber\\
g(x)=Ae^{-Cx}+B
\end{eqnarray}

The steady state solution of the Fokker Planck equation $P_{ss}(x)$ is given by:
\begin{equation}
P_{ss}(x)=\mathcal{C}_0 \frac{ \ e^{-\Phi(x)}}{g(x)} 
\label{eq-Pss}
\end{equation}
where 
\begin{equation}
\Phi(x)=-\frac{2}{\Delta L} \int_0^{x}\frac{f(x')}{g(x') } \ dx'=\frac{2}{\Delta L}\bigg{[}\frac{2}{C} \ \ell og  \bigg{(} \frac{ \  {A}+{B} e^{{C} {x}}} {{A}+{B}}\bigg{)}-x\bigg{]}
\label{phix}
\end{equation}
and the normalization constant $\mathcal{C}_0$ is given by
\begin{eqnarray}
&\mathcal{C}_0 =\bigg{[}{\int_0^{L_m} \frac{ \ e^{-\Phi(x')}}{g(x')} \ dx'}\bigg{]}^{-1}\nonumber\\
&=\bigg{[}{A (C +\frac{2}{{\Delta L}})}\bigg{]}^{-1} 
\bigg{[}e^{{L_m} \left(C+{2}/{{\Delta L}}\right)} \left(\frac{A+B e^{C {L_m}}}{A+B}\right)^{-{4}/{(C {\Delta L})}} \, _2F_1\left(\sigma_a,\sigma_b;\sigma_c;\sigma_{d1}\right)-\, _2F_1\left(\sigma_a,\sigma_b;\sigma_c;\sigma_{d2}\right)  \bigg{]}
\label{Const}
\end{eqnarray}
where $_2F_1(a ,b ; c ;z)$ represents the Gauss hypergeometric function and \\ \\ $\sigma_a=1, \ \sigma_b=\frac{C-2}{C}; \ \sigma_c=\frac{2}{C}+2;\ \sigma_{d1}=-\frac{B}{A}, \ \sigma_{d2}=-\frac{B}{A}e^{C {L_m}}$. \\

Substituting the expressions (\ref{Const}) and (\ref{phix}) for $\mathcal{C}_0$ and $\Phi(x)$, respectively, into the expression (\ref{eq-Pss}) for $P_{ss}(x)$, we get 
\begin{eqnarray}
& P_{ss}(x)=\bigg{[}{A (C +\frac{2}{{\Delta L}})}\bigg{]} \bigg{[}\frac{1}{A+B} \ e^{[{x \left(C+{2}/{{\Delta L}}\right)}]} \left(\frac{A+B e^{C x}}{A+B}\right)^{-{4}/{(C {\Delta L})}-1}\bigg{]} \bigg{/} \nonumber\\
&\bigg{[}e^{{L_0} \left(C+{2}/{{\Delta L}}\right)} \left(\frac{A+B e^{C {L_m}}}{A+B}\right)^{-{4}/{(C {\Delta L})}} \, _2F_1\left(\sigma_a,\sigma_b;\sigma_c;\sigma_{d1}\right)-\, _2F_1\left(\sigma_a,\sigma_b;\sigma_c;\sigma_{d2}\right)  \bigg{]}
\end{eqnarray}

Hence, 
\begin{eqnarray}
\langle x \rangle_{ss}&=\frac{1}{\mathcal{C}_0}\frac{{\Delta L}}{A (C {\Delta L}+2)^2}
\bigg{[}{\Delta L} \left(\frac{A+B}{A}\right)^{{4}/{(C {\Delta L)}}}  {_3F_2}\left(\eta_a,\eta_a,\eta_b;\eta_c,\eta_c;\eta_{d1} \right)  
\nonumber\\
&+e^{{L_m} \left(C+{2}/{\Delta L}\right)}(A+B)^{{4}/{(C \Delta L)}} \left(A+B e^{C {L_m}}\right)^{-{4}/{(C \Delta L)}} \left(1+\frac{B e^{C {L_m}}}{A}\right)^{{4}/{(C {\Delta L})}}
\nonumber\\
&\bigg{ \{ }{L_m} (2+C {\Delta L}) \, _2F_1\left(\eta_a,\eta_b;\eta_c;\eta_d\right)
-{\Delta L} {_3F_2}\left(\eta_a,\eta_a,\eta_b;\eta_c,\eta_c;\eta_{d2} \right)  
\bigg{\}}\bigg{]}
\end{eqnarray}
and
\begin{eqnarray}
&\langle x^2 \rangle_{ss}  =[\frac{1}{\mathcal{C}_0}]\frac{{\Delta L}}{A (C {\Delta L}+2)^3}\bigg{[}-2 {\Delta L}^2 \left(\frac{A+\text{B}}{A}\right)^{{4}/{(C {\Delta L})}} \,_4F_3\left(\eta_a,\eta_a,\eta_a,\eta_b;\eta_c,\eta_c,\eta_c;\eta_{d1}\right)\nonumber\\
&+e^{{L_m} \left(C+{2}/{{\Delta L}}\right)} (A+{B})^{{4}/{(C {\Delta L})}} \left(A+{B} e^{C {L_m}}\right)^{-{4}/{(C {\Delta L})}} \left(\frac{{B} e^{C {L_m}}}{A}+1\right)^{{4}/{(C {\Delta L})}} \nonumber\\
& \bigg{\{ } {L_m}^2 (C {\Delta L}+2)^2  _2F_1\left(\eta_a,\eta_b;\eta_c;\eta_{d2}\right)-2 {\Delta L} \ {L_m} (C {\Delta L}+2) \ \, _3F_2\left(\eta_a,\eta_a,\eta_b;\eta_c,\eta_c;\eta_{d2}\right) \nonumber\\ 
& + 2 {\Delta L}^2 \, _4F_3\left(\eta_a,\eta_a,\eta_a,\eta_b;\eta_c,\eta_c,\eta_c;\eta_{d2}\right) \bigg{\} } \bigg{] }
\end{eqnarray}
where $\eta_a=\frac{2}{C}+1, \ \eta_b=\frac{4}{C}+1, \ \eta_c=\frac{2}{C}+2, \ \eta_{d1} = -\frac{B}{A} $ and $\eta_{d2}= -\frac{B}{A} e^{C {L_m}}$

\section{Timescales}
\label{sec-Timescales}

In terms of A,B and C, the coupled differential equations become :
\begin{eqnarray}
&\frac{d[L(t)]}{dt}&=\frac{N(t)}{N_{max}}A \ e^{-C L(t)}-B \nonumber\\
 \nonumber\\
&\frac{d[N(t)]}{dt}&=\omega^+\bigg{[} 1-\frac{N(t)}{N_{max}} \bigg{]}-\omega^-N(t)-\frac{d[L(t)]}{dt}
\end{eqnarray}
where $A=J \Omega_e$,  $B=(1-\rho)^2 \Gamma_r$, and $C=2k/v$. \\
First we need to calculate the fixed point for the system. The system has one fixed point ($L^*,N^*$), 
given by
\begin{equation}
L^*=\frac{1}{C} \ell og \bigg{[} \frac{A}{B} \ \frac{\omega^+}{\omega^-+{(\omega^+/{N_{max}})}}\bigg{]}
\end{equation}
and
\begin{equation}
N^*=\frac{\omega^+}{\omega^-+{(\omega^+/{N_{max}})}},
\end{equation}
which is the steady state of the system. 
Let $\frac{d[L(t)]}{dt}=f_L(L,N)$ and  $\frac{d[N(t)]}{dt}=f_N(L,N)$. 

Introducing the matrix 
\begin{equation}
\left(
\begin{array}{cc}
 \frac{\partial f_L}{\partial L}  & \frac{\partial f_L}{\partial N}  \\
      \frac{\partial f_N}{\partial L}  & \frac{\partial f_N}{\partial N}  \\
\end{array}
\right)=\left(
\begin{array}{cc}
 -AC  e^{-CL}  N & A e^{-C \ L} \\
 AC  e^{-CL}  N  & -A e^{-C L}-\omega^--\frac{\omega^+}{N_{max}} \\
\end{array}
\right)
\end{equation}
and evaluating its elements at ($L^*,N^*$), we get
\begin{equation}
\left(
\begin{array}{cc}
 -C \ B& \  B({{\frac{\omega^-}{\omega^+}+\frac{1}{N_{max}}}}) \\
 C \ B & -B({{\frac{\omega^-}{\omega^+}+\frac{1}{N_{max}}}})-{\omega^-} -\frac{\omega^+}{N_{max}}\\
\end{array}
\right)
\label{mat-matrix}
\end{equation}
The matrix (\ref{mat-matrix})has two eigenvalues ($\lambda_+, \lambda_-$) and two corresponding eigenvectors ($V_+,V_-$); the eigenvalues are
\begin{eqnarray}
\lambda_{\pm}=\frac{1}{2}\bigg{[}-\bigg{ \{ } B({{\frac{\omega^-}{\omega^+}+\frac{1}{N_{max}}}}) +{\omega^-} +\frac{\omega^+}{N_{max}} +BC \bigg{ \} }  \nonumber\\  \pm   \sqrt{ \bigg{ \{ } B({{\frac{\omega^-}{\omega^+}+\frac{1}{N_{max}}}}) +{\omega^-} +\frac{\omega^+}{N_{max}} +BC \bigg{ \} }^2-4BC (\omega^-+\frac{\omega^+}{N_{max}})}  \bigg{]}
\label{EV_e}
\end{eqnarray}
and the corresponding eigenvectors $V_{\pm}$ are 
\begin{eqnarray}
\left(
\begin{array}{cc}
\frac{1}{2BC}\bigg{[} B({{\frac{\omega^-}{\omega^+}+\frac{1}{N_{max}}}}) +{\omega^-} +\frac{\omega^+}{N_{max}} -BC    \pm   \sqrt{ \bigg{ \{ } B({{\frac{\omega^-}{\omega^+}+\frac{1}{N_{max}}}}) +{\omega^-} +\frac{\omega^+}{N_{max}} +BC \bigg{ \} }^2-4BC (\omega^-+\frac{\omega^+}{N_{max}})}  \bigg{]}\\
1\\
\end{array}
\right) \nonumber\\
\end{eqnarray}
To simplify the expressions, that also helps in more transparent physical interpretations, we introduce the symbols 
\begin{eqnarray}
&\zeta_1 &=-B({{\frac{\omega^-}{\omega^+}+\frac{1}{N_{max}}}})-{\omega^-} -\frac{\omega^+}{N_{max}}\nonumber\\
&\zeta_2 &= BC\nonumber\\
&\zeta_3 &=4BC ( \omega^-+\frac{\omega^+}{N_{max}})
\end{eqnarray}
In terms of $\zeta_1, \zeta_2$ and $\zeta_3$, the eigenvalues and eigenvectors can be recast as 
\begin{equation}
\lambda_{\pm}=\frac{1}{2}\bigg{[}-(\zeta_1+\zeta_2)\pm \{ (\zeta_1+\zeta_2)^2-\zeta_3 \} \bigg{]}
\end{equation}
and
\begin{equation}
V_{\pm}= \left(
\begin{array}{cc} \frac{1}{2\zeta_2}\bigg{[}(\zeta_1-\zeta_2)\pm \{ (\zeta_1+\zeta_2)^2-\zeta_3 \} \bigg{]}\\
1\\
\end{array}
\right)
\end{equation}

Note that the dependence of $\lambda_{\pm}$ on the three parameters $\omega^{+}$, $\omega^{-}$ and $N_{max}$, which together characterize the population kinetics of the precursor pool, have been shown explicitly in (\ref{EV_e}). The composite parameters $A$, $B$ and $C$, as stated before, characterize the flagellar elongation, timer relaxation and flagellar shrinkage, respectively. The inverse of these two eigenvalues indicate the two timescales of relaxation of small excursions away from the steady-state. Since $\lambda_-$ is the larger of the two eigenvalues, the associated timescales $\tau_{\pm} = 1/ \lambda_{\pm}$ satisfy $\tau_+>\tau_-$.

\begin{figure}
\begin{center}
\includegraphics[width=0.40\textwidth]{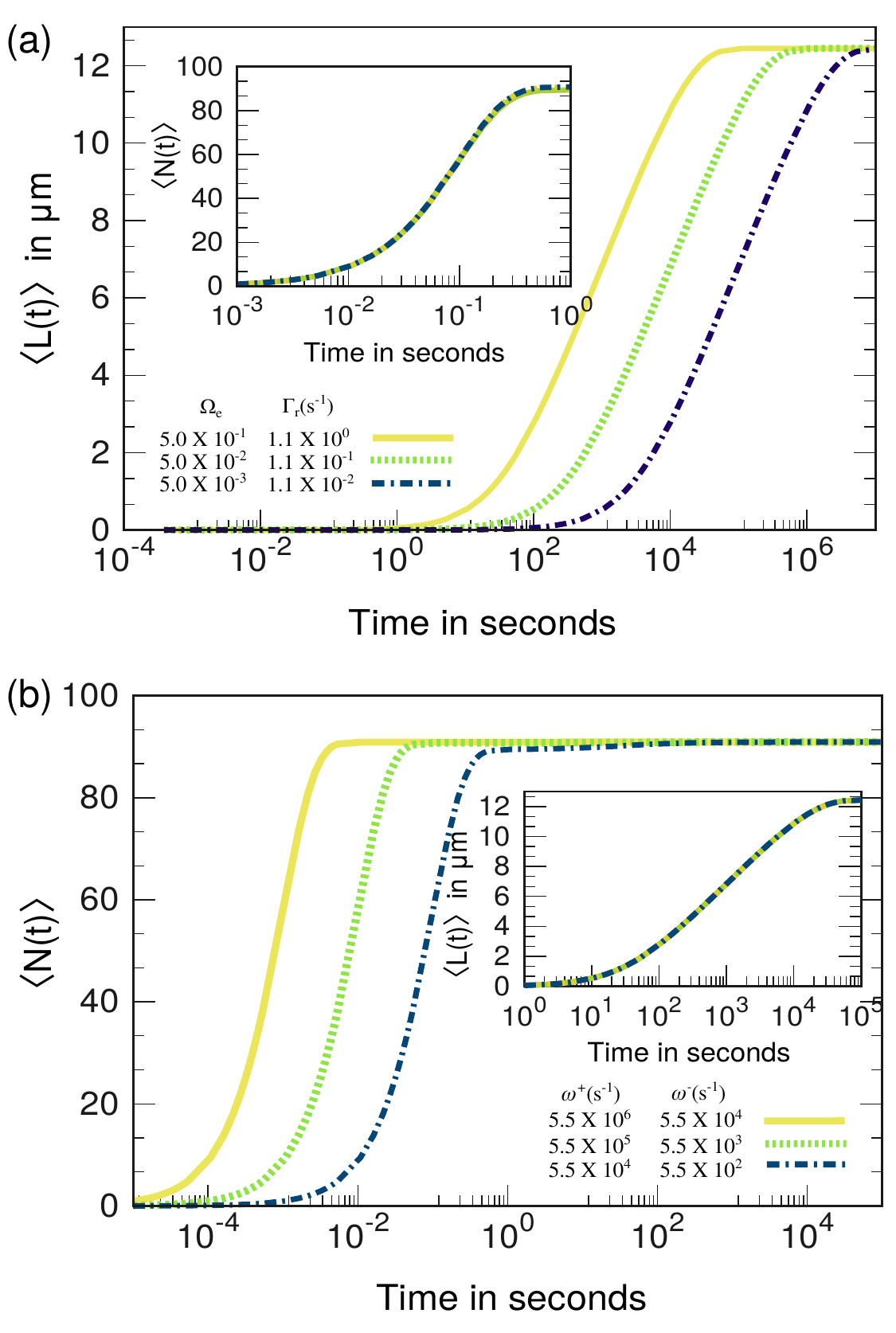}
\end{center}
\caption{ \textbf{Timescales:} (a)$\langle L(t) \rangle$ vs $t$. Inset: $\langle N(t) \rangle$ vs $t$. (b)$\langle N(t) \rangle$ vs $t$. Inset: $\langle L(t) \rangle$ vs $t$. \\
Parameters:
(a) $\rho=0.1$,  $J=0.09$, $v=0.9$, $k=1.8 \times 10^{-3}$, $\omega^+=0.5$, $\omega^-=5.0\times 10^{-3}$, $N_{max}=1000$, $\delta t=9.0 \times 10^{-6} s$, $L(0)=N(0)=0$ \\
(b) $\rho=0.1$,  $J=0.09$, $v=0.9$, $k=1.8 \times 10^{-3}$, $\Omega_e=0.5$, $\Gamma_r=1.0\times 10^{-5}$, $N_{max}=1000$, $\delta t=9.0 \times 10^{-6} s$, $L(0)=N(0)=0$}
\label{EV}
\end{figure}

\begin{figure}
\begin{center}
\includegraphics[width=0.40\textwidth]{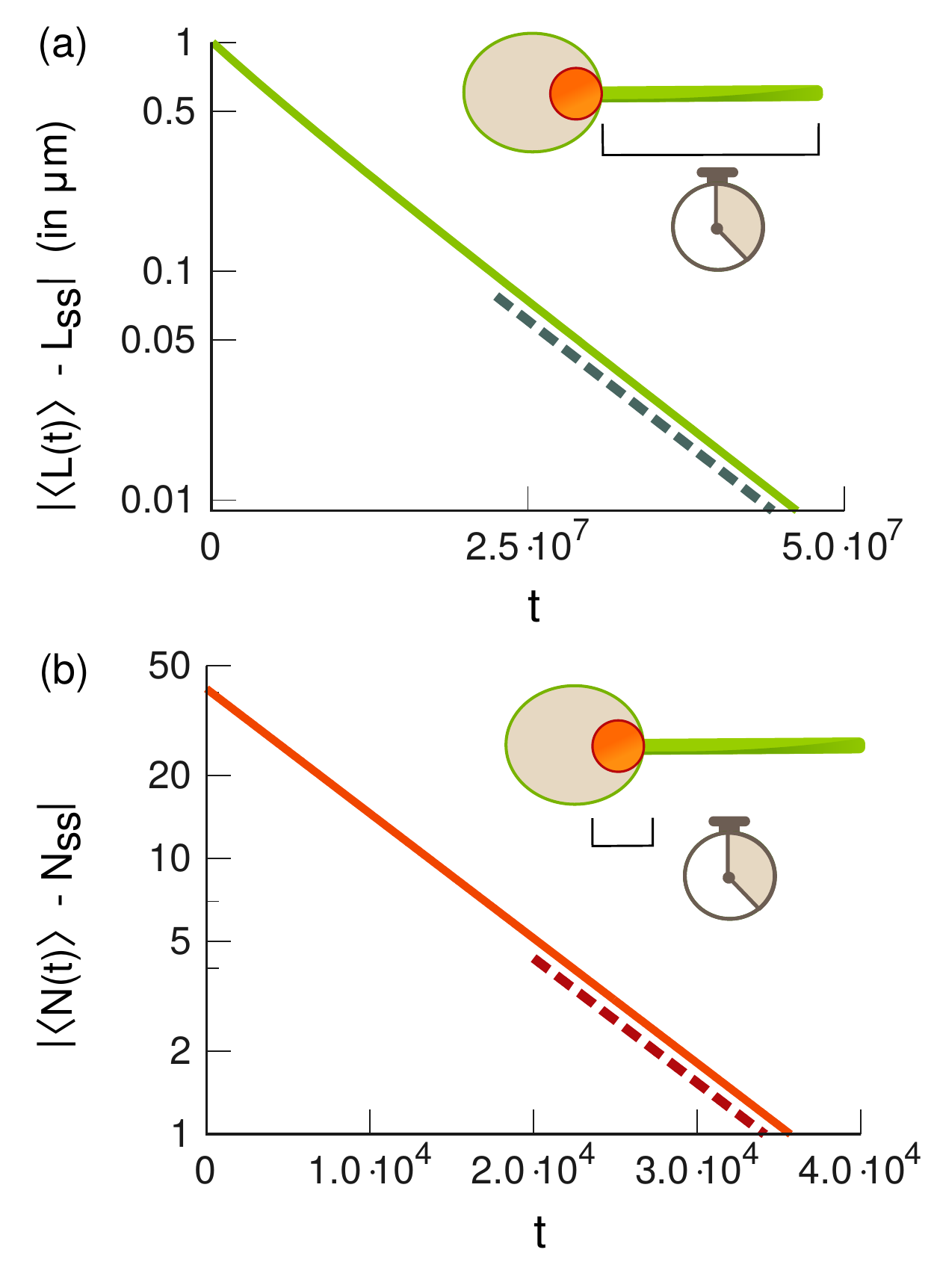}
\end{center}
\caption{ \textbf{Slight Deviation from fixed point:} \\ (a) Solid line - $|\langle L(t) \rangle-L_{ss}|$ vs $t$. Dashed line - $C^{L}exp[-\lambda_+t]$.
\\ (b) Solid line - $|\langle N(t) \rangle-N_{ss}|$ vs $t$. Dashed line - $C^{N}exp[-\lambda_-t]$. \\
where $C^L$ and $C^N$ are constants.
  \\ Parameters:\\ $\rho=0.1$,  $J=0.09$, $v=0.9$, $k=0.0011$, $\Omega_e=0.5$, $\Gamma_r=5.0\times 10^{-5}$, $\omega^+=2.0\times 10^{-2}$, $\omega^-=10^{-4}$\\
Other quantities: \\ $L_{ss}$=12.28 $\mu$m, $N_{ss}=191$ (for both (a) and (b)) \\
Initial Conditions: \\
(a) $|\langle L(t=0) \rangle-L_{ss}|$=1.0 $\mu$m and $\langle N(t=0) \rangle=N_{ss}$ \\
(b)$|\langle N(t=0) \rangle-N_{ss}|$=50 and $\langle L(t=0) \rangle=L_{ss}$ }
\label{EV_II}
\end{figure}

In figure \ref{EV}(a), we plot the $\langle L(t) \rangle $ vs $t$ for different pairs of $\Omega_e$ and $\Gamma_r$. Note that $\langle L_{ss} \rangle$ remain same for all the different sets of parameters because the ratio of $\Omega_e$ and $\Gamma_r$ is kept same for all the cases. It can be checked from the expression of eigenvalues from equation (\ref{EV_e}) that the the timescale mainly depend on $\omega^-$ and $BC$. In figure \ref{EV}(a), we are vary $B$ over three decades by changing $\Gamma_r$ over three decades and we observe that the time in which the $\langle  L(t) \rangle $ mature vary by three decades as well. 

Substituting the parameters used to plot figure \ref{EV}(a) into the expressions (\ref{EV_e}) for $\lambda_{\pm}$, we observe that $\lambda_+$ varies by three orders for the three different cases whereas $\lambda_-$ remains practically unchanged. It indicates that $\lambda_+$ determines the timescale associated with the flagellar length $L(t)$. Moreover, from the inset of figure \ref{EV} (a), we see that the timescale in which $N(t)$ attains $\langle N_{ss} \rangle$ is very small compared to the timescale in which $L(t)$ attain $\langle L_{ss} \rangle$. Besides, the curves for $\langle N_{ss} \rangle$ corresponding to the three sets of parameter values are almost identical. 
Hence, in this case, $\langle L(t) \rangle$ is given by the approximate expression (\ref{eq-expL}) which clearly shows that, in this limit, the timescale in which $\langle L(t) \rangle$ attains $\langle L_{ss} \rangle$ is $\tau=1/(BC)$. For consistency, we have also extracted the numerical value of $\lambda_+$ for all the three cases plotted in figure \ref{EV} (a) and found it to be, indeed, approximately equal to $[BC]$. Hence, we conclude that $1/\lambda_+$ governs the timescale in which length $\langle L(t) \rangle$ attains $\langle L_{ss} \rangle$.

Similarly, to understand the timescale with which $\langle N(t) \rangle$ approach $N_{ss}$, we plot $\langle N(t) \rangle$ for different values of $\omega^+$ and $\omega^-$ in figure \ref{EV}(b). However, as the ratio of $\omega^+$ and $\omega^-$ is kept same in all the three cases, $N_{ss}$ is also same for all the cases. As we vary $\omega^-$ over three decades, the time over which $\langle N(t) \rangle$ attain steady state $N_{ss}$ varies over three decades (figure \ref{EV} (b)) while the corresponding $\langle L(t) \rangle$ attains steady state value $\langle L_{ss}\rangle$ in same time interval irrespective of the time taken by the pool to achieve steady value (see figure \ref{EV}(b) inset). 

To understand this observation, we computed the $\lambda_{\pm}$ using the corresponding parameter values used to plot figure \ref{EV}(b). We observed $\lambda_-$ varies by three orders for the three different cases whereas $\lambda_+$ remains unchanged. For plotting figure \ref{EV}(b), $\Gamma_r$ is very small compared to $\omega^+$ or $\omega^-$ (see the caption of figure \ref{EV}). Therefore, under the approximation $B \simeq 0$, the formula for eigenvalues (\ref{EV}), can be approximated as 
\begin{eqnarray}
\lambda_{-}=[\omega^-+\frac{\omega^+}{N_{max}}]
\end{eqnarray}
Hence, we conclude that the timescale associated with the $N(t)$ dynamics is $\tau_-=1/\lambda_- \approx 1/[\omega^-+(\omega^+/N_{max})]$.

Finally, for futher check of consistency, we perturbed the system slightly awayfrom the fixed point (i.e., from the steady state) and observing how the perturbations died out with time. In the first case (see Fig.\ref{EV_II}(a)), we have monitored the relaxation of the initial state 
$L(0) = \langle L_{ss} \rangle - \Delta L(0), N(0)=\langle N_{ss} \rangle$; the  slope of the straight line on the semi-log plot is, indeed, $\lambda_{+}$. Similarly, in the second case, we chose the initial condition 
$L(0) = \langle L_{ss} \rangle, N(0)=\langle N_{ss} \rangle - \Delta N(0)$; the slope of the straight line on the semi-log plot in Fig.\ref{EV_II}(b) is also found to be $\lambda_{-}$.

\section{Master equations for biflagellates}
\label{appendix-masterBIFLAG}

The master equations are given by  
\begin{widetext}
\begin{eqnarray}
 \frac{dP_{L_i}(j,t)}{dt}& =& 
  \underbrace{[e^{-2k(j-1)/v} \sum_{n=0}^{N_{max}}\frac{n}{N_{max}}P_N(n,t) ]J \Omega_e \  P_{L_i}(j-1,t)  -[e^{-2kj/v} \sum_{n=0}^{N_{max}}\frac{n}{N_{max}}P_N(n,t) ]J \Omega_e \  P_{L_i}(j,t)}_{\text{Probabilitic assemby of the flagellar tip by the flux of full IFT particles }} \nonumber\\ \nonumber\\
 &+&\underbrace{(1-\rho)^2\Gamma_r P_{L_i}(j+1,t)-(1-\rho)^2\Gamma_r P_{L_i}(j,t)}_{\text{Stochastic disassembly of the tip when not occupied by any IFT particle  }}
\ \ \ \ \text {where $i$=1,2}
\end{eqnarray}
and the master equation governing the population of precursors in the pool is 
\begin{eqnarray}
& \frac{dP_N(n,t)}{dt}  = \nonumber\\ \nonumber\\
& \underbrace{\omega^+(1-\frac{(n-1)}{N_{max}})P_N(n-1,t)-\omega^+(1-\frac{n}{N_{max}})P_N(n,t)}_{\text{Population dependent synthesis of flagella precursor by the cell .}}  \nonumber\\ \nonumber\\
& \underbrace{+[\omega^-{(n+1)})P_N(n+1,t)-\omega^-({n})P_N(n,t)]}_{\text{Population dependent degradation of flagella precursor by the cell.}}   \nonumber\\ \nonumber\\
&\underbrace{+[\sum_{j=0}^{L_{max}}J \Omega_e e^{-2k j/v}P_{L_1}(j,t)][\frac{(n+1)}{N_{max}} P_N(n+1,t)-\frac{(n)}{N_{max}}P_N(n,t)]}_{\text{Contribution of pool towards assembly of the first flagellum. }} \nonumber\\ \nonumber\\
&\underbrace{+[\sum_{j=0}^{L_{max}}J \Omega_e e^{-2k j/v}P_{L_2}(j,t)][\frac{(n+1)}{N_{max}} P_N(n+1,t)-\frac{(n)}{N_{max}}P_N(n,t)]}_{\text{Contribution of pool towards assembly of the second flagellum. }} \nonumber\\ \nonumber\\
&  \underbrace{+2(1-\rho)^2 \Gamma_r[P_N(n-1,t)-P_N(n,t)]}_\text{Addition of the precursor back to the pool when chipped from the tip of both the flagella during their disassembly.  }
\end{eqnarray}

\end{widetext}

\section{Steps for simulating the model}
\label{appendix-MCsteps}
We simulate our model using Monte Carlo methods. At a given instant of time $t$ let the flagellar length be denoted by $L(t)$ and the pool population by $N(t)$. $L(t)$ can take discrete values  $j=0,...,L_{max}$ and $N(t)$ can take discrete values  $n=0,...,N_{max}$. We have chosen $L_{max}>>L_{ss}$.

At each Monte-Carlo time step, we update the values of $L(t)$ and $N(t)$ according to the following rules:\\

\textbf{Updating the flagellar length:} We generate a random number $rn$ between 0 and 1. If at the current time step the flagellar  length is $L(t)=j$, we update the length to $L(t)=j+1$ if $rn<\lambda_{j,j+1}$ or update the length to $L(t)=j-1$ if $\lambda_{j,j+1}<rn<\lambda_{j,j+1}+\mu_{j,j+1}$. While updating the length to $L(t)=j$ to $L(t)=j+1$ we update the value of $N(t)$ from $N(t)=n$ to $N(t)=n-1$. If there is no precursor in the pool ($N(t)=0$), the length cannot be increased. Similarly, while updating the length to $L(t)=j$ to $L(t)=j-1$ we update the value of $N(t)$ from $N(t)=n$ to $N(t)=n+1$.

\textbf{Updating the pool population:}
We generate a random number $rn$ between 0 and 1. If at the current time step the pool population is $N(t)=n$, we update the pool population to $N(t)=n+1$ if $rn<[\omega^+(1-n/N_{max})]$ or update the pool population to $N(t)=n-1$ if $[\omega^+(1-n/N_{max})]<rn<[\omega^+(1-n/N_{max})+\omega^-n] $.

\end{widetext}


 \end{document}